%Paper: hep-th/9406055
%From: klemm@nxth04.cern.ch
%Date: Thu, 9 Jun 94 19:13:45 +0100
%Date (revised): Sun, 10 Jul 94 17:23:49 +0100

%%%%%%%%%%%%%%%%%%%%%%%%%%%%%%%%%%%%%%%%%%%%%%%%%%%%%%%%%
%      Mirror Symmetry, Mirror Map and Applications
%       to Complete Intersection Calabi-Yau Spaces
%
%                          by
%
%      S. Hosono, A. Klemm, S.Theisen and S. T. Yau
%
%      Program INSTANTON (1.0) appended at the end of the
%      TeXfile.
%
%%%%%%%%%%%%%%%%%%%%%%%%%%%%%%%%%%%%%%%%%%%%%%%%%%%%%%%%%
\input harvmac
\baselineskip=12pt

\def\frac#1#2{{#1\over#2}}
\def\coeff#1#2{{\textstyle{#1\over #2}}}

\def\journal#1&#2(#3){\unskip, #1~\bf #2 \rm(19#3) }
\def\andjournal#1&#2(#3){\sl #1~\bf #2 \rm (19#3) }

\def\det{{\rm det}}
\def\exp{{\rm exp}}

\def\lim{{\rm lim}}
\def\log{{\rm log}}
\def\p{\partial}

\catcode`\@=11
\def\slash#1{\mathord{\mathpalette\c@ncel{#1}}}
\overfullrule=0pt
\def\steepslash{\c@ncel}
\def\frac#1#2{{#1\over #2}}

\def\inbar{\,\vrule height1.5ex width.4pt depth0pt}
\def\IQ{\relax\,\hbox{$\inbar\kern-.3em{\rm Q}$}}
\def\IB{\relax{\rm I\kern-.18em B}}
\def\IC{\relax\hbox{$\inbar\kern-.3em{\rm C}$}}
\def\IP{\relax{\rm I\kern-.18em P}}
\def\IR{\relax{\rm I\kern-.18em R}}
\def\ZZ{\relax\ifmmode\mathchoice
{\hbox{Z\kern-.4em Z}}{\hbox{Z\kern-.4em Z}}
{\lower.9pt\hbox{Z\kern-.4em Z}}
{\lower1.2pt\hbox{Z\kern-.4em Z}}\else{Z\kern-.4em Z}\fi}

\catcode`\@=12

%                      Zeitschriften:
\def\npb#1(#2)#3{{ Nucl. Phys. }{B#1} (#2) #3}
\def\plb#1(#2)#3{{ Phys. Lett. }{#1B} (#2) #3}
\def\pla#1(#2)#3{{ Phys. Lett. }{#1A} (#2) #3}
\def\prl#1(#2)#3{{ Phys. Rev. Lett. }{#1} (#2) #3}
\def\mpla#1(#2)#3{{ Mod. Phys. Lett. }{A#1} (#2) #3}
\def\ijmpa#1(#2)#3{{ Int. J. Mod. Phys. }{A#1} (#2) #3}
\def\cmp#1(#2)#3{{ Commun. Math. Phys. }{#1} (#2) #3}
\def\cqg#1(#2)#3{{ Class. Quantum Grav. }{#1} (#2) #3}
\def\jmp#1(#2)#3{{ J. Math. Phys. }{#1} (#2) #3}
\def\anp#1(#2)#3{{ Ann. Phys. }{#1} (#2) #3}
\def\prd#1(#2)#3{{ Phys. Rev.} {D\bf{#1}} (#2) #3}

\def\a{\alpha}
\def\b{\beta}
\def\d{\delta}

\def\cL{{\cal L}}

\def\{{\lbrace}
\def\}{\rbrace}
\def\({\lbrack}
\def\){\rbrack}
\def\t1{\theta_1}
\def\t2{\theta_2}
\def\t3{\theta_3}
\def\t4{\theta_4}
\def\t5{\theta_5}

\def\tK{{\tilde K}}

\def\ta#1{ \theta_{a_{#1}} }

\overfullrule=0pt

\def\cicy#1(#2|#3)#4{\left(\matrix{#2}\right|\!\!
                     \left|\matrix{#3}\right)^{{#4}}_{#1}}
\def\pa#1{{\partial\over\partial\rho_{#1}}}

\def\ds{\displaystyle}

%%%%%%%%%%%%%%%%%%%%Hosono macros%%%%%%%%%%%%%%%%%%%

\def\b{\beta}
\def\d{\delta}

\def\t{\theta}

\def\{{\lbrace}
\def\}{\rbrace}
\def\Blb{\Bigl\{}
\def\Brb{\Bigr\}}

\def\ninn{\noindent}

\def\ta#1{\theta_{a_#1}}
\def\la#1{l_A^{(#1)}}
\def\a#1{a^{{#1}}}

\footline{\hss\tenrm--\folio\--\hss}
\rightline{HUTMP-94/02}
\rightline{CERN-TH.7303/94}
\Title{LMU-TPW-94-03}
{\vbox{ \centerline{Mirror Symmetry, Mirror Map and Applications}
        \vskip2pt
        \centerline{to Complete Intersection Calabi-Yau Spaces} } }

\centerline{
S. Hosono$^{\dagger}$,
A. Klemm$^{*\triangle}$,
S.Theisen$^{\diamond}$ and S.-T. Yau$^*$\footnote{}
{E-mail: hosono@nqs07.ccis.toyama-u.ac.jp, klemm@nxth21.cern.ch,
theisen@mppmu.mpg.de and yau@harvard.math.edu}}

\bigskip

\centerline{$^\dagger$Department of Mathematics, Toyama University}
\centerline{Toyama 930, Japan}
\medskip
\centerline{$^\triangle$CERN, TH-Division, CH-1211 Geneva 23}
\medskip
\centerline{$^*$Department of Mathematics, Harvard University}
\centerline{ Cambridge, MA 02138, USA }
\medskip
\centerline{$^{\diamond}$Sektion Physik der Universit\"at M\"unchen}
\centerline{Theresienstra\ss e 37, D - 80333 Munich, FRG}
\vskip .5in

\centerline{\bf Abstract:}
\ninn
We extend the discussion of mirror symmetry, Picard-Fuchs equations,
instanton-corrected Yukawa couplings, and the topological
one-loop partition function to the case of complete intersections
with higher-dimensional moduli spaces. We will develop a new method
of obtaining the instanton-corrected Yukawa couplings through a study
of the solutions of the Picard-Fuchs equations. This leads to
closed formulas for the prepotential for the K\"ahler moduli fields
induced from the ambient space for all complete intersections in
non singular weighted projective spaces. As examples we treat part
of the moduli space of the phenomenologically interesting
three-generation models that are found in this class. We also apply
our method to solve the simplest model in which a topology change was observed
and discuss examples of complete intersections in singular ambient
spaces.

\Date{CERN-TH.7303/June 1994}

\vfill\eject

%%%%%%%%%%%%%%%%%%%%%%%%%%%%%%%%%%%%%%%%%%%%%%%%%%%
\newsec{Introduction:}
%%%%%%%%%%%%%%%%%%%%%%%%%%%%%%%%%%%%%%%%%%%%%%%%%%%

Complete intersection Calabi-Yau manifolds embedded in
products of projective spaces CICYs are the most
prominent candidates for the compactification of the heterotic string.
They have been intensively studied.
Let us point out briefly the main results;
a detailed account from a physical point of view
can be found in
\ref\hubsch{T. H\"ubsch,
            {\sl Calabi Yau manifolds: A bestiary for physicists},
            World Scientific, Singapore (1992)}.
Using the $c_1=0$ condition and curve identities it was recognized in
\ref\gh1{P. Green and T. H\"ubsch, \cmp113(1987)99}
that all CICYs can be described by finitely
many configurations of polynomials in
products of projective spaces. Each configuration leads to a family
of Calabi-Yau spaces whose generic member is smooth.
By a computer classification, 7868 configurations with Euler
numbers between $-200$ and $0$ were found in
\ref\cdls{P. Candelas, A. M. Dale, C. A. L\"utken and R. Schimmrigk,
          \npb298(1988)493} and
\ref\cls{P. Candelas, C.A. L\"utken and R. Schimmrigk,
\npb306(1988)113}.
In
\ref\ghl{P. Green, T. H\"ubsch and C. A. L\"utken, \cqg6(1989)105}
all their
Hodge numbers were calculated. There occur 265 different combinations.
Application of a theorem of Wall
\ref\wall{C.T.C. Wall, Invent. Math. 1 (1966) 355},
which states that the homotopy types of Calabi-Yau threefolds $X$
can be classified by their Hodge numbers, their topological
triple couplings $K^0_{ijk}\equiv\int_X J_i\wedge J_j\wedge J_k$
and $c_2\cdot J_i\equiv\int_X c_2\wedge  J_i$, reveals
that there are at least 2590 topologically different models in this
class
\ref\hc{A. He and P. Candelas, Commun. Math. Phys. 135 (1990) 193}.
Green and H\"ubsch have shown in
\ref\gh2{P. Green and T. H\"ubsch, \cmp119(1988)441}
that all families of CICYs
are connected by the process, already described in \cdls, of
contracting a family to a nodal configuration
and performing a small resolution of the latter.
Certain quotients of them by discrete groups were constructed in
\ref\TY{S. T. Yau, In {\sl Anomalies, geometry and topology},
        Argonne Symposion on anomalies, geometry and topology,
        Eds. W. A. Bardeen and A. R. White, World Scientific,
        Singapore 1985, with Appendix by G. Tian and S. T. Yau}
and
\ref\rolf{R. Schimmrigk, \plb193(1987) 175},
which have Euler number $\chi=-6$. Two of them, discussed below,
have a non-trivial fundamental group ($\pi_1(X)=\ZZ_3$)
and give rise to heterotic
string compactifications with three generations and a natural
option to break the $E_6$ gauge group by Wilson lines.

In this paper we want to extend the analysis of the Picard-Fuchs
equations, the construction of the mirror map and the calculation of
the instanton-corrected Yukawa couplings, which were performed for
hypersurfaces with one modulus in a (weighted) projective space in
\ref\cdgp{P. Candelas, X. De la Ossa, P. Green
     and L. Parkes, \npb359(1991)21}
\ref\morrison{D. Morrison, {\sl Picard-Fuchs equations and
     mirror maps for hypersurfaces}, in {\sl Essays on mirror
manifolds} Ed. S.-T. Yau, (Int. Press, Hong Kong, 1992)}
\ref\ktone{A. Klemm and S. Theisen, \npb389(1993)153}
\ref\font{A. Font, \npb391(1993)358}
and generalized to higher-dimensional moduli spaces of general
hypersurfaces in toric varieties in
\ref\cdfkm{P. Candelas, X. de la Ossa, A. Font, S. Katz and D.
Morrison, \npb416(1994)481;\hfill\break
P. Candelas, A. Font, S. Katz and D. Morrison,
{\sl Mirror Symmetry for Two Parameter Models II},
UTTG-25-93, IASSNS-HEP-94/12, OSU-M-94-1 (hep-th/9403187)}
\ref\hkty{S. Hosono, A. Klemm, S. Theisen and S.-T. Yau:
          {\sl Mirror symmetry, mirror map and applications
           to Calabi-Yau hypersurfaces},
           HUTMP-93/0801, LMU-TPW-93-22 (hep-th/9308122), to be
           published in Commun. Math. Phys.},
to the class of complete intersections in products of
weighted projective spaces.
It is natural to focus on the derivation of the prepotentials
for the complex structure and the
K\"ahler structure deformations, which encode all information
of theses two (topological) subsectors of the theory at tree level.
The knowledge of the prepotentials is quite relevant for the
low-energy phenomenology of the subsector of the moduli--
and the ${\bf 27}$, $\overline {\bf 27}$
matter fields in the effective field theory.

We will show that there is a subsystem
of the complex structure deformations, which is easily solvable,
that is by the mirror map, associated to the quantum cohomology
of the subset of the elements in $H^{1,1}(X,\ZZ)$
induced from the K\"ahler forms of the projective spaces.
We focus on this subsystem and present results in a closed
form for the instanton-corrected cohomology of this system.
In general the moduli space is enlarged due to exceptional divisors
coming from the resolution of possible singularities.
This situation we have treated quite generally for
hypersurfaces in \hkty. Here we will give two simple but
illustrative examples of how to deal with the exceptional divisors in
the complete intersection case.

Even in the smooth case there are usually more elements in
$H^{1,1}(X,\ZZ)$ than the ones
mentioned above. For the counting and the calculation of
intersections of the latter see \hubsch.
If they correspond, under the mirror map, to complex structure
deformations, which can be represented as deformations of the
polynomials by vector monomials, they can in principle be
incorporated. This is for instance the situation for the Tian-Yau manifold.
If a Landau-Ginzburg prescription of the model is available,
also those perturbations that are not of this kind,
can be represented as roots of monomial deformations,
as was suggested in
\ref\bcdfhjq{P. Berglund, P. Candelas, X. de la Ossa, A. Font,
    T. H\"ubsch, D. Jancic and F. Quevedo, {\sl Periods for Calabi-Yau
    and Landau-Ginzburg Vacua}, preprint CERN-TH. 6865/93,
    HUPAPP-93/3, NEIP 93-004, NSF-ITP-93-96, UTTG-13-03
    (hepth/9308005)}.

Our approach covers all cases treated in
\cdgp \morrison \ktone \libtei{A. Libgober and J. Teitelbaum,
     Duke Math. Jour.,Int. Math. Res. Notices 1 (1993) 29}
\ref\ktthree{A. Klemm and S. Theisen, {\sl Mirror maps and
     instanton sums for complete intersections in weighted
     projective space}, \mpla~(to be published)}
\ref\bs{V. Batyrev and D. van Straten, {\sl Generalized
 hypergeometric functions and rational curves on Calabi-Yau
 complete intersections in toric varieties},
 Essen Preprint (1993)}.
Due to threefold isomorphisms, it also covers the two two-moduli
Calabi-Yau spaces which were solved in \cdfkm~and among others
in \hkty, namely the degree $8$ hypersurface in $\IP^4[2,2,2,1,1]$ and the
degree $12$ hypersurface in $\IP^4[6,2,2,1,1]$.
We will see moreover that the described subsystems provide
simple examples of higher-dimensional Calabi-Yau moduli spaces,
the simplest case being the subsystem of the Tian-Yau manifold.
Apart from it being phenomenologically interesting, there  also exists
a detailed mathematical study on the rational curves of this
manifold
\ref\sommervoll{D. E. Sommervoll, {\sl Rational curves of low degree
on a complete intersection Calabi-Yau threefold in $P^3\times P^3$},
Oslo preprint ISBN 82-553-0838-5}.

We organize the material as follows. In section 2 we shortly
review how to calculate classical intersections
numbers. In the third section we explain our method of deriving
the system of Picard-Fuchs equations.
In section 4 we exhibit the general structure of the solutions
of the Picard-Fuchs equations. This, when combined with the mirror
hypothesis,
leads to the main result of this section, namely a concise
formula for the prepotential. This solves the problem of determining
the moduli dependence of the instanton-corrected Yukawa couplings and
the Weil-Peterson metric for the states associated to the
K\"ahler forms of the ambient space for the class of $(2,2)$
compactifications on CICYs in the large radius limit
completely\foot{Instead of presenting a lengthy
list of examples, we distribute the Mathematica program
INSTANTON, which calculates
the Yukawa couplings and counts the numbers of rational curves
for any complete intersection Calabi-Yau manifold discussed in
\cdls,\cls~ and other examples in singular ambient spaces. It is appended
to this paper in the hep-th bulletin board: 9406055.}.

We demonstrate our method with some selected examples in section 5.
In section 6 we discuss the connection
of certain CICY manifolds with rational
superconformal field theories. We can this way also explain
the occurrence of identical invariants for the rational and
elliptic curves on some pairs of hypersurfaces and complete
intersections by an identity of conformal field theories.
Following the approach of \ref\bcovI{M. Bershadsky, S. Ceccotti,
H. Ooguri and C. Vafa, \npb405(1993)279} we extend in section 7 the
analysis to the topological one-loop partition function, which is a
moduli-dependent quantity describing the difference between the threshold
corrections to the $E_8$ and $E_6$ gauge couplings
\ref\bcovII{M. Bershadsky, S. Ceccotti, H. Ooguri and
C. Vafa {\sl Kodaira-Spencer theory of gravity and exact results
for quantum string amplitudes}, HUTP-93/A025, RIMS-
946, SISSA-142/93/EP}
(see also
\ref\lk{J. Louis and V. Kaplunovski, in preparation}).
The expansion
of the one-loop partition function has a conjectural
interpretation in terms of Gromov-Witten invariants
for elliptic curves. As the geometrical understanding of
these predictions is in a somewhat preliminary state,
we found it useful to use our data to calculate them explicitly
for various one-, two- and three-moduli examples of different types.
In the final section we discuss some open problems and possible
avenues for future work.

%%%%%%%%%%%%%%%%%%%%%%%%%%%%%%%%%%%%%%%%%%%%%%%%%%%
\newsec{Calculation of the classical topological data of CICYs}
%%%%%%%%%%%%%%%%%%%%%%%%%%%%%%%%%%%%%%%%%%%%%%%%%%%

We consider in the following complete intersections of $l$
hypersurfaces in products of $k$  projective spaces.
Since most formulas allow for an incorporation of
weights we will state them for the general case.
Denote by $d_j^{(i)}$ the degree of the coordinates of
$\IP^{n_i}[\vec w^{(i)}]$
in the $j$-th polynomial $p_j$ ($i=1,\dots,k;\,\,j=1,\dots,l$).
We will use the notation
\eqn\notation{
\cicy\chi(\IP^{n_1}[w_1^{(1)},\ldots,w^{(1)}_{n_1+1}]\cr
              \vdots\cr
              \IP^{n_k}[w_1^{(k)},\ldots,w^{(k)}_{n_k+1}]|
              d_1^{(1)}&,\dots, & d_l^{(1)}\cr
              \vdots &       & \vdots \cr
              d_1^{(k)}&,\ldots,& d_l^{(k)}){h^{1,1}}}
for a configuration.
Let us associate to the K\"ahler forms
induced from the $i$'th
projective space the formal variable $J_i$ and consider the map
$\Pi:\ZZ[J_1,\ldots,J_k]\rightarrow \ZZ$ defined on the generators
$J_s$ as
\eqn\map{
\Pi(J_s)=\left.
\left(\prod_{r=1}^k {\partial_{J_r}^{n_r}\over n_r !}\right)
\left({\prod_{i=1}^k \prod_{j=1}^{n_i+1} (1+w_j^{(i)} J_i)
\over
\prod_{j=1}^l (1 + \sum_{i=1}^k d_j^{(i)} J_i)}\right)
\left({\prod_{j=1}^l \sum_{i=1}^k d^{(i)}_j J_i\over
\prod_{i=1}^k\prod_{j=1}^{n_i+1} w^{(i)}_j}\right)
J_s
\right|_{J_1=\ldots\,=J_k=0}\, .}
It follows from the adjunction formula that the term in the
second bracket on the right-hand side yields,
by formal expansion, the total
Chern class
\eqn\chernclass{
{\rm c}(X)=1+ c_1 + c_2 + c_3=1 + c_1^a J_a + c_2^{ab} J_a J_b+
c_3^{abc}J_a J_b J_c
}
of the Calabi-Yau manifold $X$. The  coefficients are given by
\eqn\chern{\eqalign{
c_1^a&=\sum_{i=1}^{n_a+1} w_i^{(a)}-\sum_{i=1}^l d_i^{(a)}\equiv 0\cr
c_2^{ab}&={1\over 2} \left[
-\delta^{a b}\,\sum_{i=1}^{n_a+1}\Bigl(w_i^{(a)}\Bigr)^2+
\sum_{i=1}^l d_i^{(a)} d_i^{(b)}\right]\cr
c_3^{abc}&={1\over 3} \left[ \delta^{abc}\,
\sum_{i=1}^{n_a+1} \Bigl(w_i^{(a)}\Bigr)^3-
\sum_{i=1}^l d_i^{(a)} d_i^{(b)} d_i^{(c)}\right] .
}}
Here we have  enforced the vanishing of the first Chern class.
The numerator of the third term in \map~is the top Chern class of
the normal bundle of $X$ and the denominator is a normalization
of the volume of the weighted projective space.
Applying $\Pi$ to a monomial of the $J_i$'s is equivalent
to the integration of the wedge product of the corresponding
$(1,1)$ forms, also denoted by $J_i$,
wedged with the Chern class of dual form degree over $X$.
We have therefore
\eqn\top{
\chi=\int_X c_3= \Pi(1), \quad \int_X c_2\wedge J_{m}=\Pi(J_m),\quad
K^0_{ijk}=\int_X J_{i}\wedge J_{j} \wedge J_{k}
=\Pi(J_i J_j J_k).}
Note that these formulas, with the
exception of the first line in \chern, are valid only for the case in
which
$X$ has no singularities.
For the singular case the triple intersection gets
modified  to $\int_X h_{i}\wedge h_{j} \wedge h_{k}=
\Pi(J_i J_j J_k) n_0^{(i)} n_0^{(j)} n_0^{(k)}$, where
$n_0^{(i)}$ is the least common multiple of the orders of the
isotropy groups of all fixed points of the manifold
under $\IC^*$-actions in
the $i$-th weighted projective space. The modification of the
first two integrals is more involved. For example for complete
intersections in one weighted projective space one has
\ref\dk{D. Dais and A. Klemm:
{\sl On the invariants of minimal desingularizations of
$3$-dimensional Calabi-Yau weighted complete
intersection varieties}, in preparation}
\eqn\cIIh{\int_X c_2 \wedge J = {12\over n_0 !}
\left(\partial\over \partial J\right)^{n_0}
\left.
{\prod_{i=1}^l (1- J^{d_i})\over \prod_{i=1}^{n+1}(1-J^{w_i})}
\right|_{J=0}-2 \int_X J\wedge J \wedge J.}
In section 4 we will provide a more illuminative
generalization of formulas \chern~to the
case of hypersurfaces and complete intersections with
desingularized quotient
singularities.

For example, in a product of two (weighted) projective spaces we have
a smooth configuration
\eqn\exampleI{\cicy-252(\IP^4\(3,1,1,1,1\)\cr\IP^1|6&1\cr 0&2)2\,:
\quad  \eqalign{&K^0 = 4\,J_1^3 + 2\,J_1^2 J_2,\cr
                 &c_2\cdot J_1=42,\,\,  c_2 \cdot J_2=24,}}
where we use the notation of \hkty~to display the
intersection numbers.

%%%%%%%%%%%%%%%%%%%%%%%%%%%%%%%%%%%%%%%%%%%%%%%%%%%
\newsec{Derivation of the Picard-Fuchs equations}
%%%%%%%%%%%%%%%%%%%%%%%%%%%%%%%%%%%%%%%%%%%%%%%%%%%

We will begin this section by deriving some general
results on the Picard-Fuchs equations as they apply to the
models considered here, using the toric data of the manifolds.
This approach gives one period by explicit
integration and a holonomic system of linear differential
operators, which are satisfied by all periods but which allow
for additional solutions. Among the finitely many solutions the
periods can be singled out by the requirement
that the monodromy acts irreducibly on them. Technically
the problem of specifying them is solved here
by factorizing the differential operators.

Let us first show how the system of Picard-Fuchs equations
for a restricted set of $k$ complex structure
deformation parameters can be obtained from toric data of
the (mirror) manifolds according to \bs.
This system of Picard-Fuchs
equations is equivalent to the first-order Gauss-Main differential
system, which describes the variation of the Hodge structure of
$H^3(X,\ZZ)$, restricted to the holomorphic and antiholomorphic
$(3,0)$ and $(0,3)$ forms and $k$ $(2,1)$ and $(1,2)$
forms. By the mirror map we will identify the $k$
complex structure deformations with the K\"ahler deformations
in the restricted basis of K\"ahler forms specified in
the previous section.

Here we consider only configurations in which the complete intersection does
not intersect with singular loci  of the ambient space
$\IP^{n_1}\(\vec w^{(1)}\)\times\cdots \times \IP^{n_k}\(\vec
w^{(k)}\)$ (for an example of the general situation see (5 $viii$)).
Without further restricting the generality we may choose
$w_{n_i+1}^{(i)}=1$ for all $i$.
Each projective space $\IP^{n_i}\(\vec w^{(i)}\)$ is a toric variety,
which
can be described by a (reflexive) simplicial polyhedron $\Delta_i$
with integral vertices in $\IR^{n_i}$ (see refs.
\ref\batyrev{V. Batyrev, {\sl Dual polyhedra and mirror symmetry
for Calabi-Yau hypersurfaces in toric varieties}, to appear in
Journal Alg. Geom.; V. Batyrev, Duke Math. Journ., 69 (1993) 349}
\hkty~ for its determination).

Since the ambient space is a direct product of
all $\IP^{n_i}\(\vec w^{(i)}\)$, it is also a toric variety described
by the reflexive polyhedra $\Delta=\Delta_1\times\cdots\times\Delta_k$
in $\IR^{n_1}\times\cdots\times\IR^{n_k}$.

In refs. \bs~
\ref\borisov{L. Borisov, {\it Towards the mirror symmetry for
Calabi-Yau complete intersections in Gorenstein toric fano
varieties}, Ann Arbor preprint, 1993 (alg-geom/9310005)}
it was conjectured that the mirror manifold of the CICY of
type \notation~ is given by the CICY constructed from the
combinatorial data of the dual polyhedron $\Delta^*$ of
$\Delta$. This $\Delta$ is reflexive, and the corners of the
dual $\Delta^*$ are the integral points
$\nu_{i,1}^*=(1,0,\cdots,0),\cdots,
 \nu_{i,n_i}^*=(0,\cdots,0,1)$  and
$\nu_{i,n_i+1}^*=(-w_1^{(1)},\cdots,-w_{n_i}^{(i)})$ in $\IR^{n_i}$
of $\IR^{n_1}\times\cdots\times\IR^{n_k}$. These vertices satisfy
the relations $\sum_{j=1}^{n_i+1}w_j^{(i)}\nu_{i,j}^*=0 ,\;
(i=1,\cdots,k)$.
We group the vertices $\nu_{i,j}^*$ into $l$ ($=$ number of defining
polynomials) sets (a so-called as nef-partition)
\eqn\set{
\{ \nu_{i,j}^* \}_{1\leq i\leq k, 1\leq j\leq n_i+1}
= E_1\cup\cdots\cup E_l
}
by defining $E_m \; (1\leq m\leq l)$ so that it contains $d_m^{(i)}$
vertices from $\{ \nu_{i,j}^* \}_{1\leq j\leq n_i+1}$ for each
$i=1,\cdots,k$. We extend each vertex $\nu_{i,j}^*$ of $E_m$ to
$\bar \nu_{i,j}^*= (\vec e^{\,(m)},\nu_{i,j})$ in $\IR^l\times
\IR^{n_1}\times\cdots\times\IR^{n_k}$, with $\vec e^{\,(m)}$ being
the unit vector in the $m$-th direction of $\IR^l$.
Adding the additional vertices
$\bar\nu^*_{0,p}=(\vec e^{\,(p)},\vec 0)$ one finds, as a
consequence of the first relation in \chern, $k$ independent
linear relations between the $\sum_{i=1}^l (n_i+2)$ vertices
$\bar\nu^*_{i,j}$ of the form
\eqn\linearrelationI{\sum l^{(s)}\bar\nu^*_{i,j}=\vec 0.}
The $l^{(s)}\,(s=1,\dots,k,\,j=1,\dots,l)$ are given by
\eqn\linearrelations{
l^{(s)}=(-d^{(s)}_1,\dots,-d^{(s)}_l;
\dots,0,w^{(s)}_1,\dots,w^{(s)}_{n_s+1},0,\dots)
\equiv (\{l^{(s)}_{0j}\};\{l_i^{(s)}\}).
}
The mirror manifold $X^*$ can then be written conjecturally as
the complete intersection of the vanishing locus of the
following Laurent polynomials in the variables
$X_{m,n},\,m=1,\dots,k,\,n=1,\dots,n_m$ (using the notation
of ref. \hkty) :
\eqn\laurant{
P_r=a_{r}-\sum_{\nu_{i,j}^*\in E_r} a_{i,j} X^{\nu^*_{i,j}}\qquad
(r=1,\dots,l),
}
where the sum is over the (unextended) vertices in the $m$-th set
$E_m$.
The vanishing loci of \laurant~ are considered in a toric variety
$\IP_{\Delta^*_{(1)}+\cdots+\Delta^*_{(l)}}$, $\Delta^*_{(i)}$
being a convex hull of $\{0\}$ and the set $E_i$.
In ref. \borisov~ the combinatorial aspects of this
construction have been nicely formulated.

Choosing a cycle $\Gamma_0$ determined by $|X_{m,n}|=1$ $\forall\,m,n$,
the corresponding period integral
(see \batyrev~and \hkty)
\eqn\periods{
w_0 (a)= \int_{\Gamma_0} {a_1\cdots a_l\over P_1\cdots P_l}
\prod_{m=1}^k\prod_{n=1}^{n_m} {d X_{m,n}\over X_{m,n}}
}
can be performed explicitly by expanding the integrand in a multiple
power
series in $1/a_m$, and using the residue formula.
After introducing the variables
\eqn\goodvar{
z_s={a_{s,1}^{ w_1^{(s)} } \cdots a_{s,n+1}^{ w_{n_s+1}^{(s)} } \over
     a_1^{d_1^{(s)}}\cdots a_l^{d_l^{(s)}}}\equiv a^{l^{(s)}}
   \quad (s=1,\cdots,k)}
it can be easily verified that the period $w_0(a)$ is given
by\foot{Here and
in the following we denote by $\t,z,n$ and $\rho$ the
$k$-tuples $\t_1,\ldots,\t_k,z_1,\ldots , z_k,n_1,\ldots, n_k$
and $\rho_1,\ldots,\rho_k$. We use obvious abbreviations such as
$ z^{ n}\, :=\prod_{s=1}^k z_s^{n_s}$,
${ n}!\, :=\prod_{s=1}^k n_s! $ etc.}
\eqn\powersol{
w_0(z)=\sum_{n_s\geq 0}c(n) z^{n}
}
with
\eqn\coeff{
c(n)={
\prod_{j}\left(-\sum_{s=1}^k l_{0j}^{(s)} n_s \right)!
\over
\prod_{i}\left(\sum_s l_i^{(s)}n_s \right)!}
={\prod_{j=1}^l\left(\sum_{i=1}^k n_i d^{(i)}_j\right)!
\over\prod_{i=1}^k \prod_{j=1}^{n_i+1}(w_j^{(i)}n_i)!}.
}
It satisfies the generalized hypergeometric system of
Gelfand-Kapranov-Zelevinski \ref\GKZI{I. M. Gelfand, A. V. Zelevinski
and M. M. Kapranov, Functional Anal. Appl. {\bf 23},2 (1989) 12;
English Trans. p. 94} with the $k$ linear differential operators
\eqn\lindiffop{\eqalign{
{\cal L}_s=&\prod_{j=1}^{n_s+1}\Bigl(w_j^{(s)}
\theta_s\Bigr)
\Bigl(w_j^{(s)}\theta_s-1\Bigr)\cdots
\Bigl(w_j^{(s)}\theta_s-w_j^{(s)}+1\Bigr)\cr
&\qquad -\prod_{j=1}^l\Bigl(\sum_{i=1}^k d_j^{(i)}\theta_i\Bigr)
\cdots\Bigl(\sum_{i=1}^k d_j^{(i)}\theta_i-d_j^{(s)}+1\Bigr)z_s}}
associated to the vector of the coefficients of the linear
relations $l^{(s)}$, $s=1,\ldots,k$ given in \linearrelations.
Here the $\theta_i$ are
logarithmic derivatives $\theta_i=z_i {\partial\over \partial z_i}$.
Similarly as in \hkty~one can show that these equations are satisfied
for all periods $w_j$ as they reflect the infinitesimal symmetries
of \periods, independent of the chosen cycle.
This system is holonomic, which means that the left ideal $I$
generated by \lindiffop~in the ring of linear partial
differential operators $D$ is of finite rank ${\rm rk}(I)$.  This
implies
the existence of ${\rm rk}(I)$ independent solutions
\ref\my{M. Yoshida, {\sl Fuchsian differential equations},
(Vieweg, Braunschweig 1987)}, where ${\rm rk}(I)$ is always larger
than the expected number of periods.

We are interested in a subset of solutions of the
system \lindiffop, which corresponds to period integrals
over the $2(k+1)$ cycles
dual to the restricted basis of $H^3(X,\ZZ)$.
These solutions can be characterized by the requirement
that the monodromy acts irreducibly on them
\ref\griffith{P. A. Griffiths, Bull. Amer. Math. Soc. 76 (1970) 228}.
Solving the Riemann-Hilbert problem for the reduced monodromy
leads to a reduced system of (lower-degree)
differential operators $L_1,\dots,L_h$, where
$h$ denotes the number of Picard-Fuchs equations; cf. below.
In many examples such a system can
be specified by factorizing differential operators
from \lindiffop~in the form $p_i(\t) L_i:=
\sum_{j=1}^k q_j(\t){\cal L}_j$, where $p(\t)$ and
$q_j(\t)$ are polynomials in $\t$. We remark that this is however
not the generic situation (see \hkty).
The system $L_1,\dots,L_h$ is again holonomic and generates a left ideal
whose rank is $2(k+1)$. In fact this is our criterion to check that
a given system of PF equations is complete (cf. below).

Let us demonstrate the derivation of the system $L_1,\ldots,L_h$ for
the following complete intersection Calabi-Yau manifold
\eqn\schimm{
\cicy-54(\IP^3\cr\IP^2|3&1\cr 0&3)8.}
The toric description of the mirror manifold was already given in \bs.
One has two reflexive simplicial polyhedra $\Delta_1^*,\Delta_2^*$ with
vertices in
$\IR^3\times\IR^2$:
\eqn\polyhedra{
\eqalign{
\nu^*_{1,1}&=(1,0,0;0,0),\,\nu^*_{1,2}=(0,1,0;0,0),\,
\nu^*_{1,3}=(0,0,1;0,0),\,\nu^*_{1,4}=(-1,-1,-1;0,0);\cr
\nu^*_{2,1}&=(0,0,0;1,0),\,\nu^*_{2,2}=(0,0,0;0,1),\,
\nu^*_{2,3}=(0,0,0;-1,-1).}
}
We now group the vertices into two sets $E_1=\{\nu^*_{1,1},\nu^*_{1,2},
\nu^*_{1,3}\}$ and
$E_2=\{\nu^*_{2,1},\nu^*_{2,2},\nu^*_{2,3},\nu^*_{1,4}\}$
and define the extended vertices $\bar\nu^*=(\vec e_{1,2};\nu^*)$ in
$\IR^7$
where we choose $\vec e_1^{\,(1)}=(1,0)$ and $\vec e^{\,(2)}=(0,1)$ for the
vertices in the first and second sets, respectively.
The results derived in the following are independent
of how we group the vertices into two sets, as long as the first set
contains three vertices $\nu^*_{1,i}$ and the second set contains
the remaining vertices.
After adding the two vertices $\nu^*_{0,k}=(\vec e^{\,(k)},\vec 0)$,
$k=1,2$,
this leads by \laurant~to the following  two Laurent  polynomials:
\eqn\laurenta{
\eqalign{P_1&=a_1-a_{1,1} X_1-a_{1,2} X_2-a_{1,3} X_3\cr
         P_2&=a_2-a_{2,1} Y_1-a_{2,2} Y_2-{a_{2,3}\over Y_1 Y_2}-
         {a_{1,4}\over X_1 X_2 X_3}.}
}
We now have independent linear relations\foot{The components of
the $l^{(k)}$ refer to vertices $\bar \nu^*_{i,j}$ with $i,j$ in
lexicographic
order.}  $\sum l^{(k)}\bar\nu^*_{i,j}=0$,
$k=1,2$, between the vertices, namely
\eqn\relations{
l^{(1)}=(-3,-1;1,1,1,1,0,0,0)\,\quad{\rm and}\quad
l^{(2)}=( 0,-3;0,0,0,0,1,1,1).
}
They are of the form \linearrelations~ and define the variables
$z_1={a_{1,1} a_{1,2} a_{1,3} a_{1,4}\over a_1^3 a_2},
\,z_2={a_{2,1} a_{2,2} a_{2,3} \over a_2^3}$ via \goodvar.
The parameters $a_{i,j}$ correspond to trivial automorphisms
of \laurenta~ and can be set to one.
Using \lindiffop~we can associate
the following GKZ system to the $l^{(i)}$'s
\eqn\GKZsystem{
\eqalign{
\cL_1&=\t_1^4-3\,\t_1\, (\t_1+3\,\t_2 )\,(3\,\t_1 -1)\,(3\,\t_1-2)\,z_1
\cr
\cL_2&=\t_2^3-(\t_1 +
3\,\t_2)\,(\t_1+3\,\t_2-1)\,(\t_1+3\,\t_2-2)\,z_2\,. \cr}
}
In addition to the power series solution the system has eleven logarithmic
ones.
To obtain an irreducible
subsystem we factorize in the following way:
$\cL_1=:\t_1 L_1$ and $\cL_1+ 27 \cL_2=: \t_1 (\t_1+\t_2) L_2$,
which leads to the reduced system
\eqn\rsI
{\eqalign{
L_1&=\t_1^3-3\,(\t_1+3\,\t_2 )\,(3\,\t_1 -1)\,(3\,\t_1-2)\,z_1 \cr
L_2&=(\t_1^2-3\,\t_1\t_2+9\,\t_2^2)-3\,
(3\,\t_1-1)(3\,\t_1
-2)\,z_1-27\,(\t_1+3\,\t_2-1)\,(\t_1+3\,\t_2-2)\,z_2.}
}
The fact that the number of equations equals the number of moduli
is special to the example here. In fact we will see in example
(5 $vii$) that the numbers of linear differential operators
describing the Picard-Fuchs system locally can vary
in the different patches of the moduli space.

Let us finally comment on some generic features
of the moduli space of the  mirror manifold and its
compactification, as far as we will need them for fixing the
holomorphic ambiguity in section 7.
It is always easy to find the invariance group of the Laurent
polynomials, which acts by phase multiplication on the
parameter $a_i$ and the coordinates $X_i,Y_i,\ldots $.
In our example one finds a $\ZZ_9$ invariance group of \laurenta~acting
by $(X_i,Y_j,a_1,a_2)\mapsto (\alpha^k X_i,\alpha^{6 k} Y_i,\alpha^k a_1,
\alpha^{6 k} a_2)$, with $\alpha= \exp{2 \pi i\over 9}$ and $k\in \ZZ$.
Because of this invariance of the parameter space of \laurenta,
we have to define the moduli space of the mirror manifold as the
quotient ${\cal M }= \tilde {\cal M}/\ZZ_9$, where
$\tilde {\cal M}$ is parametrized by $a_1,a_2$.
The fact that  \goodvar~are invariant under such group actions
and hence well defined on the quotient can be seen in general\foot{This
generalizes the reasoning, which leads for the quintic to the
consideration of $z=1/a^5$ \cdgp~ as the good variable.}.

The singularity of the moduli space of the Laurent polynomials
at $a_i=0$ due to the phase symmetries can now always be
described by toric geometry. In general one has to
consider as a second step also the quotient with respect to
invariances of the Laurent polynomials, which are not acting
simply by phase multiplication.
In our example such symmetries are not present and
we see that $\tilde n_1=a_1^9$, $\tilde n_2=a_1^3 a_2$ and
$\tilde n_3=a_2^3$ generate the
multiplicative semigroup of invariant monomials under the $\ZZ_9$, but
satisfy the relation $\tilde n_1 \tilde n_3= \tilde n_2^3$, describing
a rational $A_2$ double point. It is now straightforward to give a
toric description of the moduli space
using the secondary fan construction of ref.
\ref\odapark{L. Billera, P. Filliman and B. Sturmfels,
Adv. Math. 83 (1990) 155;\hfill\break
T. Oda and H.S. Park,
T$\hat o$hoku Math. J. 43 (1991) 375.}.
The secondary fan,
whose dimension equals the number of K\"ahler moduli,
contains the K\"ahler cone. Since it is a complete fan, it
gives a toric description of the compactification of K\"ahler moduli space;
for a review, see also
\ref\agm{P.S. Aspinwall, B. R. Greene and D. Morrison,
\plb303(1993)249,\npb416(1994)414}.
If we define a matrix $B$ whose rows are the generators $l^{(s)}$
of the Mori cone, then the columns of $B$ generate the one-dimensional
cones of the secondary fan. Their minimal generators for the  model
considered here
are the vectors $e_1,\,e_2,\,-(e_1+3 e_2)$ and $-e_1$,
where $e_1,\,e_2$ generate a square lattice. The cone
\foot{Here the conventions
are as in \ref\hkt{S. Hosono, A. Klemm and S. Theisen
{\sl Lectures on mirror
symmetry}, HUTMP-94/01,LMU-TPW-94-02}.}
$\langle e_1,e_2\rangle$ is the K\"ahler cone and the cone
$\langle -e_1,-(e_1+3e_2)\rangle$ describes the $A_2$ double point.
This description of the $A_2$ double point is related to that given,
e.g. in \hkt, by a change of basis $e_1\to -e_2,\,e_2\to-e_1+e_2$.

The general theory, due to Hironaka, tells us that the
compactified moduli space can be resolved in such a way that all
singularities of the Picard-Fuchs equation are regular divisors with normal
crossings and part of the resolutions necessary to achieve this
goal can be described by toric resolution if we start as above. Finding
such a resolution is necessary in order to analyse the full
modular group. A thorough analysis of the modular groups for
Calabi-Yau compactifications was given for one-modulus examples
in \cdgp\ktone\font\ and for two types of two-moduli examples in \cdfkm.

\vfill\eject

%%%%%%%%%%%%%%%%%%%%%%%%%%%%%%%%%%%%%%%%%%%%%%%%%%%
\newsec{Local behaviour of the solutions, mirror map and
instanton-corrected Yukawa couplings}
%%%%%%%%%%%%%%%%%%%%%%%%%%%%%%%%%%%%%%%%%%%%%%%%%%%

In this section we calculate the singular locus of the Picard-Fuchs
equations and discuss some essential parts of the local behaviour of their
solutions.
We explain how to introduce canonical coordinates and
fix a canonical form of the period vector.
This leads to simple expressions of the
instanton-corrected Yukawa couplings and the
K\"ahler potential in terms of the solutions.
We will see that all topological data of the Calabi-Yau manifold
appear naturally in this period vector.

It was shown in \ref\deligne{P. Deligne, {\sl Equations
diff\'erentielles
\`a point singulieres r\'egulieres} Lecture Notes of Mathematics 163,
(Springer Heidelberg 1970)} that general Picard-Fuchs systems
have only regular singular points, i.e. locally the solutions
are given by power series or series involving
finite powers of logarithms in $z_i$.

Let us now describe the singular locus (cf. \my).
Denote the linear partial differential operators of degree $m$, defined
in a neighbourhood $U$ of $z\in\cal M$
(the subspace of the complex structure moduli space to which our
analysis applies),
by $L_i=\sum_{|p|\le m} a_i^p(z) \left({d\over d z}\right)^p$.
They define a left ideal $I$  in the ring of partial
differential operators on $U$.
We now introduce the symbol of $L_i$:
$\sigma(L_i)=
\sum_{|p|=m} a_i^p(z) \xi_1^{p_1}\cdots \xi_k^{p_k}$, where
$\xi_1,\ldots,\xi_k$ is a coordinate system in the fibre of the
cotangent bundle $T^*U$ at $z$. The ideal of symbols
is defined by $\sigma(I)=\{\sigma(L)|L\in I\}$.
The singular locus is $S(I)=\pi({\rm Ch}(I)- U\times \{0\})$,
where the characteristic variety ${\rm Ch}(I)$ is the
subvariety in $T^*U$ specified by the ideal of symbols and
$\pi$ is the projection along the fibre of $T^*U$.
The fact that $\sigma(I)$ is generated by $\sigma(L_i)$ is
a special property of Picard-Fuchs systems. This follows for instance
from the way the Picard-Fuchs equations are derived by the
Griffith-Dwork-Katz reduction method and simplifies
the calculation of $S(I)$.
Let us demonstrate this for $L_1,L_2$ given in eq. \rsI.
The symbols are
\eqn\symbols{
\eqalign{
\sigma(L_1)&=z_1^3(\xi_{1}(1-27 z_1)-81{z_2}\xi_{2})\xi_{1}^2\cr
\sigma(L_2)&={z_1}^2(1-27{z_1}-27{z_2})\xi_{1}^2-3 {z_1} {z_2}
(1+54 {z_2}) \xi_{1} \xi_{2}+9 {z_2}^2 (1-27 {z_2}) \xi_{2}^2.}
}
A case-by-case analysis reveals that ${\rm Ch}(I)$
decomposes into the following components:
\eqn\discriminant{
\eqalign{
{\rm Ch}(I)=
&\{81 {z_2} \xi_{2}-(1-27{z_1})\xi_{1}=
(1-27{z_1})^3-27{z_2}=0\}\cup\{{z_1}={z_2}=0\} \cr
&\cup\{\xi_{1}=(1-27 {z_2})=0\}
\cup\{\xi_{1}={z_2}=0\}\cup \{\xi_{2}={z_1}=0\}.}
}
Denoting the projections of the components
on $U$ by $\Delta_i$ we have
$S(I)=\prod_{i=0}^3 \Delta_i$, with
$\Delta_0=(1-27{z_1})^3-27{z_2}$,
$\Delta_1=(1-27 z_2)$,
$\Delta_2=z_1$,
$\Delta_3=z_2$.

The singularities $\Delta_i=0$ detected so far correspond to
the discriminant locus in the moduli space on which the defining
polynomials
cease to be transverse, i.e. where $p_1=\dots=p_l=0$ and
$dp_1\wedge\cdots\wedge dp_l=0$ or, equivalently,
where the Laurent polynomials fail to be $\Delta^*$ regular \batyrev.
We calculate for comparison the discriminant of \laurenta~ at the
end of section (5{\it i}).
To study further singularities we have to compactify the moduli space.
We can describe the compactification by the toric variety corresponding
to the secondary fan constructed in the previous section.
Then we find, with
the same method as above, a singular locus of the Picard-Fuchs equation
at the origin of the patch defined by $\sigma_3$, which is
due to the $\ZZ_9$ identification of the parameter space discussed
above.
Such an analysis can be made for the general case, and the
singular locus determined this way is used in section 7 in
order to fix the holomorphic ambiguity for various examples.

{}From the definition of the (unnormalized)
Yukawa couplings (coefficients of the
cubic form $\Xi$ in \ref\bg{R. L. Bryant and P. A. Griffith, in
Progress
in Mathemathics {\bf 36} {\sl Arithmetic and Geometry}, (Birkh\"auser,
Basle 1983) page 77-102}) and the variational property of the $(3,0)$-form
\bg~
one finds (see also
\ref\cd{P. Candelas and X. de la Ossa, \npb355(1991)455})
that the vanishing of the symbols at degree
three gives relations for the Yukawa-couplings by the simple
replacement $\xi_i\xi_j\xi_k\to K_{z_i z_j z_k}$;
e.g. from $\sigma(L_1)=0$
one has $K_{z_1z_1z_2}={(1-27 z_1)\over 81 z_2}K_{z_1z_1z_1}$. This
determines the Yukawa couplings
up to a gauge-dependent overall function. The gauge in which
the Picard-Fuchs equations are derived is defined by \periods, and
the  corresponding gauge-dependent function can be calculated, up to a
constant, by the methods outlined in \cdgp\hkty.

These unnormalized Yukawa couplings have singularities on the
discriminant
of the Calabi-Yau manifold. From the derivation of the Yukawa
couplings in \bg\hkty, it follows that there is always
one component which will appear in the denominator of all
the Yukawa-couplings, which we call the general component $\Delta_0$,
following \ref\GKZII{I. M. Gelfand, M. M. Kapranov and A. V. Zelevinski,
Leningrad Math. J. 2 (1991) 449}.

In the following we will demonstrate that the calculation
of those Yukawa couplings on $X^*$, which are functions of the complex
structure moduli is not necessary for the purpose of
determining the number of rational and elliptic curves: one can
directly compute the instanton-corrected couplings on $X$.

Because of the fact that the symbols of $I$ generate the ideal of
symbols it is a rather simple algebraic problem to write down the
associated first order-Pfaffian system, equivalent to the
Gauss-Manin connection. However we found that some
local properties of the solutions are most easily obtained directly
from $L_1,\ldots,L_h$. Most important is the
analysis of the solutions around $z=0$, which is obviously
a singular point in every system derived by factorization from
\lindiffop. For the analysis it is useful to
introduce the ring ${\cal R}$, which we define as the polynomial
ring $\IC[\t]$ modulo the
ideal ${\cal I}$ generated by the principal parts $I_s$
of the operators $L_s$
\eqn\ideal{
I_s({ \t})=\lim_{z \to 0} L_s(\t,z),
}
i.e. ${\cal R}=\IC[\t]/{\cal I}.$
First we note that the $I_s(\t)$ are homogeneous polynomials and
the solution to $I_s=0$, $\forall \, s$ is
$2(k+1)$-times degenerate at $\theta=0$;
$\IC[\t]$ has a natural
vector space structure with the
monomials as orthogonal basis. We can choose
representatives of ${\cal R}$ as homogeneous
polynomials orthogonal to ${\cal I}$.
The subspace $R$ of $\IC[\t]$ spanned by them
has in all cases dimensions $\{1,k,k,1\}$ at degrees $\{0,1,2,3\}$.
As we will now show, it can be identified with the vector space
of solutions  to $L_1,\ldots,L_h$.
The grading translates
to the fact that we have one pure power series solution,
$k$ solutions with a part linear in logarithms of $z$,
$k$ solutions with a part quadratic in the logarithms of
$z$, and 1 solution which has a part cubic in the logarithms.
This identifies $z=0$ as the point where all but one
cycle in $H_3(X)$ are degenerate \ref\todorov{T. Todorov,  private
communication},
which is also referred to as a
point of maximally unipotent monodromy\ref\morrisonII{D. Morrison,
{\sl Compactifications of moduli spaces inspired by mirror symmetry},
Duke 1993 Preprint DUK-M93-06}
and provides precisely
the structure of solutions one needs for the mirror map.

Extending the definition of $x!=\Gamma(x+1)$ to $x\in\IR$,
we define the coefficient $c( n+  \rho)$ for arbitrary values
of the $k$ parameters $ \rho_i$ and define the
$\rho$-dependence of
\powersol~as
\eqn\wnullrho{w_0( z,  \rho)=
\sum_{n_i\ge 0} {c( n+ \rho)}
 z^{ n+  \rho}.}
By the method of Frobenius, the logarithmic solutions
are obtained by taking linear combinations
of derivatives $D_{ \rho}=\sum (b_{ n}/n !)
\p_{ \rho}^{ n}$ of $w_0( z, \rho)$ evaluated at $ \rho= 0$.
Here we define
$\p_{\rho_i}:={1\over 2 \pi i}{\p\over \p_{\rho_i}}$. The factor $2\pi i$
will make the monodromy matrix around $z=0$ integer.
As $[L_s,\p_{\rho_i}]=0$, it is then sufficient to check whether
\eqn\cond{\left.
D_{ \rho}\left(L_s w_0( z,
\rho)\right)\right|_{ \rho= 0}=0,\,\, \forall s
}
to establish $D_{ \rho} w_0( z, \rho)|_{ \rho= 0}$
as a solution. By consideration of the explicit form of the
series \wnullrho, one can show that the conditions for
vanishing of the constant terms in \cond
\eqn\condI{
\left. D_{ \rho} (I_s( \rho)
c( 0, \rho) z^\rho )\right|_{ \rho= 0}=0,\,\, \forall\, s,
}
are in fact also sufficient.

We introduce an identification of the ring $\IC[\theta]$
with the ring of the partial derivatives w.r.t. $\rho$:
\eqn\corres{\varphi:
\sum_{n_i\ge 0} b_n \theta^n
\mapsto
\sum_{n_i\ge 0} {b_{ n}\over   n!} \p_{ \rho}^{ n},}
which induces  an isomorphism
between the vector space $R$ and the vector space of solutions
to $L_1,\ldots,L_k$, as can be seen in the following.
An element $r\in R$ is orthogonal to all
polynomials of the form $m(\theta) I_s$, where
$m(\theta)$ is monomial in $\theta$. As $r$ and
$I_s$ are homogeneous, one has to check
orthogonality only for monomials $m(\theta)$
of degree ${\rm deg}\,(r)-{\rm deg}\,(I_s)$.
Suppose $r$ is of lower degree, then $I_s$
\condI~will obviously hold for
$D_\rho=\varphi(r)$, because all terms in
$D_{ \rho} (I_s( \rho) c( 0, \rho) z^\rho )$ have
positive degree in $\rho$ and will vanish
after setting $\rho$ to zero.
If ${\rm deg}\,( r) = {\rm deg}\,(I_s) + n$,
then ${\rm deg}\,(r)-n$ derivatives of
$D_\rho=\varphi(r)$ have to act on $I_s$ to give a
non-vanishing term. By our choice of the factorials the
vanishing of these contributions is equivalent to
$r\perp m(\theta) I_s=0$ for any $m(\theta)$
with ${\rm deg}\,(  m(\theta))=n$. Hence $r\in R$ iff
$\varphi(r)w_0(z,\rho)|_{\rho=0}$ is a solution.

Besides the unique power-series solution \powersol, we choose for
the $k$ solutions linear in the logarithms the basis
$\p_{\rho_i} w( z, \rho)|_{ \rho =  0}$.
They are given by
\eqn\logperiods{
w_i(z)=\sum_{n_i\geq0}d_i(n)z^n + w_0(z)\,{\log z_i\over 2 \pi i}\, ,
}
with
$$
d_i(n)={1\over 2\pi i}{\p\over\partial\rho_i}
{c( n+ \rho)}\Bigg|_{ \rho= 0}\,.
$$

The top element of $R$ is unique, up to a constant, which
we will fix in a moment, and so is the solution cubic in the
logarithms.

Defining now \eqn\defroder{
D_{i}^{(1)}:=\p_{\rho_i},\,\,\,\,\,\,
D^{(2)}_i:={1\over2}\tilde K_{ijk}\p_{\rho_j}\p_{\rho_k}
\,\,\,\,\, {\rm  and} \,\,\,\,\,
D^{(3)} := -{1\over6}\tilde K_{ijk}\p_{\rho_i}\p_{\rho_j}\p_{\rho_k} ,}
where the summation is implicit,
we have a natural basis for the period vector
\eqn\pI{\Pi(z)=\left(\matrix{w_0(z)\cr
D^{(1)}_i w_0(z,\rho)|_{\rho=0}\cr
D^{(2)}_i w_0(z,\rho)|_{\rho=0}\cr
D^{(3)} w_0(z,\rho)|_{\rho=0}}\right).}
When one actually performs the derivatives w.r.t. $\rho_i$, one
has to be careful when treating the poles of the gamma-function and
its derivatives. We clear up these technical
details in Appendix A.

So far we have only dealt with the complex-structure deformation
parameters $z_i$. To describe the mirror map between the subsectors
of the theory depending only on the complex structure parameters and
the complexified K\"ahler structure parameters\foot{
Recall that the moduli describing the variation of the Ricci flat
K\"ahler
metric and the antisymmetric tensor background field can be
parametrized by
$\delta g_{i\bar\jmath}dz^i\wedge d\bar z^{\bar\jmath}=
\sum_{i=1}^{k} \delta \hat t^i J_i$ and $\delta
B_{i\bar\jmath}dz^i\wedge d\bar
z^{\bar\jmath}=\sum_{i=1}^{k}\delta \tilde t^i J_i$ where ${\hat t_i}$
and $\tilde t_i$ can be combined to
$t^i=\tilde t^i+i\hat t^i$,  the so-called complexified
K\"ahler structure parameter \cd}, respectively,
we will briefly review their common structure.
They can be identified with topological field theories defined
by the BRST operators $Q^C=G^+ + \bar G^-$ and $Q^K=G^+ + \bar G^+$,
respectively. The $G^\pm,\bar G^\pm$ in $Q^C$ and $Q^K$ are related to
the zero modes of the superpartners of the energy-momentum tensor of the
underlying $N=2$ superconformal field theory by two different kinds of
twist procedures which make either $Q^C$ or $Q^K$ to scalar operators
\ref\nIItopa{E. Witten, \cmp117(1988)353,\cmp118(1988)411; T. Eguchi
and S. K. Yang, \ijmpa5(1990)1693;\hfill\break
R. Dijkgraaf, H. Verlinde, E Verlinde, \npb352(1991)59}\bcovII.
At string tree level all relevant information is encoded in
two prepotentials (free energy) $F^{C},F^{K}$,
which are sections of holomorphic line
bundles $L^2$ over the complex and K\"ahler structure
moduli spaces respectively.

There exists a coordinate choice, so-called inhomogeneous special
coordinates $t^i$, such that the Yukawa couplings (structure constants) are
ordinary third derivatives ($\p_i={\p\over\p t^i}$)
\eqn\yuk{K_{ijk}=\p_i\p_j\p_k F}
of the prepotential. For general coordinate choices the derivatives
have to be covariantized w.r.t. the connection of the
line bundle $L$ and the metric on the moduli space. Unlike for
topological models on manifolds with $c_1>0$ \ref\mk{M. Kontsevich
and Yu. Manin {\sl Gromov-Witten Classes, Quantum Cohomology,
and enumerative Geometry\/}  Max-Planck-Institut f\"ur
Mathematik, Bonn preprint 1994}, the prepotential $F$ for Calabi-Yau
threefolds cannot be derived from the associativity of the structure
constants alone. For these cases one has on the other hand an
additional geometrical structure, known as special geometry,
which was discovered first in the context of
$N=2$ supergravity theories \ref\dlv{B. de Wit and A. Van Proeyen,
\npb245(1984)89, B. de Wit, P. Lauwers and A. Van Proeyen, \npb255(1985)569}
and derived for the K\"ahler and complex structure moduli spaces
of Calabi-Yau and/or $N=2$ string compactification in
\ref\seiberg{N. Seiberg, \npb303(1988)286}
\ref\ferraraetal{S. Cecotti, S. Ferrara and L. Girardello,
\ijmpa4(1989)2475; S. Ferrara, Nucl. Phys. (Proc. Suppl.) 11 (1989) 342}
\ref\dkl{L. Dixon, V. Kaplunovsky and J. Louis, \npb329(1990)27}\cd.
It implies that also the K\"ahler potential for the Weil-Peterson metric
derives from the prepotential (hence the name) as
\eqn\kaehler{K=-\log \left((t^i-\bar t^i)
(\partial_i F+\bar\partial_i \bar F)+
2\,\bar F- 2\,F\right).}
On the complex-structure side
the Yukawa couplings are expressed for a choice
of the holomorphic three-form $\Omega$ as \bg~
$K_{ijk}=\int \Omega \partial_i\partial_j\partial_k \Omega$
and the K\"ahler metric
is given by\foot{$\Sigma=\pmatrix{0&-\sigma
\cr\sigma&\phantom{-}0\cr}$, $\sigma=\pmatrix{0&1\cr1&0\cr}$}
$K=-\log({\Pi^{\dag}} \Sigma \Pi)$ where $\Pi$
is the period vector for a symplectic basis of $H_3(X)$.
As the holomorphic three-form $\Omega$ is defined only up to
gauge transformations $\Omega\rightarrow f(z)\Omega$ ($f(z)$
holomorphic) it takes values in a holomorphic line bundle $L$
over the moduli space. The transformation properties of the
quantities introduced are: $\Omega\in L$, $F\in L^2$,
the Yukawas $K_{ijk}$ transform as components
of elements in $L^2\otimes {\rm Sym} ((T^*_{\cal M})^{\otimes 3})$ and
$e^{-K}\in L\otimes \bar L$;
$e^{2K}g^{i\bar\imath}g^{j\bar\jmath}g^{k\bar k}K_{ijk}
\bar K_{\bar\imath\bar\jmath\bar k}$ is then invariant under K\"ahler
and coordinate transformations. The physical Yukawa
couplings appearing in the effective Lagrangian are those for
canonically normalized matter fields.

To relate the solution of the Picard-Fuchs equations directly to $F$,
we may use the form of the Picard-Fuchs differential operators in
special coordinates in the K\"ahler gauge
\ref\cfll{A. Ceresole, R. D'Auria, S. Ferrara,
W. Lerche and J. Louis,\ijmpa8(1993)79},
\eqn\pfspecial{
\sum_{i=1}^k \p_j \p_p (K_r^{-1})^{li}\p_r\p_i.}
The period vector is then trivially expressed in terms of $F$:
$\Pi(t)=(1,t^i,\p_i F,2\,F-t^i\p_i F)$ (here and below we again use
summation convention).
The relation of the coordinates  in  K\"ahler gauge $t^i$  to the homogeneous
special coordinates $(X^0,X^i)$,  in which the period vector reads
$(X^0,X^i,(\p {\cal F}/\p X^i), (\p {\cal F}/\p X^0))$, is given by
$t^i=X^i/X^0$ with ${\cal F}=(X^0)^2 F$.

Guided by the mirror hypothesis,
we should have the same structure for the
K\"ahler side and therefore, for a formal large radius expansion of $F^{K}$
\eqn\prepot{F^{K}= {1\over 6} K^0_{ijk} {t^i t^j t^k}+ {1\over 2}
             a_{ij} t^i t^j + b_i t^i  +
            {1\over 2} c +F_{\rm inst.}}
around ${\rm Im }(t^i)\rightarrow \infty$, the ``period vector''
\eqn\pIII{
X^0\left(\matrix{1\cr
                 t^i\cr
                 {1\over 2}  K^0_{ijk} t^j t^k + a_{ij}t^j +b_i
                 +\p_{t_i}F_{\rm inst.}\cr
                -{1\over 6} K^0_{ijk} t^i t^j t^k+b_i t^i+ c+
                 (2 F_{\rm inst.}-t^i \partial_{t_i} F_{\rm inst.})}
\right)=X^0\Pi(t).}
Here $K^0_{ijk}$ are the classical intersections calculated in section
2 and $F_{\rm inst.}$ is
a power series in $q_i=e^{2 \pi i t^i}$, which
encodes the instanton corrections.

Starting from \pI~ in $z_i$ coordinates we now have to make a coordinate
transformation to suitable inhomogeneous coordinates, which are defined by
ratios of solutions of the Picard-Fuchs equations as
\eqn\newcoord{ t^i(z)= {w_i(z)\over w_0(z)}.}
The reason for picking this quotient is that the
invariance of the theory under the mo\-no\-dromies around $z_i=0$
can be identified with the invariance of the topological sigma
model under real shifts $t^i\rightarrow t^i+1$
of the antisymmetric background field or, equivalently, of
the effective field theory under discrete Peccei-Quinn
symmetries. As will become clear below, this choice identifies
the form of the prepotential for the complex structure moduli space on
$X^*$ with the one expected for the K\"ahler structure moduli space on
$X$. It strongly resembles the elliptic curve case, where
the ratios of two arbitrary solutions of the period equation
and its inverse satisfy a third-order non-linear
differential equations introduced by Schwarz. Inserting the solutions
$z(t)$ in the $J$-invariant of the elliptic curve gives  the wellknown
expansion $J({at + b\over ct + d})={1\over 1728}(q^{-1}+ 144+\ldots)$ of the
$J$ function in terms of $q=e^{2 \pi i ({at + b\over c t +d})}$.
The choice of the logarithmic solution for $w_1$ and the power
series solution for $w_0$ at the point of maximal unipotent
monodromy then corresponds to $a=d=1,b=c=0$. For Calabi-Yau spaces
one likewise finds that the mirror maps $z_i(q_1,\ldots,q_h)$
always have an integral expansion, which is a necessary
condition for obtaining integer instanton numbers.
One therefore expects that the mirror map plays an
important r\^ole for the theory of modular forms on the
moduli space of Calabi-Yau space, which however seems to be
much more intricate as the modular group acts in general non-arithmetically
on the geometrical parameters.

Inserting the inverse map of \newcoord~into \pI~we obtain the
transformed period vector
\eqn\pIV{
\tilde \Pi(t)=w_0(t)\left(\matrix{1\cr
                 t^i\cr
                 {1\over 2}  \tilde K^0_{ijk} t^j t^k + \hat b_i
                 +\p_{t_i}\tilde F_{\rm inst.}\cr
                -{1\over 6} \tilde K^0_{ijk} t^i t^j t^k-
                \hat b_i t^i+ \hat c+
                 (2 \tilde F_{\rm inst.}-t^i \partial_{t_i}
                  \tilde F_{\rm inst.})}
\right)\,.}
By comparison with \pIII~ we can view all quantities marked with a
tilde,
up to one overall normalization, as predictions of mirror symmetry.
Especially we have to identify $K^0_{ijk}$ with
$\tilde K^0_{ijk}$. In fact, for all complete intersection
cases the top element of the ring $R$ encodes the intersection numbers
in
the integral basis of divisors coming from the ambient spaces.
This turns out to be true also for the hypersurfaces discussed
in \hkty, for the basis of divisors that generate $H_2(X,\ZZ)$.

After fixing the normalization, we see that the instanton-corrected
Yukawa coupling $K_{ijk}$ can be uniquely expressed by the solutions
of the Picard-Fuchs equations as
\eqn\icy{
K_{ijk}(t)=\p_{t_i}\p_{t_j}{D^{(2)}_k w_0|_{\rho=0}\over w_0}(t)}
or, equivalently, by the derivatives of the prepotential
\eqn\prepot{
F(t)={1\over2}\left({1\over w_0}\right)^2
\Blb w_0 D^{(3)}w_0+D^{(1)}_l w_0 D^{(2)}_l w_0 \Brb(t)
  \mid_{\rho=0}. }
We will use formulas \icy, \prepot, which does not require
the evaluation of the Yukawa couplings on the mirror manifold,
to compute a convergent expansion for the instanton-corrected Yukawas
and the prepotential in the large radius limit, and to predict
the number of rational curves for various manifolds.

So far we have seen that the natural choice for the
period vector \pI~matches the leading terms in $t$ of the
components of \pIII~ and leads to a prediction
of the instanton corrections. Let us now compute also the
lower-order terms in $t$, i.e. the constants $\hat b_i,\hat c$. For this
purpose one has to take the derivatives w.r.t. $\rho$ explicitly
\eqn\derivatives{
\eqalign{
\pa{i} c(0)&
=-\ds{\left(\sum_{\alpha} l^{(i)}_{0\alpha}+
\sum_{\alpha} l_\alpha^{(i)}\right)\Gamma'(1)
=: c_1^i\Gamma'(1)=0}
\cr
\pa{i}\pa{j} c(0)&
=\ds{ {\pi^2\over 6}\left(\sum_{\alpha} l^{(i)}_{0\alpha} l^{(j)}_{0\alpha}-
\sum_{\alpha}l^{(i)}_\alpha l^{(j)}_\alpha\right)
=: {\pi^2\over3} c_2^{ij} }\cr
\pa{i}\pa{j}\pa{k} c(0)&
=\ds{2\left(
\sum_{\alpha} l^{(i)}_{0\alpha} l^{(j)}_{0\alpha} l^{(k)}_{0\alpha}+
\sum_{\alpha} l^{(i)}_\alpha l^{(j)}_\alpha l^{(k)}_\alpha\right)\zeta(3)
=: 6 c_3^{ijk}\zeta(3)},}}
where the coefficients $c_n$ coincide for complete intersections
in products of non-singular weighted projective spaces
with the ones given in \chern.
Using the Gauss-Bonnet formula we get
from this the following remarkable identities
\eqn\identities{
\eqalign{
\int_X c_2 J_i&=\ds{\int_X c_2^{jk} J_j J_k J_i=
{3 \over \pi^2}\int_X J_j J_k J_i\pa{j} \pa{k} \, c(0)=
-24 D^{(2)}_i c(0)}\cr
\chi&=\ds{\int_X c_3^{ijk}J_i J_j J_k={1\over 6\zeta(3)}
\int_X J_i J_j J_k\, \pa{i}\pa{j}\pa{k}c(0)=
i{(2\pi )^3\over  \zeta(3)} D^{(3)} c(0)},}}
with $D^{(2)}_i$ and $D^{(3)}$ defined in \defroder, which express
the Euler number and $\int_M c_2 J_i$ in terms of the intersection numbers
and the generators of the Mori cone $l^{(i)}$. These identities hold
also for the canonically resolved manifolds, if the complete
intersection had canonical quotient singularities.
The linear terms in the third row of \pIV~
are thus $\hat b_i=-{1\over 24}\int_M c_2 J_i$ and in the last row
$\hat c=-i {\zeta(3)\over (2 \pi )^3 }\chi$.

The imaginary part of the constant $\hat c$
can be identified with a  $\sigma$-model loop contribution
and is proportional to the Euler number of the manifold. The
constant of proportionality seems to be universal and has been
calculated
explicitly for the quintic hypersurface in $\IP^4$
and other one modulus cases ~as
$-i {\zeta(3)\over (2\pi)^3}$.  It is a necessary condition in order to
have a continuous Peccei-Quinn symmetry $t_j\to t_j+\alpha_j,\,\alpha_j$
real, which is broken by instanton corrections to discrete shifts
$t^i\to t^i+1$, that ${\rm Im}(a_{ij})={\rm Im}(b_i)=0$,
as we mentioned before.

While the real constants $b_i$, $a_{ij}$ in \pIII~are irrelevant for
physical quantities, they are important for fixing an integral
symplectic
basis for the period vector. In fact we find that the constant
$\hat b_i$ for the one modulus cases \ktone\font~is also correctly
reproduced
in the third line. The constants in front of the first subleading terms
in \pIV~do not correspond to a choice of an integral basis, as can be
seen by comparison with (4.13).
Especially the constants $a_{ij}$
do not seem to be directly related
to topological numbers; in fact there are no further topological numbers
at our disposal.
These constants have to be fixed by analysing the monodromy matrices
themselves.
The monodromy operation $T_i:t^i\rightarrow t^i+1$ on \pIII~is
obvious. The requirement that it is integral and
symplectic yields restrictions on $a_{ij}$,
e.g. for the one modulus cases $a=({K^o\over 2}\, {\rm mod}\, \ZZ)$ (and
$2 b=((1-{K^o\over 6})\, {\rm mod }\, \ZZ)$). The integer part
is irrelevant as it can be absorbed by an $SP(4,\ZZ)$ transformation,
hence
the basis can be specified. Although we have no general proof, it is
tempting to conjecture that the occurrence of the topological numbers
$K_{ijk}^0$, $\chi$ and, up to $SP(2 h+2,\ZZ)$ transformations, also
$\int_X c_2 J_i$ in $F$ at the point of maximal unipotent monodromy
in a integral sympletic basis, is a general feature.

The instanton part of the Yukawa couplings comes from stationary
points of the classical string action, which correspond to holomorphic
mappings from $\IP^1$ to the CY manifold.
In the coordinates  $t^i$ defined in \newcoord, it enjoys
the following expansion
\eqn\icye{
\eqalign{
K_{klm}
&=\int_X J_k \wedge J_l \wedge J_m+\sum_d \int_{{\cal M}_{C_d}} d\mu
\,\, {e^{2 \pi i\int_{C_d} K(X)}
\over 1-e^{2 \pi i \int_{C_d} K(X)}}\cr
&=K^0_{klm}+\sum_{d_1,\ldots,d_k} {n^r_{d_1,\ldots,d_k} d_k\, d_l\,
d_m\over
1-\prod_{i=1}^k q_i^{d_i}} \prod_{i=1}^k q_i^{d_i}\, .}}
Here we have defined the degree of the curve as $d_i=\int_C J_i$,
which is an integer for a basis  $J_i\in H^{1,1}(\hat X,\ZZ)$.
The denominator $(1-\exp{\int_{C_d} K(x)})$
gives the correct combinatorical contributions from multi-covers of the
curves such that the integral $d\mu$ of the Euler class over the
compactified
moduli space ${\cal M}_{C_d}$ of holomorphic maps of multi-degree
$d$ from $\IP^1$ \ref\witten{E. Witten, in {\sl Essays on mirror
manifolds\/} Ed. S.T. Yau, (Int. Press. Hong Kong 1992)}
can be taken over single-cover curves only. The resulting
invariants $n^r_{d_1,\ldots,d_k}$ are expected to be integers: for
isolated curves they just count their numbers. In \hkty~examples
of negative invariants $n^r$ were found. They admit only the
interpretation that there are non-isolated singular curves at
the corresponding degree.
The occurrence of the terms $d_k\,d_l\,d_m$ in the second line is due to
the
integral of the part of the moduli space describing the
reparametrization of $\IP^1$, as was explained for isolated curves in
\ref\am{P. Aspinwall and D. Morrison \cmp151(1993)45}.

%%%%%%%%%%%%%%%%%%%%%%%%%%%%%%%%%%%%%%%%%%%%%%%%%%%%%%%%%%%
\newsec{Selected examples}
%%%%%%%%%%%%%%%%%%%%%%%%%%%%%%%%%%%%%%%%%%%%%%%%%%%%%%%%%%

\noindent({\sl i}) As our first example, we will calculate the
topological invariants $n^r$ for the manifold \schimm.
The ideal ${\cal I}$ is generated by
$I_1=\t_1^2-3\,\t_1\t_2+9\,\t_2^2$ and $I_2=\t_1^3$,
so $R$ is spanned by
\eqn\idealexample{
\eqalign{1&\cr \t_1,\,\, &\,\,\t_2\cr 9\, \t_1\t_2+3\, \t_2^2,\,\, &
\,\, 9\,\t_1^2+ 3\, \t_1\t_2
\cr 9\,  \t_1^2 \t_2 &+ 3\, \t_1 \t_2^2.}
}
By \corres~this translates to a basis of solutions
for $L_1,L_2$  as may be verified.
The fact that the top element of $R$
coincides, up to scaling, with $K^0 = 9\,J_1^2 J_2 + 3\,J_1 J_2^2$,
calculated
by \top, is a non-trivial check on the mirror hypothesis.
Using \icy~one obtains concise
formulas for the instanton-corrected intersection numbers:
\eqn\conciseformula{
K_{i,j,1}(t)=\p_{t_i}\p_{t_j} {
(9\,  \p_{\rho_1}\p_{\rho_2}+{3\over 2}\p_{\rho_2}^2)w_0|_{\rho=0}
\over w_0}(t),\quad
K_{i,j,2}(t)=\p_{t_i}\p_{t_j} {
({9\over 2} \p_{\rho_1}^2 + 3\,\p_{\rho_1}\p_{\rho_2})w_0|_{\rho=0}
\over w_0}(t)
}
from which the topological invariants $n^r$ follow by comparison
with \icye. Ee display them below for multi-degrees
$(d_{J_1},d_{J_2})$, $d_{J_1}+d
_{J_2}\leq 6$.

$$
\vbox{\offinterlineskip
\hrule
\halign{ &\vrule# & \strut\quad\hfil#\quad\cr
\noalign{\hrule}
height1pt&\omit&   &\omit&\cr
&(1,0)&& 243  &\cr
&(2,0)&& 243  &\cr
&(3,0)&&  54 &\cr
&(4,0)&& 243  &\cr
&(5,0)&& 243  &\cr
&(6,0)&&  54  &\cr
height1pt&\omit&   &\omit&\cr
\noalign{\hrule}}
\hrule}
\,\,
\vbox{\offinterlineskip
\halign{ &\vrule# & \strut\quad\hfil#\quad\cr
\noalign{\hrule}
\noalign{\hrule}
height1pt&\omit&   &\omit&\cr
&(0,1)&& 63  &\cr
&(0,2)&& 63  &\cr
&(0,3)&& 54 &\cr
&(0,4)&& 63  &\cr
&(0,5)&& 63  &\cr
&(0,6)&& 54  &\cr
height1pt&\omit&   &\omit&\cr
\noalign{\hrule}}
\hrule}
\quad
\vbox{\offinterlineskip
\halign{ &\vrule# & \strut\quad\hfil#\quad\cr
\noalign{\hrule}
\noalign{\hrule}
height1pt&\omit&   &\omit&\cr
&(1,1)&&  972 &\cr
&(2,2)&&  156249  &\cr
&(3,3)&&  60018786 &\cr
height1pt&\omit&   &\omit&\cr
\noalign{\hrule}
\noalign{\hrule}
height1pt&\omit&   &\omit&\cr
&(2,1)&& 15309  &\cr
&(4,2)&& 111401163  &\cr
\noalign{\hrule}
height1pt&\omit&   &\omit&\cr
&(3,1)&&  179901  &\cr
height1pt&\omit&   &\omit&\cr
\noalign{\hrule}
\noalign{\hrule}
height1pt&\omit&   &\omit&\cr
&(4,1)&&  1558845  &\cr
height1pt&\omit&   &\omit&\cr
&(5,1)&&  11558295  &\cr
height1pt&\omit&   &\omit&\cr
\noalign{\hrule}}
\hrule}
\,\,
\vbox{\offinterlineskip
\halign{ &\vrule# & \strut\quad\hfil#\quad\cr
\noalign{\hrule}
\noalign{\hrule}
height1pt&\omit&   &\omit&\cr
&(1,2)&&   3402 &\cr
&(2,4)&&   4803867 &\cr
height1pt&\omit&   &\omit&\cr
\noalign{\hrule}
\noalign{\hrule}
height1pt&\omit&   &\omit&\cr
&(3,2)&&  4830597 &\cr
height1pt&\omit&   &\omit&\cr
\noalign{\hrule}
\noalign{\hrule}
height1pt&\omit&   &\omit&\cr
&(1,3)&&  9720 &\cr
height1pt&\omit&   &\omit&\cr
\noalign{\hrule}
\noalign{\hrule}
height1pt&\omit&   &\omit&\cr
&(2,3)&&  977589  &\cr
height1pt&\omit&   &\omit&\cr
\noalign{\hrule}
\noalign{\hrule}
height1pt&\omit&   &\omit&\cr
&(1,4)&&   25515 &\cr
height1pt&\omit&   &\omit&\cr
\noalign{\hrule}
\noalign{\hrule}
height1pt&\omit&   &\omit&\cr
&(1,5)&&  61236 &\cr
height1pt&\omit&   &\omit&\cr
\noalign{\hrule}}
\hrule}
$$
In fact, $J_i$ is a basis of a subspace of $H^2(X,\ZZ)$
in which the K\"ahler cone is simply given by
\eqn\Kaehlercone{
\sigma(K)=\left\{\sum_i t_i J_i| t_i>0\right\}
}

In \rolf~a three-generation model $(\chi=-6$) was constructed, by
dividing the manifold \schimm~by a $G=\ZZ_3\times \ZZ_3$ symmetry and
resolving the singular quotient.
The group $G$ is generated, on the homogeneous
coordinates of $\IP^3\times \IP^2$, by
$(x_0,x_1,x_2,x_3,y_1,y_2,y_3)\mapsto
(x_0, x_1,x_2, x_3,\alpha y_1,y_2,\alpha^{-1}y_3)$
($\alpha=\exp{2\pi i\over 3}$), with three fixed tori $T_i=\{x\in
\IP^3|x_i=0,p_1=0\}$,
$i=1,2,3$,
and a freely acting cyclic permutation
$(x_0,x_1,x_2,x_3,y_1,y_2,y_3)\mapsto (x_0,x_2,x_3,x_1,y_2,y_3,y_1)$.
This action is only compatible with a subspace of the
moduli space of complex structure
deformations, namely with the following perturbations
\eqn\schim{\eqalign{
           p_1&=\sum_{i=1}^3 x_i y_i^3 - \tilde a_1 x_0 y_1 y_2 y_3\cr
           p_2&=\sum_{i=0}^3 x_i^3  - \tilde a_2 x_1 x_2 x_3.}}
To relate this description to \laurant, one relates
the variables $X_{1,i}\;(i=1,\cdots,4)$ and $X_{2,j}\;(j=1,2,3)$
in \laurant~ to the following Laurent monomials of the homogeneous
coordinates of $\IP^3\times\IP^2$,
\eqn\map{
\eqalign{
&
X_{1,1}={y_1^3x_1 \over x_0y_1y_2y_3}\;,\;\,\,\,
X_{1,2}={y_2^3x_2 \over x_0y_1y_2y_3}\;,\;\,\,\,
X_{1,3}={y_3^3x_3 \over x_0y_1y_2y_3}\;,\;\,\,\, \cr
&
X_{1,4}={x_0^3 \over x_1x_2x_3}\;,\;\,\,\,
X_{2,1}={x_1^3 \over x_1x_2x_3}\;,\;\,\,\,
X_{2,2}={x_2^3 \over x_1x_2x_3}\;,\;\,\,\,
X_{2,3}={x_1^3 \over x_1x_2x_3}\;,\; \cr} }
by which we can identify \laurenta~ with \schim~ and the parameter $a_i$
with
${\tilde a}_i$ ($a_{ij}=1$).
Furthermore we see that \laurenta~
is invariant under $G$, which implies that
the subsystem of the moduli space we are considering is in the
invariant sector of $G$ and part of the moduli space of the
three-generation model.

The mirror manifold can also be constructed as the quotient of
\schim~w.r.t. the group of order $27$ generated by
$(x_0,x_1,x_2,x_3,y_1,y_2,y_3;a_1,a_2)\mapsto
(x_0, \beta^{3(m+n)} x_1,\beta^{6 m} x_2,\beta^{6 n} x_3,$
$\beta^{-m-n} y_1, \beta^{-8 m} y_2,\beta^{-8 n} y_3;a_1,a_2)$
with $\beta=\exp{2 \pi i\over 9}$ $m,n\in \ZZ$. The full invariance
group of \schim~is generated by the above transformations and
$(x_0,x_1,x_2,x_3,y_1,y_2,y_3;a_1,a_2)\mapsto
(x_0, \beta^{3 k} x_1, x_2,x_3,
\beta^{-k} y_1, y_2, y_3;$ $\beta^k a_1,\beta^{6 k} a_2)$, $k\in \ZZ$,
i.e. we have the same identification of the parameter space of the
mirror manifold as we found for \laurenta.

We can always find a map analogous to \map, which maps the Laurent
polynomials $P_1,\dots, P_l$ to quasi-homogeneous polynomials
$p_1,\dots,p_l$. In terms of those the fundamental period $w_0$ can be
obtained as the integral \bcdfhjq
\eqn\pcd{w_0=\prod_{i=1}^l (-a_i) \int {1\over p_1\cdots p_l}
dx_{1,1},\ldots dx_{k,n_k+1}\, ,}
with integration contours $|x_{i,j}|=1$.
While we can always define a map
such that $w_0$ is given by this expression, in most
of the cases however this map cannot be chosen such that the
polynomials $p_1,\ldots,p_l$ define a transverse configuration
in $\IP^{n_1}\(\vec w^{(1)}\)\times\cdots \times \IP^{n_k}\(\vec
w^{(k)}\)$.
An example of such a case will be given at the end of section
6.

Let us demonstrate the direct calculation of the discriminant
locus for the manifold defined in \laurenta~ by examining the
conditions of $\Delta$-regularity which reads that for all faces
$\Theta_1^*$ and $\Theta_2^*$ of $\Delta_1$ and $\Delta_2^*$
we need $P_{\Theta_1^*}=P_{\Theta_2^*}=0$ and that all
$2\times 2$ sub-determinants of the matrix whose two rows are
$X_{i,j}{\partial P_{\Theta_1^*}\over\partial X_{i,j}}$ and
$X_{i,j}{\partial P_{\Theta_2^*}\over\partial X_{i,j}}=0$ (no sum)
vanish simultaneously.

For convenience we rename the parameters $a_i$, $a_{i,j}$
occurring in $P_1$ and $P_2$ as $a_0,\ldots,a_4$ and  $b_0,\ldots,b_3$.
If we introduce variables $U_i\equiv U_{1,i}$ and $V_j\equiv U_{2,j}$
such that $X_{i,j}={U_{ij}\over U_{i0}}$
the conditions for non-regularity become for $\Theta_1^*=\Delta_1^*$
and $\Theta_2^*=\Delta_2^*$:

\eqn\bb{
\eqalign{\sum_{i=0}^3 a_i U_i=0\,,&\quad \sum_{j=0}^3 b_j V_j +a_4
U_4=0\cr
{a_4 U_4\over V_0 U_0}(a_i U_i-a_j U_j)=0\,,\quad&\quad
{a_i U_i\over U_0 V_0}(b_3 V_3-b_k V_k)=0\,,\quad i,j=1,2,3\,,\quad
k=1,2\, .}
}
Note that ${U_1 U_2 U_3 U_4\over U_0^3 V_0}={V_1 V_2 V_3\over
V_0^3}=1$.
There are several ways to satisfy the second set of equations.
For instance if we
choose to satisfy them by equating the expressions in parentheses to
zero,
we find, using the variables $z_1={a_1 a_2 a_3 a_4\over a_0^3 b_0},\,
z_2={b_1 b_2 b_3\over b_0^3}$ and introducing $A=a_4 U_4$ and $B=b_0
V_0$,
$z_1=-{1\over 27}{A\over B}$ and $z_2={1\over 27}{(A+B)^3\over B^3}$,
which satisfy $(1-27\, z_1)^3-27\, z_2=0$.
Another possible choice is to set $a_4 U_4=0$ and $(b_3 V_3-b_k
V_k)=0$. This leads to $1-27\, z_2=0$.
Other components of the discriminant can be obtained by considering
other (lower-dimensional) faces of $\Delta_1^*$ and $\Delta_2^*$.

\vskip 5 mm
\noindent({\it ii}) Our next example is the Tian-Yau manifold
given by the configuration
\eqn\aa{
\cicy-18(\IP^3\cr \IP^3|3& 0& 1\cr 0&3 &1){14}\,:
\quad  \eqalign{&K^0 = 9\,J_1^2 J_2+9\, J_1 J_2^2,\cr
                &c_2 J_1=c_2 J_2=36.}}
The quotient of this manifold w.r.t. the $\ZZ_3$ group acting
by $(x_0,x_1,x_2,x_3,y_0,y_1,y_2,y_3)\mapsto
(x_0,x_1,x_2,\alpha x_3,y_0,y_1,y_2,\alpha^2 y_3)$
($\alpha=e^{2\pi i\over 3}$) on the homogeneous coordinates
of $\IP^3\times \IP^3$ yields a simple realization of a three
generation compactification, which is diffeomorphically equivalent
to the one discussed in 5 ({i}).
A preliminary phenomenological discussion was given in
\ref\sm{S. Kalara and R. N. Mohapatra,\prd36(1987)3474}
and
\ref\gkmr{B. R. Greene, K. H. Kirklin,
          P. J. Miron and G. G. Ross, \npb278(1986)667}.
For the above configuration all complex structure
deformations can be represented as monomials.
Hence the full moduli dependence of all Yukawa
couplings and the K\"ahler potential for $27$'s and
$\overline{27}$'s can in principle be calculated
by a straightforward, although very tedious, application of the methods
described in \hkty~and \cdfkm.

The vertices of the dual polyhedra are now
$\nu^*_{1,1}\!=\!(1,0,0;0,0,0),\dots,$
\hfill\break
$\nu^*_{2,4}\!=\!(0,0,0;-1,-1,-1)$,
which we group into three sets as
$E_1=\{\nu^*_{1,1},\nu^*_{1,2},\nu^*_{1,3}\},\,
E_2=\{\nu^*_{2,1},\nu^*_{2,2},\nu^*_{2,3}\}$ and
$E_3=\{\nu^*_{1,4},\nu^*_{2,4}\}$
\foot{The results are, of course, again independent of the way we group
the
vertices into three sets, as long as the first and second sets contain
three vertices pertaining to the first and the the second $\IP^3$,
respectively.}.
This corresponds to the three Laurent  polynomials
\eqn\bb{
\eqalign{P_1&=a_0-a_1 X_1-a_2 X_2-a_3 X_3\,,\qquad
         P_2=b_0-b_1 Y_1-b_2 Y_2-b_3 Y_3\,,\cr
         &\qquad\qquad\qquad P_3=c_0-{c_1\over X_1 X_2 X_3}-
         {c_2\over Y_1 Y_2 Y_3}\,.}}
The period \powersol~follows straightforwardly by
performing the integral (see \periods)
\eqn\periodintegral{
w_0(z_1,z_2)=
\int_{|X_i|=|Y_j|=1}{a_0 b_0 c_0\over P_1 P_2 P_3}\prod{dY_i\over Y_i}
{dX_i\over X_i}\, ,
}
which gives $w_0(z_1,z_2)$ as in eq. \powersol~
with $z_1={a_1 a_2 a_3 c_1\over a_0^3 c_0}$ and
$z_2={b_1 b_2 b_3 c_2\over b_0^3 c_0}$.
This corresponds to introducing  extended vertices
$\bar\nu^*=(\vec e_i,\nu^*)$,
where $\vec e_1=(1,0,0),\,\vec e_2=(0,1,0)$ and $\vec e_3=(0,0,1)$
for the vertices in the three sets. The linear relations between the
$\bar\nu^*$ are then in accordance with
the general formula \lindiffop~and read
\eqn\phIlattice{
l^{(1)}=(-3,0,-1; 1,1,1,1,0,0,0,0)\, ,  \quad
l^{(2)}=( 0,-3,-1; 0,0,0,0,1,1,1,1).
}
The associated differential operators
\eqn\ccc{
\eqalign{
&\cL_1=\t_1^3 - 3\, (\t_1+  \t_2)\,(3\,\t_1-1)\,(3\,\t_1-2)\,z_1\cr
&\cL_2=\t_2^3 - 3\, (\t_1 + \t_2)\,(3\,\t_2-1)\,(3\,\t_2-2)\,z_2\cr}}
can be factorized $\cL_1 + \cL_2\equiv(\t_1+\t_2)L_2$ with the
second-order operator ($L_1=\cL_1$)
\eqn\ddd{
L_2=(\t_1^2-\t_1\t_2+\,\t_2^2)-3\,
(3\,\t_1-1)(3\,\t_1 -2)\, u - 3\, (3\, \t_2 -1)\, (3\, \t_2-2)\,z_2.}
The Yukawa couplings are
\eqn\eee{
\tK^{(3,0)}={1\over  z_2^2\Delta_0\Delta_1^2},\quad
\tK^{(2,1)}={1 \over 27\, z_1^2 z_2\Delta_0 \Delta_1},}
where the general component of the discriminant surface is
$\Delta_0=1-27 z_1 - 27 z_2$ and a second component
reads $\Delta_1=(1-27 z_1)$.
For symmetry reasons $\tK^{(0,3)}$, $\tK^{(1,2)}$ are
given by the above expressions, but with $z_1$ and $z_2$
exchanged. Due to this symmetry  we
list below the invariants $n^r_{d_1,d_2}$ only for
$d_1\leq d_2$.

%{{{0, 1}, 81}, {{1, 0}, 81}, {{1, 1}, 729}, {{1, 2}, 2187},
%{{1, 3}, 6885}, {{1, 4}, 18954}, {{1, 5}, 45927}, {{1, 6}, 305226954},
%{{2, 1}, 2187}, {{2, 3}, 300348}, {{2, 5}, 61909596}, {{3, 1}, 6885}
%{{3, 2}, 300348}, {{3, 4}, 76452057}, {{4, 1}, 18954}, {{5, 1},45927},
%{{6, 1}, 92241}, {{0, 2}, 81}, {{2, 0}, 81}, {{2, 2}, 33534},
%{{2, 4}, 1708047}, {{4, 2}, 1708047}, {{0, 3}, 18}, {{3, 0}, 18},
%{{3, 3}, 5433399}, {{0, 4}, 81}, {{4, 0}, 81}, {{0, 5}, 81},
%{{5, 0}, 81}, {{0, 6}, 18}, {{6, 0}, 18}

$$
\vbox{\offinterlineskip
\hrule
\halign{ &\vrule# & \strut\quad\hfil#\quad\cr
\noalign{\hrule}
height1pt&\omit&   &\omit&\cr
&(0,1)&& 81 &\cr
&(0,2)&& 81 &\cr
&(0,3)&& 18 &\cr
&(0,4)&& 81 &\cr
&(0,5)&& 81 &\cr
&(0,6)&& 18 &\cr
height1pt&\omit&   &\omit&\cr
\noalign{\hrule}}
\hrule}
\quad
\vbox{\offinterlineskip
\halign{ &\vrule# & \strut\quad\hfil#\quad\cr
\noalign{\hrule}
\noalign{\hrule}
height1pt&\omit&   &\omit&\cr
&(1,1)&&  729 &\cr
&(2,2)&&  33534 &\cr
&(3,3)&& 5433399  &\cr
height1pt&\omit&   &\omit&\cr
\noalign{\hrule}
\noalign{\hrule}
height1pt&\omit&   &\omit&\cr
&(1,2)&& 2187  &\cr
&(2,4)&& 1708047 &\cr
height1pt&\omit&   &\omit&\cr
\noalign{\hrule}}
\hrule}
\quad
\vbox{\offinterlineskip
\halign{ &\vrule# & \strut\quad\hfil#\quad\cr
\noalign{\hrule}
\noalign{\hrule}
height1pt&\omit&   &\omit&\cr
&(1,3)&&  6885 &\cr
height1pt&\omit&   &\omit&\cr
\noalign{\hrule}
\noalign{\hrule}
height1pt&\omit&   &\omit&\cr
&(1,4)&&  18954&\cr
height1pt&\omit&   &\omit&\cr
\noalign{\hrule}
\noalign{\hrule}
height1pt&\omit&   &\omit&\cr
&(1,5)&&  45927 &\cr
height1pt&\omit&   &\omit&\cr
\noalign{\hrule}
\noalign{\hrule}
height1pt&\omit&   &\omit&\cr
&(2,3)&&  300348 &\cr
\noalign{\hrule}}
\hrule}
$$
\vskip 3 mm

The number of curves up to  degree 3 agree with those calculated
by algebraic counting methods in \sommervoll. This calculation
confirms the lines $(0,1)$ and the degree $(0,2)$, $(2,1)$ curves
on this CICY.
According to \sommervoll~ there are no $(0,3)$ curves and $567$ degree
$(1,1)$ curves, leaving aside the possibility of nodal cubics
and degenerate rational curves respectively. More recently, all
entries up to degree 3 have been confirmed
\ref\sommervollII{D. E.
Sommervoll, private communication}.
The invariant
$n^r_{0,3}=18$ was calculated as the Euler number of the tangent
bundle of the moduli space of a family of nodal cubics.

In Appendix B we present general formulas for the predicted numbers
of lines for all possible manifolds of type (5 {\it i}) in
$\IP^3\times\IP^2$
and of type (5 {\it iii}) in $\IP^3\times\IP^3$. In deriving these
formulas,
we utilized formula \icy~ for the instanton-corrected Yukawa couplings.

Next we  will treat two examples, which involve a somewhat more
complicated
factorization procedure, but skip in the following most of the details.

\vskip 5 mm
\noindent({\it iii})
For the  manifold defined by the configuration matrix
$$
\cicy-120(\IP^3\cr \IP^3|3&1&0\cr 1&1&2){12}
$$
the linear differential operators \lindiffop~ are, after trivial
factorization,
of fourth and third order, respectively.
We get a second-order operator by factorising
$27 \cL_1+ (40 \t_1 +13 \t_2) \cL_2= (\t_1+\t_2)(3\t_1+\t_2)L_1$ with
\eqn\fff{
L_1= ( 9 \t_1- 12 \t_1 \t_2+ 13 \t_2^2)-
27(3\t_1+\t_2-1)(3\t_1+\t_2-2)\,z_1-2(2\t_2-1)(40\t_1+13\t_2)\, z_2.}
Because of the Yuakwa couplings complexity, we refrain from giving them
explicitly. The expansion of the prepotential by \prepot~does not
require
their knowldege. From the first few terms in the
expansion of the prepotential
\eqn\ggg{
\eqalign{F=&
{2\over 6} (t^1)^3+
{6\over 2} t^1 (t^2)^2 +
{8\over 2}(t^1)^2 t^2 +
{44\over 24}t^1+
{48\over 24}t^2+120 {i\zeta(3)\over {(2 \pi )^3}}
+180 q_1+{405\over 2} q_1^2+{380\over 3} q_1^3
\cr &+
48 q_2+876 q_1 q_2+9772 q_1^2 q_2+
46 q_2^2 + 3536 q_1 q_2^2  + {16\over 9} q_2^3+O(q^4),}}
we can read off the number of lines of degrees  $(1,0)$ and $(0,1)$ as
$180$ and $48$, respectively.
Other curves of low degree come with multiplicity $n^r_{2,0}=180$,
$n^r_{0,2}=40$,
$n^r_{1,1}=876$, $n^r_{1,2}=3536$ and $n^r_{2,1}=9672$.

\vskip 3 mm
\noindent({\it iv})
As the last two-moduli example we choose a model with five bilinear
constraints
$$\cicy-32(\IP^4\cr \IP^4|1&1&1&1&1\cr 1&1&1&1&1){12}$$
Starting from
\eqn\hhh{
\cL_1 =\t_1^5-(\t_1+\t_2)^5 z_1,\qquad
\cL_2 =\t_2^5-(\t_1+\t_2)^5 z_2,}
we factorize in five steps, namely
($1$) $\cL_1+\cL_2=(\t_1+\t_2)\cL_3$,
($2$) $\t_2 \cL_3-5 \cL_2=(\t_1+\t_2)\cL_4$,
($3$) $2 \cL_3+ \cL_4=(\t_1+\t_2)\cL_5$,
($4$) $2 \cL_3-\t_1 \cL_5=(\t_1+\t_2)\cL_6$,
($5$) $ \cL_5 + 2 \cL_6=(\t_1+\t_2)\cL_7$ and
choose for example $\cL_5$ and $\cL_7$ as the third- and
second-order differential operators $L_1$ and $L_2$, respectively.
The prepotential as derived from \prepot~reads
\eqn\iii{
\eqalign{F=&{5\over 6} \sum_i (t^i)^3 +
{10\over 2}\sum_{i,j}' t^i (t^j)^2 +
{20\over 24}\sum_i t^i +
32 {i\zeta(3)\over (2 \pi)^3}+ \sum_i \left(
50 q_i + {25\over 4} q_i^2+{25\over 32} q_i^4\right)
\cr & +
\sum_{i,j}'\left({650\over 2}q_i q_j + 1475 q_i^2 q_i+ 650 q_i^3
q_j+{117725\over 8} q_i^2 q_j^2\right) + O(q^5)\, ;}}
here the summation indicated with $^\prime$ is
over distinct indices only.

Hence we have in total $100$ lines  and the non-zero
invariants  of curves up  to degree four are
$n^r_{1,1}=n^r_{1,3}=650$, $n^r_{1,2}=1475$ and
$n^r_{2,2}=29350$.

\vskip 3 mm
\noindent({\it v})
As a three-moduli example we consider
$$\cicy-48(\IP^3\cr \IP^1\cr \IP^1|3&1\cr 0&2\cr 0&2)9
\,:\quad\eqalign{
&K^0= 6\,J_1^2 J_2+6\, J_1^2 J_3 + 3 J_1 J_2 J_3,\cr
&c_2 J_1=36,\, c_2 J_2= c_2 J_3=24.}$$
for which we have, from \lindiffop:
\eqn\jjj{
\eqalign{
\cL_1&=\t_1^3 - 3 (3 \t_1-2)(3 \t_1-1) (\t_1+2 \t_2+2 \t_3)\, z_1 \cr
\cL_2&=\t_2^2 - (\t_1 + 2 \t_2 + 2 \t_3 -1)(\t_1+ 2 \t_2+ 2 \t_3)\, z_2
\cr
\cL_3&=\t_3^2 - (\t_1 + 2 \t_2 + 2 \t_3 -1)(\t_1+ 2 \t_2+ 2 \t_3)\,
z_3.}}
An independent second-order differential operator $L_4$
($L_i=\cL_i,\,i=1,2,3$)
can be factorized from the system in the following way
\eqn\kkk{
\cL_1+(16\t_3-4\t_1)\cL_2+(16\t_2-4\t_1)\cL_3=(\t_1+2\t_2+2\t_3)L_4.}
Due to their symmetry, we have
$n^r_{d_1,d_2,d_3}=n^r_{d_1,d_3,d_2}$
and will thus list the non-zero
invariants only for $d_2\leq d_3$
in the range $d_1+d_2+d_3\leq 6$.
$$
\vbox{\offinterlineskip
\halign{ &\vrule# & \strut\quad\hfil#\quad\cr
\noalign{\hrule}
\noalign{\hrule}
height1pt&\omit&   &\omit&\cr
&(0,0,1)&&    18   &\cr
height1pt&\omit&   &\omit&\cr
\noalign{\hrule}
\noalign{\hrule}
height1pt&\omit&   &\omit&\cr
&(0,1,1)&&    60    &\cr
&(0,2,2)&&    48   &\cr
&(0,3,3)&&    60 &\cr
height1pt&\omit&   &\omit&\cr
\noalign{\hrule}
\noalign{\hrule}
height1pt&\omit&   &\omit&\cr
&(0,1,2)&&    18 &\cr
%&(2,4,0)&&     0 &\cr
height1pt&\omit&  &\omit&\cr
\noalign{\hrule}
\noalign{\hrule}
height1pt&\omit&   &\omit&\cr
&(0,2,3)&&    18 &\cr
%&(2,4,0)&&     0 &\cr
height1pt&\omit&  &\omit&\cr
\noalign{\hrule}
\noalign{\hrule}
height1pt&\omit&   &\omit&\cr
&(0,3,4)&&    18 &\cr
%&(2,4,0)&&     0 &\cr
height1pt&\omit&  &\omit&\cr
\noalign{\hrule}}
\hrule}
\,\,
\vbox{\offinterlineskip
\halign{ &\vrule# & \strut\quad\hfil#\quad\cr
\noalign{\hrule}
\noalign{\hrule}
height1pt&\omit&   &\omit&\cr
&(1,0,0)&&    216  &\cr
&(2,0,0)&&    216  &\cr
&(3,0,0)&&     48  &\cr
&(4,0,0)&&    216  &\cr
&(5,0,0)&&    216  &\cr
&(6,0,0)&&     48   &\cr
height1pt&\omit&   &\omit&\cr
\noalign{\hrule}
\noalign{\hrule}
height1pt&\omit&   &\omit&\cr
&(1,0,1)&&    216 &\cr
&(2,0,2)&&    216 &\cr
&(3,0,3)&&     48 &\cr
height1pt&\omit&  &\omit&\cr
\noalign{\hrule}}
\hrule}
\,\,
\vbox{\offinterlineskip
\halign{ &\vrule# & \strut\quad\hfil#\quad\cr
\noalign{\hrule}
\noalign{\hrule}
height1pt&\omit&   &\omit&\cr
&(1,1,1)&&    1512 &\cr
&(2,2,2)&&    621000 &\cr
height1pt&\omit&   &\omit&\cr
\noalign{\hrule}
\noalign{\hrule}
height1pt&\omit&   &\omit&\cr
&(1,1,2)&&    1512 &\cr
height1pt&\omit&   &\omit&\cr
\noalign{\hrule}
\noalign{\hrule}
height1pt&\omit&   &\omit&\cr
&(1,1,3)&&    216 &\cr
height1pt&\omit&   &\omit&\cr
\noalign{\hrule}
\noalign{\hrule}
height1pt&\omit&   &\omit&\cr
&(1,2,2)&&    7128 &\cr
height1pt&\omit&   &\omit&\cr
\noalign{\hrule}
\noalign{\hrule}
height1pt&\omit&   &\omit&\cr
&(1,2,3)&&    7128 &\cr
height1pt&\omit&   &\omit&\cr
\noalign{\hrule}
\noalign{\hrule}
height1pt&\omit&   &\omit&\cr
&(2,0,1)&&    2106 &\cr
&(4,0,2)&&    414720 &\cr
height1pt&\omit&   &\omit&\cr
\noalign{\hrule}
\noalign{\hrule}
height1pt&\omit&   &\omit&\cr
&(2,1,1)&&    28232 &\cr
height1pt&\omit&   &\omit&\cr
\noalign{\hrule}}
\hrule}
\,\,
\vbox{\offinterlineskip
\halign{ &\vrule# & \strut\quad\hfil#\quad\cr
\noalign{\hrule}
\noalign{\hrule}
height1pt&\omit&   &\omit&\cr
&(2,1,2)&&    85212 &\cr
height1pt&\omit&   &\omit&\cr
\noalign{\hrule}
\noalign{\hrule}
height1pt&\omit&   &\omit&\cr
&(2,1,3)&&    28232 &\cr
height1pt&\omit&   &\omit&\cr
\noalign{\hrule}
\noalign{\hrule}
height1pt&\omit&   &\omit&\cr
&(3,0,1)&&     17856 &\cr
height1pt&\omit&   &\omit&\cr
\noalign{\hrule}
\noalign{\hrule}
height1pt&\omit&   &\omit&\cr
&(3,0,2)&&     17856 &\cr
height1pt&\omit&   &\omit&\cr
\noalign{\hrule}
\noalign{\hrule}
height1pt&\omit&   &\omit&\cr
&(3,1,1)&&    656952 &\cr
height1pt&\omit&   &\omit&\cr
\noalign{\hrule}
\noalign{\hrule}
height1pt&\omit&   &\omit&\cr
&(3,1,2)&&    2984904 &\cr
height1pt&\omit&   &\omit&\cr
\noalign{\hrule}
\noalign{\hrule}
height1pt&\omit&   &\omit&\cr
&(4,0,1)&&    95094   &\cr
height1pt&\omit&   &\omit&\cr
\noalign{\hrule}
\noalign{\hrule}
height1pt&\omit&   &\omit&\cr
&(4,1,1)&&    8757828 &\cr
height1pt&\omit&   &\omit&\cr
\noalign{\hrule}
\noalign{\hrule}
height1pt&\omit&   &\omit&\cr
&(5,0,1)&&    414720   &\cr
height1pt&\omit&   &\omit&\cr
\noalign{\hrule}
\noalign{\hrule}}
\hrule}
$$
In section 7 we discuss also the invariants associated to the
elliptic curves of this three\-moduli example.

The closed formulas \icy\ and \prepot\ can be easily
evaluated for general toric varities with higher-dimensional
moduli space provided the generators of the Mori cone $l^{(i)}$
(cf. \hkty) and the intersection numbers in the corresponding basis
are known. By \icye~ they become a very useful tool for enumerative
geometry.

We will demonstrate this in the following for a complete
intersection with a six-dimensional K\"ahler moduli space.

\vskip 3 mm
\noindent({\it vi})
For the six-moduli complete intersection
$$\cicy-48(\IP^1\cr\IP^1\cr \IP^1\cr \IP^1\cr \IP^1\cr \IP^1|
        0& 1& 1  \cr
        0& 1& 1  \cr
        1& 1&  0  \cr
        1& 1&  0  \cr
        1&  0& 1  \cr
        1&  0& 1 ){6}$$
straightforward evaluation of \prepot~ gives
immediately the prepotential, which reads up to order four in the $q_i$
\eqn\lll{
\eqalign{F=&{1\over 3}\sum'_{i,j,k} t^i t^j t^k+ \sum_it^i +
48{i\zeta(3)\over (2 \pi
)^3}+\sum_i\Bigl(16\,q_i+2\,q_i^2+{16\over27}\,q_i^3+{1\over4}q_i^4
\Bigr)
\cr & +
\sum'_{i,j}{1\over 2} \Bigl(8\,q_i q_j+q_i^2 q_j^2\Bigr)+
{8\over 3} \sum'_{i,j,k}q_i q_j q_k+ {10\over 3}
\sum'_{i,j,k,l}q_i q_j q_k q_l+
O(q^5);}}
here again the primed sums are to be taken over distinct indices only.
{}From this expression we can readily predict the number of lines
$n^r_{1,0,0,0,0,0}=8$, curves of multidegree $n^r_{1,1,0,0,0,0}=8$,
$n^r_{1,1,1,0,0,0}=16$ and $n^r_{1,1,1,1,0,0}=80$. Other invariants
follow by permutation. There are no curves of degree $(2,0,0,0,0,0)$,
$(3,0,0,0,0,0)$ and $(4,0,0,0,0,0)$.
The expressions for the Yukawa couplings and the
Weil-Peterson metric in the large radius expansions follow from
\yuk~ and \kaehler.

\vskip 2 mm

We use our next  examples to demonstrate that the analysis of the
large complex respectively K\"ahler structure limit
in the previous section and formulas \icy,\prepot~have
an application also to embbedings in toric varieties with
Gorenstein singularities.

\vskip 3 mm
\noindent({\it vii})
We  consider first the hypersurface of degree $18$ in the weighted
projective space $\IP^4[6,6,3,2,1]$;
in the  short-hand notation \notation~this reads
$\cicy-144(\IP^4[6,6,3,2,1]|18)7$.
This model was discussed qualitatively in the context of
topology changes by flops in \agm,
but it was not solved.

Generally, for quasi-smooth hypersurfaces and complete intersections
in weighted $\IP^n$, the singularities of the
ambient space intersect the Calabi-Yau
hypersurface $X$ in sets of codimension two and three, see e.g.\hkt\dk.

In our example  we have a singular curve $C_1$ in $X$ with a $\ZZ_2$-action
on its normal bundle, which lies inside the singular
stratum where the third and the fifth homogeneous
coordinates of the weighted projective space are set to zero.
The resolution introduces a $\IP^1$ fibration
over that curve, which gives rise to one exceptional divisor $D_1$.
Similarly, the singular stratum of the weighted projective space
where $z_4=z_5=0$ intersects $X$ in a curve $C_2$ with a
$\ZZ_3$-action on the normal bundle, whose resolution leads
to a fibration with a $\IP^1\wedge\IP^1$ sphere-tree over that curve,
hence two irreducible
divisors $D_2,D_3$ on the resolved space.
The singular curves meet in three $\ZZ_6$ singular points inside
the stratum $z_3=z_4=z_5=0$, whose resolutions support one
exceptional divisor for each point, $\tilde D_4,\tilde D_5,\tilde D_6$.
Hence we have, including the divisor $D_0$ from the Picard group of
the singular space, a seven-moduli case.

The toric description of the mirror pair in terms
of reflexive polyhedra was given in \batyrev, and reviewed in
\hkty\agm. Associated to the manifold $X$
and its mirror $X^*$ are the simplicial polyhedra
$\Delta$ and $\Delta^*$, which are defined as the
convex hull of the following points:
\eqn\mmm{
\Delta={\rm conv}
\left(\matrix{
\nu_1&=(\phantom{-}2,-1,-1,-1)\cr
\nu_2&=(-1,\phantom{-}2,-1,-1)\cr
\nu_3&=(-1,-1,\phantom{-}5 ,-1)\cr
\nu_4&=(-1,-1,-1,\phantom{-}8)\cr
\nu_5&=(-1,-1,-1,-1)}\right),\quad
\Delta^*={\rm conv}
\left(\matrix{
\nu^*_1&=(\phantom{-}1,\phantom{-}0,\phantom{-}0,\phantom{-}0)\cr
\nu^*_2&=(\phantom{-}0,\phantom{-}1,\phantom{-}0,\phantom{-}0)\cr
\nu^*_3&=(\phantom{-}0,\phantom{-}0,\phantom{-}1,\phantom{-}0)\cr
\nu^*_4&=(\phantom{-}0,\phantom{-}0,\phantom{-}0,\phantom{-}1)\cr
\nu^*_5&=( -6,-6,-3,-2)}\right).}
Five points in the dual polyhedron $\Delta^*$, namely
the point in its interior; $\nu_0^*=(0,0,0,0)$ and the points
on edges of codimension three: $\nu_6^*=(-3,-3,-1,-1)$,
$\nu_7^*=(-2,-2,-1,0)$, $\nu_8^*=(-4,-4,-2,-1)$ and on the face
of codimension two: $\nu_9^*=(-1,-1,0,0)$ can be identified with the
(exceptional) divisors of the Calabi-Yau hypersurface $X$.
Only for these divisors can the toric describtion of the K\"ahler
cone and its dual,
the Mori cone,  be applied\foot{This technical complication could
have been avoided if we had instead considered the model
$\cicy{-360}(\IP^4[1,2,3,12,18]|36)5$,
which has same type of singularities, but in which case
all divisors can be described
by toric geometry in this case.}. We prefer to work with the Mori cone
as its generators are directly related to the expressions
for the PF equations and the definition of the local
coordinates for the large complex structure limit.

The $\ZZ_6$ singularity of type $\IC^3/G$ with
$G={\rm diag }(\alpha,\alpha^2,\alpha^3)$ ($\alpha=\exp({2 \pi i\over
6}$) is described in this construction by the three-dimensional
cone $\Sigma$, spanned from the origin
$\nu^*_0$
by $\{\vec \nu^*_3,\vec \nu^*_4,\vec \nu^*_5\}$. The cone
$\Sigma$ is not basic, i.e. the three vectors
$\{\vec \nu^*_3,\vec \nu^*_4,\vec \nu^*_5\}$ do not generate the
lattice $\Sigma\cap M$, with $M=N=\{(n_1,n_2,n_3,n_4)|n_i\in \ZZ\}$.
The volume of
$\sigma\cap \Delta^*={\rm conv}(\nu_0^*,\nu^*_3, \nu^*_4,\nu^*_5)$ is six,
which is the order of the defining group for this singularity;
$\Sigma \cap \Delta^*$  contains the point $\nu_6^*$ on the
edge $\{\nu_3^*,\nu_5^*\}$, which can be identified with the
exceptional divisor $D_1$ in the resolution over the curve $C_1$,
as well as the points $\nu_7^*,\nu_8^*$ on the edge
$\{\nu_4^*,\nu_5^*\}$, which correspond to $D_2,D_3$.
On the other hand, the point $\nu_9^*$ inside the triangle
with corners $\{\nu_3^*,\nu_4^*,\nu_5^*\}$ corresponds
to only one divisor $D_4$ in the resolution of three $\ZZ_6$ singular points.
In the formula for the dimension of $H^{1,1}(X)$ \batyrev~
the multiplicity
three is taken into account by an additional term,
which multiplies the interior points in the three\-dimensional cone
$\Sigma$ with the number of points on the interior of its two-dimensional
dual, which in this case is spanned by the origin in $\Delta$ and $\{\vec
\nu_1,\vec \nu_2\}$. So it contains two points
$\delta_1=(1,0,-1,-1)$ and  $\delta_2=(0,1,1,-1)$.

There are five different canonical resolutions of the $\IC^3/\ZZ_6$
singularity, corresponding to the five possible subdivisions of the
cone $\Sigma$ into six basic cones
of volume one, which are spanned from $\nu_0^*$ by:
\def\drei(#1,#2,#3){\{\vec \nu_#1^*,\vec \nu_#2^*,\vec \nu_#3^*\}}
\eqn\nnn{
\eqalign{
{\rm A}&: \quad
\drei(3,6,9),\,
\drei(6,5,8),\,
\drei(8,9,6),\,
\drei(8,7,9),\,
\drei(7,4,9),\,
\drei(4,3,9)
\cr
{\rm B}&: \quad
\drei(3,6,8),\,
\drei(6,5,8),\,
\drei(8,9,3),\,
\drei(8,7,9),\,
\drei(7,4,9),\,
\drei(4,3,9)
\cr
{\rm C}&:\quad
\drei(3,6,9),\,
\drei(6,5,9),\,
\drei(5,8,9),\,
\drei(8,7,9),\,
\drei(7,4,9),\,
\drei(4,3,9)
\cr
{\rm D}&:\quad
\drei(3,6,9),\,
\drei(6,5,8),\,
\drei(8,7,6),\,
\drei(7,9,6),\,
\drei(7,4,9),\,
\drei(4,3,9)
\cr
{\rm E}&:\quad
\drei(3,6,9),\,
\drei(6,5,8),\,
\drei(8,7,6),\,
\drei(7,4,6),\,
\drei(4,9,6),\,
\drei(4,3,9)}
}
for the different resolutions.
The generators of the Mori cone $l^{(i)}$ and the Picard-Fuchs
system in the corresponding large complex
structure coordinates $z_i$ can be obtained for each subdivision
by the methods described in \hkty. They will of course
depend on the subdivision.

The subdivions are connected by flops, e.g. subdivision
$B$ is obtained from subdivision $A$ by the flop,
which blows down the $\IP^1$ represented by
the subcone spanned from $\nu_0^*$ by
$\{\vec \nu_6^*,\vec \nu_9^*\}$
in subdivision $A$ and subsequently blows up the
$\IP^1$ associated with the new subcone spanned from
$\nu_0^*$ by $\{\vec \nu_3^*,\vec \nu_8^*\}$ in subdivision
$B$. We therefore expect simple relations of
the generators of the Mori cones and large complex structure
coordinates for all
subdivisions to those of subdivision $A$.
After extending the vectors $\nu_i^*$ to $\bar \nu_i^*=(1,\nu_i^*)$
the generators of the Mori cone can be constructed as described
in \odapark,\hkty~to yield
\eqn\ppp{
\eqalign{l_A^{(1)}&=(-3; 1,1,0,0,-1,
\phantom{-}1,\phantom{-}0,\phantom{-}1,\phantom{-}0),\,\,
l_A^{(2)}= (\phantom{-}0  ; 0,0,1,0,0,
-1,\phantom{-}0,\phantom{-}1,-1)\cr
l_A^{(3)}&=(\phantom{-}0; 0,0,0,0,\phantom{-}1,
-1,\phantom{-}0,-1,\phantom{-}1),\,\,
l_A^{(4)}=(\phantom{-}0  ; 0,0,0,0,0,
\phantom{-}1,\phantom{-}1,-1,-1)\cr
l_A^{(5)}&=(\phantom{-}0; 0,0,0,1,\phantom{-}0,
\phantom{-}0,-2,\phantom{-}1,\phantom{-}0)\, .}}
They define the large complex structure variables as
$z^{(A)}_1=-a^{l_A^{(1)}}$ and $z^{(A)}_i=a^{l_A^{(i)}}$ for $i=2,3,4,5$.
As the next step we set up the $\bar \Delta^*$-hypergeometric system
(see e.g. \batyrev\hkty~for details).
Every linear relation among the $\bar \nu_i^*$,
not just the ones encoded in the
$l_A^{(i)}$ (cf. \linearrelationI), defines a linear
differential operator of this system, which is satisfied by all periods.
It should be mentioned again that this system is not equivalent to the
Picard-Fuchs system, the non-trivial task will be to find,
possibly after factorization,
a minimal system of differential operators which uniquely
determine the $(2k+2)$-dimensional period vector in the domain
containing the points $z^{(A,\ldots,E)}=0$. These systems are
equivalent to the first-order Gauss-Manin system
in these domains and all other differential operators are
elements of the local left ideal associated to it (cf. section 4).
As we will see, the minimal numbers of differential
operators generating this ideal can vary from domain to domain.
The consideration of the rings $R^{(A,\ldots,E)}$ will be essential
in order to find these generators.

{}From the fact that the free polynomial ring $\IC[\theta^{(A)}_z]$ has
to be truncated at degree 2
from fifteen to five elements by the principal part
of the linear differential operators, we know that there
must be ten second-order differential operators. Indeed we
find that the linear relations between the sites of the points
in $\bar \Delta^*$, which are expressed by the
following vectors $l_A^{(2)}$, $l_A^{(3)}$, $l_A^{(4)}$,
$l_A^{(5)}$,
$l_A^{(2)}+l_A^{(3)}$,
$l_A^{(2)}+l_A^{(4)}$,
$l_A^{(3)}+l_A^{(4)}$,
$l_A^{(4)}+l_A^{(5)}$,
$l_A^{(3)}+l_A^{(4)}+l_A^{(5)}$,
define directly nine second-order differential operators of the
$\bar \Delta^*$-hypergeometric system
(cf. \batyrev\hkty)
\eqn\secondorderI{
\eqalign{
L^{(A)}_1&=\ta3\ta8-\a{\la2}\ta6\ta9,\qquad \qquad \quad\,\,
L^{(A)}_2=\ta5\ta9-\a{\la3}\ta6\ta8,
\cr
L^{(A)}_3&=\ta6\ta7-\a{\la4}\ta8\ta9,\qquad \qquad \quad\,\,
L^{(A)}_4=\ta4\ta8-\a{\la5}(\ta7-1)\ta7,
\cr
L^{(A)}_5&=\ta3\ta5-\a{\la2+\la3}(\ta6-1)\ta6,\quad
L^{(A)}_6=\ta3\ta7-\a{\la2+\la4}(\ta9-1)\ta9,
\cr
L^{(A)}_7&=\ta5\ta7-\a{\la3+\la4}(\ta8-1)\ta8, \quad
L^{(A)}_8=\ta4\ta6-\a{\la4+\la5}\ta7\ta9,
\cr
L^{(A)}_9&=\ta4\ta5-\a{\la3+\la4+\la5}\ta7\ta8.
}}
The logarithmic derivatives
$\theta_{a_i}:=a_i {\partial\over \partial_{a_i}}$ can be readily
translated to logarithmic derivatives in the $z_i$
by the identity $\theta_{a_i}=\sum_{k=1}^h l^{(k)}_{i} \theta_k$.
Via this relation we also see that we can factor a $\theta_1$ operator
from the differential operator for $l_A^{(1)}+l_A^{(3)}$ to yield
the tenth second-order differential operator $L_{10}^{(A)}$
\eqn\secondorderII{\eqalign{
\theta_1 L^{(A)}_{10}&=\ta1\ta2\ta9-\a{\la1+\la3}(\ta0-3)(\ta0-2)(\ta0-1) \cr
                     &=\t_1\bigl(\t_1(\t_2-\t_3+\t_4)+3 z_1 z_2
(3\t_1+2)(3\t_1+1) \bigr).}}
While $\IC[\theta_z]$ is truncated by the ideal $I_s(\theta_z)$,
$s=1,\ldots,10$ obtained from
\secondorderI,\secondorderII\ to five elements at degree
2, one can easily check that we still have two
elements at degree three. So we need one further
independent third-order differential operator. We find e.g. that
$l^{(1)}_A$ leads, after factorization of $\theta_1$,
to the operator \eqn\thirdorder{\eqalign{
&L^{(A)}_{11}=\t_1(\t_1-\t_2-\t_3\! + \! \t_4)(\t_1+\t_2-\t_3-\t_4+\t_5)
+3 z_1 (3\t_1+2)(3\t_1+1)(\t_1-\t_3),}}
which truncates $\IC\(\theta_z\)$
it to one element at degree 3 and no element at
any higher order.
Now we can construct, as in section 4, via the ring
$R^{(A)}$ the period vector and calculate the prepotential explicitly.
The classical intersection part and the
corresponding expression for the deformed prepotential
can be found most simply from \icy~to be, up to terms of order $q^4$,
\eqn\qqq{
\eqalign{F^{(A)} &=
{18\over 3!} (t^1)^3
+ {9\over 2} (t^1)^2 t^2
+ {3\over  2} t^1 (t^2)^2
+ {21\over 2} (t^1)^2 t^3
+ 9 t^1 t^2 t^3
+ {3\over  2}(t^2)^2 t^3
+ {21\over 2} t^1 (t^3)^2
\cr &
+ {9\over 2} t^2 (t^3)^2\!
+ \! {21\over 3!} (t^3)^3 \!
+ \! {12\over 2}(t^1)^2 t^4 \!
+ \! 6 t^1 t^2 t^4 \!
+ \! 12 t^1 t^3 t^4 \!
+ \! 6 t^2 t^3 t^4 \!
+ \! {12\over 2} (t^3)^2 t^4\!
+ \! {6\over 2} t^1 (t^4)^2
\cr &
+ {6\over 2} t^3 (t^4)^2
+ {6\over 2} (t^1)^2 t_5
+ 3 t^1 t^2 t^5
+ 6 t^1t^3 t^5
+ 3 t^2t^3 t^5
+ {6\over 2 } (t^3)^2 t^5
+ 3t^1 t^4 t^5
+ 3 t^3 t^4 t^5
\cr &
+ {1\over 24}(72 t^1+ 36 t^2+78 t^3 + 48 t^4 +24 t^5)
+ 144{i\zeta(3)\over (2\pi)^3}+( 27\,q_1 + 3\,q_2 + 3\,q_3 + 3\,q_4)
\cr & + \left({{3\,{{q_2}^2}}\over 8} - {{405\,{{q_1}^2}}\over 8} +
108\,q_1\,q_3  + {{3\,{{q_3}^2}}\over 8} - 6\,q_2\,q_4 +
{{3\,{{q_4}^2}}\over 8} + 3\,q_4\,q_5 \right)
\cr &+\!
\left( 244{{q_1}^3} \! + {{{{q_2}^3}}\over 9}\! + 81{{q_1}^2}q_3
\! + 27q_1q_2q_3 \! + {{{{q_3}^3}}\over 9} \! + 27q_1q_3q_4
\! +  3q_2q_3q_4 \! + {{{{q_4}^3}}\over 9} \!
- 6q_2q_4q_5
\right).}
}
We can next use the mirror hypothesis to interpret the
coefficients of the cubic terms in $t^i$ of \qqq~
as triple intersection numbers of $H_2(X,\ZZ)$. Note
that all triple intersection numbers are positive, as expected in the
base $J_i$ which generates the K\"ahler cone.
The linear and the quadratic terms in the $t^i$ obtained by
\derivatives,~\identities~ can be compared with the classical
calculation of these topological numbers performed for instance in \dk.
These formulas relate all topological data to the
$l^{(i)}$ and provide an excellent check of our calculation.
As it stands, the
classical part of $F^{(A)}$ does not refer to the basis of divisors
$D_i$ for which these intersection numbers can be directly obtained
by the formulas for resolved complete intersections in
toric varieties summarized, for instance in \hkty\dk.
To make the comparison we can easily transform to that basis by
passing to the variables $t_{D_i}$\foot{Note that this reparametrization
is quite different from the transformations into different domains,
discussed below, because we still keep $z_i=0$ as the point around,
which we expand our solutions of the PF-equation.}
via $t^i=\delta^{i,1} t_{D_0}+\sum_{j=1}^{k-1}l^{(i)}_{A,j+5} t_{D_j}$, i.e.
from the $z^{(A)}_i$
variables to the complex structure deformation
variables which occur explicitly in the Laurent polynomial,
namely ${1\over a_0^3},a_6,a_7,a_8,a_9$.
This leads to the prediction
\eqn\rrr{
\eqalign{
F_{\rm cl}^{(A)} &=
   {18\over 3!} t_{D_0}^3 -
   {6\over 2}   t_{D_0} t_{D_1}^2 -
   {3\over 2}   t_{D_1}^3 -
   {6\over 2}   t_{D_0} t_{ D_2}^2 -
   {3\over 2}   t_{D_1}^2 t_{D_3} +
   3            t_{D_0} t_{D_2} t_{D_3} -
   {6\over 3}   t_{D_0} t_{D_3}^2-
\cr &
   {3\over 2}   t_{D_1} t_{D_3}^2 -
   {3\over 3!}  t_{D_3}^3 -
   {3\over 2}   t_{D_1}^2 t_{D_4}-
   {6\over 2}   t_{D_2}^2 t_{D_4} +
   3            t_{D_1} t_{D_3} t_{D_4}  +
   3            t_{D_2} t_{D_3} t_{D_4} -
   {3\over 2}   t_{D_3}^2 t_{D_4} -
\cr &
   {3\over 2}   t_{D_1} t_{D_4}^2 -
   {3\over 2}   t_{D_3} t_{D_4}^2 +
   {21\over 3!} t_{D_4}^3 +
   {1\over 24} ( 72 t_{D_0}+6 t_{D_1}+ 6 t_{D_3}+ 6 t_{D_4})
   + 144{i\zeta(3)\over (2\pi)^3},}
}
which is in agreement with the formulas of the
classical intersections in the basis of the $D_i$.
Note however that all intersection numbers among the divisors
on the triangle with corners $\{\nu^*_3,\nu_4^*,\nu_5^*\}$ are multiplied
by 3 in $F_{\rm cl}^{(A)}$.
This is due to the fact that the $\ZZ_6$ fixed point
has multiplicity 3 and the toric description of this singularity
by $\Sigma$ refers to the symmetric combination
$D_4=\tilde D_4+\tilde D_5  +\tilde D_6$
of the exceptional divisors over the three singular points on $X$, with
$$K^0_{\tilde D_i\tilde D_j\tilde D_k}=
\cases{6 &${\rm if}\,\,i=j=k$\cr 0 & ${\rm otherwise}$}.$$

To investigate the theory in different domains of the
moduli space we note the simple transformation property of
the $\theta_i$ operators in the Picard-Fuchs system, i.e. for
$z_i^{(B)}=\prod_j (z_j^{(A)})^{m_{i,j}}$ $\leftrightarrow$
$l_B^{(i)}=\sum_j m_{i,j} l_A^{(i)}$ the $\theta_i$ transform as
$\theta_i^{(A)}=\sum_j m_{j,i} \theta^{(B)}_j$. We are
interested in transformations that lead outside the Mori cone,
i.e. in which not all entries of the matrix $m$ are positive.
A quick look at \secondorderI~ and \secondorderII~ reveals that
the possibilities for such transformations are
rather restricted if we insist on a completely degenerate
large complex structure limit in the new domain $B$,
i.e. a ring structure $R^{(B)}$ with the properties discussed
in section 4. For instance, we cannot just invert $z_5$
(which would correspond to the replacement $l_A^{(5)}\to -l_A^{(5)}$)
without generating inhomogeneous terms in $I(\theta_z)$,
which is incompatible with the required
ring structure of $R$. It is easy to see that
the only possibilities are to invert $z_2$, $z_3$, $z_4$ and
$z_4 z_5$, accompanied by transformations of the other
variables.
These transformations
correspond to the flops leading to the coordinate patches
described by the Mori cone of subdivisions $B$, $C$, $D$ and $E$.
They form part of the secondary fan as described in \odapark.

We start with the inversion of $z_2$, which leads to subdivision $B$
for which the generators of the Mori cone read
$l_B^{(1)}=l^{(1)}_A$,
$l_B^{(2)}=-l_A^{(2)}$,
$l_B^{(3)}=l_A^{(2)}+l_A^{(3)}$,
$l_B^{(4)}=l_A^{(2)}+l_A^{(4)}$ and
$l_B^{(5)}=l_A^{(5)}.$
Note that the matrix $m$ for a single
flop squares to unity and the principal parts of
the transformed system truncate $\IC[\theta_z]$
at degree 2 to five elements, but two elements remain at
degree 3. We can remedy the situation by adding the
operator associated to $l_A^{(1)}+l_A^{(2)}+l_A^{(3)}$
to our system \secondorderI-\thirdorder:

\eqn\addI{\theta_1 L^{(A)}_{12}=
\ta1\ta2\ta3\ta8-\a{\la1+\la2+\la3}(\ta0-3)(\ta0-2)(\ta0-1)\ta6,}
which factorizes to a third-order operator $L_{12}^{(A)}$.
For subdivision $C$:
$l_C^{(1)}=l^{(1)}_A+l^{(3)}_A$,
$l_C^{(2)}=l_A^{(2)}+l_A^{(3)}$,
$l_C^{(3)}=-l_A^{(3)}$,
$l_C^{(4)}=l_A^{(3)}+l_A^{(4)}$ and
$l_C^{(5)}=l_A^{(5)}$, no further operator has to be added.
To complete our system w.r.t. subdivision $D$:
$l_D^{(1)}=l^{(1)}_A$,
$l_D^{(2)}=l_A^{(2)}+l_A^{(4)}$,
$l_D^{(3)}=l_A^{(3)}+l_A^{(4)}$,
$l_D^{(4)}=-l_A^{(4)}$
$l_D^{(5)}=l_A^{(4)}+l_A^{(5)}$
and w.r.t. $E$:
$l_E^{(1)}=l_A^{(1)}$,
$l_E^{(2)}=2 l_4^{(1)} + l_A^{(2)}+l_A^{(5)}$,
$l_E^{(3)}=l_A^{(3)}+l_A^{(4)}$,
$l_E^{(4)}=l_A^{(5)}$
$l_E^{(5)}=-l_A^{(4)}-l_A^{(5)}$ we have to add three third order
operators $L_{13}^{(A)},\,L_{14}^{(A)},\,L_{15}^{(A)}$ :
\eqn\addII{\eqalign{
\theta_1 L^{(A)}_{13}&=\ta1\ta2\ta6\ta7-\a{\la1+\la2+\la4}
                       (\ta0-3)(\ta0-2)(\ta0-1)\ta8\cr
\theta_1 L^{(A)}_{14}&=\ta1\ta2\ta4\ta6-
          \a{\la1+\la3+\la4+\la5}(\ta0-3)(\ta0-2)(\ta0-1)\ta7\cr
L^{(A)}_{15}&=\ta3\ta4\ta5-\a{\la2+\la3+2\la4+\la5}\ta8(\ta9-1)\ta9\,.}}

The system $L^{(A)}_{1}$,\dots,$L^{(A)}_{15}$ contains the information
that is necessary to extract the prepotential in all large complex
structure regions. The basis of solutions given in
\pI~can now be obtained explicitly in all regions.
As they are all solutions to the same system of Picard-Fuchs
equations, expressed in different patches of the moduli space, they are
analytic continuations of each other with trivial monodromy.
Especially the prepotential $F^{(A,B,C,D,E)}$ encoded in the $2k+2$
component of \pI~ is the same analytic function whose expansion in the
domains corresponding to the different resolutions can be
evaluated with the attached program. Also, for $F^{(B,C,D,E)}$
we obtain, after transformation via
$t^i=\delta^{i,1} t_{D_0}+\sum_{j=1}^{k-1}l^{(i)}_{j+5} t_{D_j}$
to the $D_i$ basis, the classical
intersection numbers as calculated e.g.
by the formulas of \hkty, with an enhancement factor 3 for
intersections among divisors on the triangle.

We also checked, up to order 8, that the expansions of
$F^{(A,B,C,D,E)}$ are compatible with an integer expansion for $n^r$
in \icye. For example,
from $F^{(A)}$ we obtain, up to degree eight, instanton
contributions $n^r_{i,0,0,0,0}$ with alternating sign
$$
\vbox{\offinterlineskip\tabskip=0pt
\halign{\strut\vrule#&
\hfil~~$#$~~&
\hfil~~$#$~~&
\hfil~~$#$~~&
\hfil~~$#$~~&
\hfil~~$#$~~&
\hfil~~$#$~~&
\hfil~~$#$~~&
\hfil~~$#$~~&
\hfil~~$#$~~&
\vrule#\cr
\noalign{\hrule}
&i  &1 & 2 &  3& 4   & 5   & 6     & 7     & 8       &\cr
&n^r&27&-54&243&-1728&15255&-153576&1696086&-20053440&\cr
\noalign{\hrule}}
\hrule}$$
We can read from the prepotential that
$$
n^r_{0,1,0,0,0}=n^r_{0,0,1,0,0}=n^r_{0,0,0,1,0}=n^r_{0,0,0,1,1}=3.
$$
There are no curves of degree
$(0,i,0,0,0)$, $(0,0,i,0,0)$,
$(0,0,0,i,0)$ and $(0,0,0,i,i)$ for $i>1$ and no curve $(0,0,0,0,i)$
for all $i$.
By comparing the expansions in the different regions we find
that the three rational curves with degrees
$(0,1,0,0,0),\; (0,0,1,0,0),\; (0,0,0,1,0)$ and $(0,0,0,1,1)$,
respectively, are those that are shrunk to zero volume and whose
corresponding invariant changes sign under the process of the four
possible flop operations interrelating them, starting from resolution $A$
(cf. the discussion in \agm).

For certain directions in the K\"ahler cone e.g. $(i,0,i,0,0)$
one has a periodicity in the invariants
$n_r=108,108,144,\ldots$ as
in the cases ({\sl 5 i,ii,iii,v}) before.
It is remarkable that in all cases,
where we obtain periodicity, one of the numbers,
the third in the scheme $a,a,b\ldots$, is always the negative
of the Euler number of the manifold $X$.

\vskip 3 mm
\noindent({\it viii})
So far we have considered complete intersections in non-singular
ambient spaces and, in the last  example
as well as in \hkty, hypersurfaces in ambient
spaces with Gorenstein singularities. Let us
investigate in the following the more general situation
of complete intersections of codimension $n-3$
in an $n$-dimensional singular ambient
space and take the following hypersurface as well as the following
complete intersections with two moduli as examples:
\eqn\sss{
\eqalign{
A:&\,\,\, \cicy{-168}(\IP^4[2,2,2,1,1]|8)2\cr
B:&\,\,\, \cicy{-132}(\IP^5[2,2,2,2,1,1]|4,6)2\cr
C:&\,\,\, \cicy{-112}(\IP^6[2,2,2,2,2,1,1]|4,4,4)2.}
}
They all have polynomial constraints of the Fermat type
and exhibit the simplest singular locus, namely
a singular curve with a $\ZZ_2$-action on the normal
bundle induced from the ambient space, whose resolution
gives as exceptional divisor a $\IP^1$ bundle over that curve.
For all complete intersections of the Fermat type in a weighted
space $\IP^n\(\vec w\)$ with $w_{n+1}=1$ we can define, in generalization of
\mmm, an $n$-dimensional pair of
simplicial reflexive polyhedra $\Delta,\Delta^*$ as the convex hulls
\eqn\ttt{
\Delta={\rm conv}\left(\matrix{\nu_1&\!\!\!\!\!\!\!=
\left({\sum_i d_i\over w_1}-1,-1,\ldots,-1\right)\cr
                            \vdots& \quad \vdots \cr
\nu_n&\!\!\!\!\!\!\! = \left(-1,\ldots,-1,{\sum_i d_i \over w_n}-1\right)\cr
\nu_{n+1}&\!\!\!\!\! =(-1,\ldots\ldots\ldots\ldots,-1)\quad\cr}\right),
\quad
\Delta^*={\rm conv}\left(\matrix{\nu^*_1&\!\!\!\!\!\! =(1,0\ldots,0)\qquad\cr
           \vdots&\, \vdots\cr
           \nu^*_n&\!\!\!\!\!\! =(0,\ldots,0,1)\qquad\cr
           \nu^*_{n+1}&\!\!\!\!\!\! =(-w_1,\ldots,-w_n)}\right),}
in an $n$-dimensional lattice.
For all
three examples we have, beside the origin $\nu_0^*=(0,\ldots,0)$,
exactly one additional point in $\Delta^*$, namely
$\nu^*_{n+2}=(-1,\ldots,-1,0)$. Extending the lattice by $n-3$
dimensions, as described in section 3, we find the linear
relations among the extended lattice sites $\bar \nu_i^*$,
which are summarized in the $l$-vectors:
\eqn\ltm{\eqalign{A:&\,\,\, l^{(1)}=(-4;1,1,1,0,0,1),\quad
l^{(2)}=    (0;0,0,0,1,1,-2)\cr
           B:&\,\,\, l^{(1)}=     (-2,-3;1,1,1,1,0,0,1),\quad
l^{(2)}=(0,0;0,0,0,0,1,1,-2)\cr
           C:&\,\,\, l^{(1)}=(-2,-2,-2;1,1,1,1,1,0,0,1),\quad
l^{(2)}=(0,0,0;0,0,0,0,0,1,1,-2)}}
(The above choices for the generators $l_{B,C}$ do not uniquely define
the numerical effective partition of the lattice points of $\Delta^*$.)
The associated GKZ system factorize in all cases
to a third- and second-order Picard-Fuchs equation, where
the latter has the form
\eqn\pfIItm{L_2=\t_2^2 - z_2 (2 \t_2-\t_1)(2 \t_2-\t_1+1).}
Also the principal part of the third-order operator
is universal. We find
\eqn\pfIIItm{\eqalign{
A:&\,\,\,  L_1=\t_1^2(2 \t_2-\t_1)-4 z_1 (4 \t_1+3) (4\t_1+2) (4\t_1+1) \cr
B:&\,\,\,  L_1=\t_1^2(2 \t_2-\t_1)-6 z_1(2 \t_1+1) (3\t_1+2) (3\t_1+1) \cr
C:&\,\,\,  L_1=\t_1^2(2 \t_2-\t_1)-8 z_1 (2 \t_1+1)^3 ,}}
while the terms proportional to $z_1$ signal a different
structure for the expansions of the solutions around the  singularity at
${1\over z_1},{1\over z_2}\rightarrow 0$, namely six pure power series
solutions for $A$ and $B$ and
solutions involving logarithms for $C$. The Yukawa couplings
\eqn\uuu{
  K_{111}={1\over z^3_1 \Delta_0},\,\,
  K_{112}={1 - \kappa z_1\over z_1^2 z_2 \Delta_0},\,\,
  K_{122}=-{1-2\kappa z_1\over z_1 z_2 \Delta_0\Delta_1},\,\,
  K_{222}={1-\kappa z_1+ 4 z_2-12 \kappa z_1 z_2
          \over 2 z_2^2 \Delta_0 \Delta_1^2}}
and the components of the discriminant
$\Delta_0=(1-\kappa z_1)^2-4 \kappa^2 z_1^2 z_2$, $\Delta_1=(1- 4 z_2)$
can be parametrized by $\kappa=256,108,64$ for the three
cases in turn.

Using the model specific data \ltm-\pfIIItm~ in our general
formulas, we get the following predictions of the topological
data and the invariants of the rational curves listed here
up to 3
three
$$
\vbox{\offinterlineskip\tabskip=0pt
\halign{\strut\vrule#&\hfil~$#$~&\hfil$#$
\quad&\hfil~$#$\quad&\hfil~$#$
\quad&\hfil~$#$\quad&\hfil~$#$
&\vrule#&~\hfil $#$~&\hfil $#$~&\hfil $#$~&\hfil $#$~&\hfil $#$~&
\hfil $#$~&\vrule#\cr
\noalign{\hrule}
&{\rm Model}&\chi&\int c_2 J_1& \int c_2 J_2 &
K^0_{111}&K^0_{112}&& n^r_{1,0} & n^r_{2,0} & n^r_{2,1} &n^r_{3,0}& n^r_{3,1}&
n^r_{0,1} &\cr
\noalign{\hrule}
&A:&-168&56&24& 8&4   &&   640 & 10032    &72224&288384&7539200 &4&\cr
&B:&-132&60&24&12&6   &&   360 &  2682    &17064&35472 &770280  &6&\cr
&C:&-112&64&24&16&8   &&   256 &  1248    &7232 &10496 &197632  &8&\cr
\noalign{\hrule}}
\hrule}$$
In all cases we observe that $n^r_{0,1}$ is the only
nonvanishing invariant for $n^r_{0,i}$, generally
$n_{i,j}=0$ for $j>i$ and similar to \cdfkm~
$n^r_{i,j}=n^r_{i,i-j}$ $\forall i>0,\,\, j\le [i/2]$.
Furthermore we obtain that $n^r_i=\sum_{j} n^r_{i,j}$, where
$n^r_i$ are the invariants for the rational curves
of the models $\cicy{-176}(\IP^5|4,2)1$, $\cicy{-144}(\IP^6|3,2,2)1$
and $\cicy{-120}(\IP^7|2,2,2,2)1$, and $n^r_{i,j}$ are the ones
for $A$, $B$ and $C$ respectively.
The invariants of the elliptic curves will be evaluated in section 7.
As before, we have checked that the topological numbers coincide
after the change of basis with the ones calculated in \dk.

In the general case the Picard-Fuchs equation will not
follow as easily as above by factorization of the GKZ system.
Rather the analysis of additional symmetries of the period
will be necessarily similar, as it is  described for hypersurfaces in \hkty.
On the other hand the examples indicate that our description of
the instanton-corrected Yukawa couplings also apply to the rich class of
complete intersections with Gorenstein singularities.

\vskip 3 mm
The higher-degree invariants for all non-singular complete intersection in
products
of weighted projective spaces and for all other examples discussed in this
section
can be evaluated by the program INSTANTON.

%%%%%%%%%%%%%%%%%%%%%%%%%%%%%%%%%%%%%%%%%%%%%%%%%%%%%%%%%%
\newsec{Connection with rational superconformal theories.}
%%%%%%%%%%%%%%%%%%%%%%%%%%%%%%%%%%%%%%%%%%%%%%%%%%%%%%%%%%

In this section we would like to comment on
different realizations of equivalent manifolds and their relation
to exactly solvable superconformal theories.
The sigma model on a Calabi-Yau manifold can be identified with
a $(2,2)$ superconformal two-dimensional
field theory, whose partition function and correlation functions
are sometimes known exactly, at least at a special point in
moduli space. Although more general identifications should exist,
in the known examples the SCFT is a GSO projected
tensor product of minimal $(2,2)$ superconformal
field theories
\ref\Doron{D. Gepner, \plb199(1987)380,
\npb 296 (1988) 757,\npb311(1988)191}.
The classification of the
latter follows an $ADE$ pattern and there is a
one-to-one correspondence with the classification of modality-zero $ADE$
singularities. The defining equation
of the latter can be viewed as Landau-Ginzburg potentials for
two-dimensional $(2,2)$ supersymmetric field theory having the SCFT
as its infrared limit. The partition functions \ref\partition\ref{
A. Cappelli, C. Itzykson and B. Zuber,\cmp113(1987)1; D. Gepner,
\npb 287(1987)111} of the $ADE$ superconformal models at level $k$,
as well as their coupings, are explicitly known
\ref\couplings{G. Sotkov and M. Stanishkov, \plb215(1988)647;
A. Kato and Y. Katizawa, \npb319(1989) 474; A. Klemm and
J. Fuchs, \anp194(1989)303; S. Cordes and Y. Kikuchi,
Texas A \& M preprint CTP-TAMU-92/88}.
The identification of their LG potentials is as follows:
\eqn\vvv{
\eqalign{
A_k&\sim z^{k+2},\quad\quad k\in \ZZ,\cr
D_k&\sim z^{k+2\over 2}+ z y^2,\quad k\in 2 \ZZ \cr
E_6&\sim y^3 + z^4,\quad k=10,\cr
E_7&\sim z^3 + z y^3,\quad k=16,\cr
E_8&\sim y^3+ z^5,\quad k=28}}
and for tensor product models the LG potential is simply the
sum of the corresponding LG potential terms.
The central charge $c=\sum_{i=1}^n {3 k_i\over k_i+2}$
is the sum of the central charges of the factor theories
and has to be 9 to cancel the conformal anomaly.

In \ref\fkss{J. Fuchs, A. Klemm, C. Scheich and M. Schmidt,
Ann. Phys. 204 (1990) 1} a large number of identifications between
GSO projected partition functions was found, among them
$G_1\equiv(A_2,A_2,A_2,A_6,A_6)\simeq G_2\equiv(A_2,A_2,A_2,D_6,D_6)$.
We will argue that this implies an identification
of the full string theory at a special point in the moduli
space of the  hypersurface
$X_1=\cicy-168(\IP^4[2,2,2,1,1]|8)2$
and of the complete intersection
$$X_2=\cicy-168(\IP^4\cr \IP^1|4&1\cr 0&2)2.$$

To see this one has to perform  a geometrical analog of
the GSO projection in the tensor product theory
on the LG model. One can either apply an heuristic path integral
argument due to ref. \ref\gvw{B. Greene, C. Vafa and N. Warner,
\npb 324(1989)371}\fkss, or gauge the LG model as proposed
in \ref\witten{E. Witten, \npb403(1993)159}.
Both operations involve, similar to the GSO projection in the
tensor product model, an orbifoldisation and one has
to be careful to end up with the same symmetry group as in the
SCFT. Before GSO projection one has in the SCFT a $\ZZ_{k+2}$
symmetry in each factor model for the $A_k$
theories and for the $D_k$ theories if $k\in 4\ZZ+2$.
These symmetries are readily identified with the symmetries
generated by $z\mapsto \exp{2 \pi i\over k+2} z$ and
$z\mapsto \exp{4 \pi i\over k+2}z,\,\,y\mapsto
\exp{-2  \pi i\over k+2} y$ on the LG fields of the
$A_k$ and $D_k$ models, respectively.
The GSO projection on the tensor product theory is implemented
by orbifoldisation with respect to the diagonal subgroup which in
the above case is a $\ZZ_{{\rm lcm}\{k_i+2\}}$.
The symmetry group of the GSO projected theory will therefore be
${\cal S}\times\prod_{i=1}^5 \ZZ_{k_i+2}/
\ZZ_{{\rm lcm}\{k_i+2\}}$, where ${\cal S}$ is permutation of
identical factors.
In the first argument \gvw~the orbifoldisation is replaced by
a map of the variables $z_i\rightarrow \xi_i$ with
constant Jacobian such that the LG potential becomes linear in one
or more of the $\xi_i$. They can be viewed as Lagrange
multipliers and integrating them out restricts the field
configuration to an affine patch of a product of weighted
projective spaces. For the SCFT of type $G_1$
we have $z_1=\xi_1^{1\over k_1+2},\,\,
z_i=\xi_i \xi^{1\over k_i+2}$, $i=2,\ldots,5$ with
$|{\partial \xi\over \partial x}|={\rm const.}$ precisely because $c=9$
implies $\sum_{i=1}^5 {1\over k_i+2}=1$.
Integrating out $\xi_1$ and going back to
homogenous variables yields manifolds of type $X_1$, i.e.
hypersurfaces  in $\IP^4\(\vec w\)$. The diagonal subgroup
of the phase symmetries on the $z_i$ is now trivial in $\IP^4\(\vec
w\)$,
so that we end up with the same symmetry group as in the SCFT.
Similarly, for the second type ($G_2$) we have
$z_1=\xi_1^{1\over k_1+2}$,
$z_i=\xi_i \xi_1^{1\over k_i+2}$,
$i=2,3,4,6$, $z_5=\xi_5^{1\over 2}/\xi_1^{1\over 2(k_1+2)}$,
$z_7=\xi_7\xi_5^{1 \over 2}/\xi^{1\over 2(k_1+2)}$ and
integrating out $\xi_1,\xi_5$ yields manifolds\foot{We can
interprete the LG potential $W=\sum_{i=1}^5 x_i^4 + x_4 x_6^2 +x_5
x_7^2=0$ also as four dimensional hypersurface in the five dimensional
ambient space $\IP^5(2,2,2,2,3,3)$ with $c_1=8$ and consider
as in \ref\cdp{P. Candelas, E. Derrick and L. Parkes,
\npb407(1993)115.} a restricted cohomology of this space to define the six
periods, which leads to the same prepotential and hence the same
physical theory.} of
type
$X_2$. The identification in $\IP^4$ and $\IP^1$ trivialize a $\ZZ_8$,
s.t. the remaining symmetry is again as in the GSO projected SCFT.
By the same combinatorics it is possible
to introduce one or two gauge group operations
respectively, which leave the superpotential invariant
and lead, by the argument of \witten, to the same geometrical
interpretation.

Using the basis of divisors $J,D$ for the singular hypersurface
described in \hkty~and the explicit formulas given there
we can calculate the intersection numbers in this basis.
The evaluation of the second Chern form on
$J$ is given in formula \cIIh.
These data and their analogs for the complete intersection,
calculated by \top, are displayed below.
$$
\vbox{\offinterlineskip
\hrule
\halign{ &\vrule# & \strut\quad\hfil#\quad\cr
\noalign{\hrule}
height1pt&\omit&   &\omit&\cr
&&& $X_1$&\cr
\noalign{\hrule}
height1pt&\omit&   &\omit&\cr
&$K_{JJJ}$       && 8  &\cr
&$K_{JJD}$       && 0  &\cr
&$K_{JDD}$       && -8   &\cr
&$K_{DDD}$       && -16  &\cr
&$\int c_2 h_J$  && 56  &\cr
&$\int c_2 h_D$      &&  8  &\cr
height1pt&\omit&   &\omit&\cr
\noalign{\hrule}}
\hrule}
\,\,
\vbox{\offinterlineskip
\hrule
\halign{ &\vrule# & \strut\quad\hfil#\quad\cr
\noalign{\hrule}
height1pt&\omit&   &\omit&\cr
&&& $X_2$&\cr
\noalign{\hrule}
height1pt&\omit&   &\omit&\cr
&$K_{{J_1}{J_1}{J_1}}$       && 8  &\cr
&$K_{{J_1}{J_1}{J_2}}$       && 4  &\cr
&$K_{{J_1}{J_2}{J_2}}$       && 0  &\cr
&$K_{{J_2}{J_2}{J_2}}$       && 0  &\cr
&$\int c_2 h_{J_1}$  && 56  &\cr
&$\int c_2 h_{J_2}$  && 24  &\cr
height1pt&\omit&   &\omit&\cr
\noalign{\hrule}}
\hrule}$$
The theorem of Wall, applicable for manifolds without torsion,
states that $X_1$ and $X_2$ are homotopy equivalent if these
topological numbers coincide, up to a linear transformation
of the basis. Identifying $J_1=J$ and $J_2={1\over 2}(J-D)$
we see that this is in fact the case.
Model $X_1$ has been treated in great detail in refs.\hkty~and
\cdfkm. In fact, one can prove that $X_1$ and $X_2$ are diffeomorphically
equivalent, by realising both as singular fiber spaces where the
generic fiber is a $K_3$
over $\IP^1$  with pairs of points identified on the latter.
Similary we can
find for the complete intersections $B$ and $C$ in one singular
projective space, which were discussed in ($5$ $viii$),
diffeomorphic realisations in products of nonsigular projective spaces.
We have the following equivalences
\eqn\equivalences{\eqalign{
A:&\,\,\,
{\cicy-168(\IP^4[2,2,2,1,1]|8)2}\simeq
{\cicy-168(\IP^4\cr \IP^1|4&1\cr 0&2)2} \cr
B:&\,\,\,
{\cicy-132(\IP^5[2,2,2,2,1,1]|4,6)2}\simeq
{\cicy-132(\IP^5\cr \IP^1|2&3&1\cr 0&0&2)2} \cr
C:&\,\,\,
{\cicy-112(\IP^6[2,2,2,2,2,1,1]|4,4,4)2} \simeq
{\cicy-112(\IP^6\cr \IP^1|2&2&2&1\cr 0&0&0&2)2}.}}

Let us discuss the solution for the model $X_2$. Here
we have the vertices of the dual
polyhedra $\nu^*_{1,1}=(1,0,0,0;0),\,\nu^*_{1,2}=(0,1,0,0;0),\,
\nu^*_{1,3}=(0,0,1,0;0),\,\nu^*_{1,4}=(0,0,0,1;0),\,
\nu^*_{1,5}=(-1,-1,-1,-1;0)$ and $\nu^*_{2,1}=(0,0,0,0;1),\,
\nu^*_{2,2}=(0,0,0,0;-1)$. We group them into two sets:
$E_1=\{\nu^*_{1,1},\nu^*_{1,2},\nu^*_{1,3},\nu^*_{1,4}\}$ and
$E_2=\{\nu^*_{1,5},\nu^*_{2,1},\nu^*_{2,2}\}$ and proceed as described
in section three. This leads to
\eqn\www{
l^{(1)}=(-4,-1;1,1,1,1,1,0,0)\quad,\qquad
l^{(2)}=( 0,-2;0,0,0,0,0,1,1)
}
We can also write down the Laurent polynomials. They are
\eqn\laurantII{
\eqalign{P_1&=a_0-a_1 X_1-a_2 X_2-a_3 X_3-a_4 X_4\cr
         P_2&=b_0-{b_1\over X_1 X_2 X_3 X_4}-b_2 Y_1-{b_3\over Y_1}}}
Note that due to the freedom to rescale all variables and each
polynomial, there are only two relevant parameters in $P_1$ and
$P_2$, corresponding to the two complex structure moduli
on $X_2^*$.
The period \powersol~follows straightforwardly by
performing the integral \periods.
One also finds $z_1={a_1 a_2 a_2 a_4 b_1\over a_0^4 b_0}$ and
$z_2={b_2 b_3\over b_0^2}$.
$l^{(1)}$ and $l^{(2)}$ lead, after trivial factorization,
to differential operators of orders four $(\cL_1)$
and two ($\cL_2=L_2$), respectively. A third order operator
$L_1$ can be obtained via $\cL_1-4\t_1^2\cL_2=(2\t_2+\t_1)\,L_1$.
The Yukawa couplings for the model $X_2$ are found to be
\eqn\xxx{
\eqalign{K^{(3,0)}&={1\over z_1^3\,\Delta_0}\,,\qquad
         K^{(2,1)}={1-256\,z_1\,-\,4\,z_2\over 2\,
         z_1^2 z_2 \, \Delta_0\,\Delta_1}\,,\cr
         K^{(1,2)}&={3-512\,z_1+4\,z_2z_1 z_2 \, \over\Delta_0\, \Delta_1^2}
           \,,\qquad
         K^{(0,3)}={1-256\,z_1+24\,z_2-3072\,z_1\,z_2+16\,z_2^2
                    \over2\,z_2^2\, \Delta_0\,\Delta_1^3}}
}
with
$$
\Delta_0=(1-256\,z_1)^2-4\,z_2,\quad \Delta_1=(1-4 z_2)
$$
The resulting topological invariants for the rational and
elliptic curves for the case $A,B$ and $C$ in \equivalences~ are
of course the same for both realisations. Model
$X_1$ was solved in  \hkty,\cdfkm. In fact, in \cdfkm~ the degrees of the
curves are given with respect to the basis appropriate for $X_2$.
We can find a map analogous to the one in \map~
and hence a configuration $\cicy-168(\IP^4\cr \IP^1|4&1\cr 0&2)2$,
which reproduces, via the integral \pcd, the expression for $w_0$.
However, this configuration cannot be chosen to be transverse
for generic points in the two dimensional subspace of the
moduli space under consideration. It is also interesting to note
that the coordinates used here and the one used in \hkty,\cdfkm~
are connected by a transcendental function.

A similar comparison can be made between the two models
$\cicy-252(\IP^4\(6,2,2,1,1\)|12)2$.
and
$\cicy-252(\IP^4\(3,1,1,1,1\)\cr\IP^1|6&1\cr 0&2)2$.
At special points in moduli space they correspond to
the Gepner models $(A_4,A_4,A_{10},A_{10})$ and
$(A_4,A_4,D_{10},D_{10})$, respectively.
One finds again the relations $J_1=J$ and $J_2={1\over2}(J-D)$
between the divisors and identical topological invariants.

%%%%%%%%%%%%%%%%%%%%%%%%%%%%%%%%%%%%%%%%%%%%%%%%%%%
\newsec{Topological one-loop partition function and
the number of elliptic curves}
%%%%%%%%%%%%%%%%%%%%%%%%%%%%%%%%%%%%%%%%%%%%%%%%%%%

Knowledge of the periods  and the canonical coordinates
allows, up to the difficulty of fixing an holomorphic ambiguity
at each step, to recursively calculate higher loop topological
partition funcions as was shown in
\bcovII.
To fix the ambiguity we need as gobal properties of the moduli
space $\cal M$ the singularities discussed in sec. 4.

We will focus on the one loop case and calculate the expression
defined in the $N=2$ SCFT on the torus as
\eqn\oneloop{
F_1={1\over 2} \int {d \tau \over \tau_2} {\rm Tr}\, (-1)^F
F_L F_R q^{L_0} \bar q^{\bar L_0},}
where the trace is to be taken over the left- and right moving
Ramond sectors. As shown in \bcovII~ and \lk~
this quantity is,
for the heterotic string with canonical embedding
of the spin connection into the gauge group, related to the
difference of the threshold corrections to the gauge couplings of
the $E_6$ and $E_8$, namely $12 F_1=\Delta(E_6)-\Delta(E_8)$.
Using the holomorphic anomaly equation
it was shown in \bcovI~ that it can be written as
\eqn\oneloopII{F_1=\log\left[ M(z,\bar z) |f(z)|^2\right],}
where the holomorphic-antiholomorphic mixing $M(z,\bar z)$ is
given by
\eqn\yyy{
\log M =\sum_{p,q} (-1)^{p+q}\, {p+q\over 2} {\rm Tr}_{p,q}
\log\det(g)-{1\over 12} {\rm Tr} (-1)^F ;
}
here $g$ is the $t,\bar t$ metric introduced in
\ref\cv{S. Cecotti and C. Vafa, Nucl.Phys. B367(1991)359}.
It is related to the
Weil-Peterson metric $G_{i\bar\jmath}$ by
$g_{i\bar\jmath}/g_{ 0 \bar 0}= g_{i \bar\jmath}
\exp(K)=G_{i\bar\jmath}$ with
$G_{i\bar\jmath}= \partial_i\partial_{\bar\jmath} K$.
For $\sigma$-models on Calabi-Yau spaces \oneloopII~ can be rewritten
as
\eqn\oneloopIII{F_1=\log\left[\exp\left((3 + k- {\chi\over 12})
K\right)
\det[G_{i\bar\jmath}]^{-1} |f(z)|^2\right].}
In our application we will finally understand  $F_1$ as a function
of the K\"ahler moduli $t,\bar t$ which are related to $a,\bar a$ or
$z,\bar z$ by the mirror map. To fix the holomorphic ambiguity $f(z)$
one considers the large volume limit $t,\bar t \rightarrow \infty$
for which one has the asymptotic behaviour
\eqn\oneloopasym{
\lim_{t,\bar t \rightarrow \infty} F_1=
- {2 \pi i \over 12} \sum_{i=1}^{k}
(t_i + \bar t_i) \int c_2 J_i.}
It was conjectured in \bcovI~that $F_1^{top}\equiv
\lim_{\bar t\rightarrow \infty} F_1$
has the following expansion
\eqn\oneloopexp{{F_1}^{top}={\rm const.}\!-\!{2 \pi i \over 12} \sum_{i=1}^{k}
t_i\!\int\!c_2 J_i-\sum_{n_l} \left[2 n^e_{d_1,\ldots,d_k}\log(\eta(
\prod_{i=1}^{k} q_i^{d_i}))+ {1\over 6} n^r_{d_1\ldots d_k}
\log(1-\prod_{i=1}^k {q_i}^{d_i})\right]}
in terms of the Euler numbers $n^r_{d_1,\ldots,d_h}$ and
$n^e_{d_1,\ldots,d_k}$
of the tangentbundle over the appropriate compactified moduli
space of the mappings from $\IP^1$ and $T^2$, respectively,
to the Calabi-Yau space.
In  the case of isolated curves they count the number of
rational curves and elliptic curves respectively.
Using  \pIII,\kaehler~ one gets in the general case as the
$\bar t\rightarrow \infty $ limit of \oneloopIII~
\eqn\onelooptoplim{F_1^{top}=\log \left[\left(a_1\cdots a_l\over
\omega_0\right)^{3 + k-\chi/12}{\partial(a_1\ldots a_k)\over
\partial(t_1\ldots t_k)} f(z)\right]+ {\rm const.}}
The factor $(a_1\cdots a_l)/\omega_0$ corresponds to the gauge choice
with the fundamental period normalized as in \periods\foot{$F_1^{top}$ depends
of course on $k$ parameter $a_i$ $i=1,\ldots,k$. By the $\IC^*$
symmetries, acting on the parameters of the Laurent polynomials,
we can set $l-k$ of the $a_i$ appearing in \periods~ to one,
if the number of polynomial constraints $l$ exceeds $k$.}
The holomorphic anomaly is determined by the requirement that
$F_1^{top}$ has to be a regular function everywhere in moduli space
except possibly at the components of the discriminante surface
determined by $\Delta_i=0$, which can be directly determined from the
Laurent polynomial (cf. (5{\it i})).
Besides the components $\Delta_i=0$ there appear
also other singular loci $\delta_i=0$ in systems of Picard-Fuchs
equations, which can be understood as identification singularities
of the parameters space of the Laurent polynomial as it was
discussed at the end of section three.
In the cases we discuss we have $\delta_i=a_i$.
While it is evident from \periods~that $\left(a_1\ldots a_k
\over \omega_0\right)$ is regular
at $\delta_i=0$ the Jacobian of the mirror map \map~
might have singularities at $\delta_i$.
We therefore make the following general ansatz for
$f(z)$ \eqn\ansatz{
f(z)=\Delta_0^{r_0}\ldots\Delta_m^{r_m}\delta_1^{s_1}
\ldots \delta_k^{s_k}.}
Of course the $\Delta_i=z_i$
singularities are included in this ansatz.
Inserting this ansatz in \onelooptoplim~and comparing
the leading term with \oneloopasym~yields equations for the $r_i$ and
$s_i$.
If the  manifold happens to be transverse at $a=0$ the
powers of the $\delta_i$, which in this case only have to compensate
possible singularities of the Jacobian, can in principle be determined by
analytic continuation of the periods and hence the
mirror map to the point $a=0$. In the general case we use the values
the numbers $n^e_{d_1,\ldots,d_h}$ of a few elliptic
curves of low polydegree w.r.t. an integral basis of divisors,
typically the fact they have to vanish,
to fix all parameters and predict the other numbers.

Let us first discuss as examples various one moduli cases
realized as Fermat hypersurfaces $X_{k_1}$ of degree $k_1$ or
complete intersections $X_{k_1,\ldots,k_n}$ of multidegree
$k_1,\ldots,k_n$ in a single weighted projective space.
{}From \ktone\libtei\ktthree we have one component
of the singular locus for all cases at $\Delta_0=(1- a^{\sum k_i})$.
We start with a short review of the one moduli hypersurfaces in a
weighted $\IP^4$, denoted by $X_k=\cicy{}(\IP^4[\vec w]|,k){}$ .
In this case $\Delta_0$ is the only component of the
singular locus and it was observed in
\bcovII~that $r_0=-1/6$ for all cases yielding
the following invariants for the elliptic curves:
$$
\vbox{\offinterlineskip\tabskip=0pt
\halign{\strut\vrule#&\hfil~$#$~&\hfil$#$\quad&\hfil~$#$\quad&\hfil~$
#$
&\vrule#&~\hfil $#$~~&\hfil $#$~~&\hfil $#$~~&\hfil $#$~~&\vrule#\cr
\noalign{\hrule}
&{\rm Model}&\chi&\int c_2 J&s_1 && n^e_1 & n^e_2 & n^e_3 &n^e_4 &\cr
\noalign{\hrule}
&X_5 &-200&50&0  &&   0 & 0    &609250  &37214316625&\cr
&X_6 &-204&42&0  &&   0 & 7884    &145114704  &1773044315001&\cr
&X_8 &-296&44&1  &&   0 & 41312    &21464350592&1805292092664544&\cr
&X_{10}&-288&34&1 && 280
&207680680&161279120326560&103038403740690105440&
\cr
\noalign{\hrule}}
\hrule}$$

Similarly for the one moduli complete intersections in ordinary
projective spaces (cf. \libtei) we found
$r_0=-1/6$ and the $s_1$ value indicated
in the following table. This result was obtained by requiring
$n^e_1=0$, which holds for intersections in ordinary
projective spaces, and imposing \oneloopasym. Note that these
manifolds are not transverse at $\delta_1=a=0$.
In order to compare with the results of section (5 $viii$)
we list also the rational curves for these models
$$
\vbox{\offinterlineskip\tabskip=0pt
\halign{\strut\vrule#
&\hfil~$#$~
&\hfil~$#$~
&\hfil~$#$~
&\hfil~$#$
&\vrule#&~
\hfil $#$~
&\hfil $#$~
&\hfil $#$~
&\hfil $#$~
&\hfil $#$~
&\hfil $#$~
&\vrule#\cr
\noalign{\hrule}
&{\rm Model}&\chi&\int c_2 J&s_1
                      &&       & j=1 &  j=2  &  j=3  &  j=4 & j=5 & \cr
\noalign{\hrule}
&X_{3,3} &-144&54&11  && n^r_j &  9 &  1053 &52812& 6424326&11394483834&\cr
&        &    &  &    && n^e_j &  0 & 0    &3402  &5520393&482074484&\cr
\noalign{\hrule}
&X_{4,2} &-176&56&{28\over 3}
                      && n^r_j &  8 &1280  &92288 &15655168&3883902528&\cr
&        &    &  &    && n^e_j &  0 & 0    &2560  &17407072& 24834612736&\cr
\noalign{\hrule}
&X_{3,2,2}&-144&60&{115\over6}
                      && n^r_j &  12& 720  &22428 &1611504 &168199200&\cr
&        &    &  &    && n^e_j &  0 & 0    &64& 265113&198087264&\cr
\noalign{\hrule}
&X_{2,2,2,2}&-128&64&{85\over 3}
                      && n^r_j & 16 & 512  &9728&416256&25703936&\cr
&        &    &  &    && n^e_j &  0 &     0&   0& 14752&8782848 & \cr
\noalign{\hrule}}
\hrule}$$

For the complete intesections in weighted projective
spaces (cf. \ktthree) we found that $r_0=-1/6$ likewise and $s_1$ from
\oneloopasym~gives the following integral expansion for the
$n^e_d$
$$
\vbox{\offinterlineskip\tabskip=0pt
\halign{\strut\vrule#&\hfil~$#$~~&\hfil$#$\quad&\hfil~~$#$\quad&\hfil~~$
#$
&\vrule#&
{}~\hfil $#$~~
&\hfil $#$~~
&\hfil $#$~~
&\hfil $#$~~
&\vrule#\cr
\noalign{\hrule}
&{\rm Model}&\chi&\int c_2 J&s_1 && n^e_1 & n^e_2 & n^e_3 &n^e_4 &\cr
\noalign{\hrule}
&X_{4,3} &-156&48&{67\over 6} &&   0 & 27    &16124238&38170438&\cr
&X_{6,2}&-256&52&{29\over 3} && 0 &-504&1228032&79275665304&\cr
&X_{4,4} &-144&40&11  && 0 & 1408    &6953728 &2684185380&\cr
&X_{6,4}&-156&32&{31\over 3} &&
8&258336&5966034464&1267294361302800&\cr
&X_{6,6} &-120&22&9
&&360&40691736&4956204918240&616199133057629184&\cr
\noalign{\hrule}}
\hrule}
$$
All invariants of the $X_{6,2}=\cicy{-256}(\IP^5[3,1,1,1,1,1]|6,2)1$ model
are consistent with the indentification of this model
with the one parameter subspace of
$\cicy-252(\IP^4[6,2,2,1,1]|12)2$\cdfkm, which implies
$n^e_i=\sum_{j} n^e_{i,j}$.
Especially $n_2^e=-504$, which is the only negative invariant
for a one parameter family, is in agreement with
$n^e_{2,0}=n^e_{2,2}=-492$, $n^e_{2,1}=480$ for
the two parameter hypersurface (comp. \cdfkm).
We also checked that the identification of $X_{6,4}=\cicy{-156}
(\IP^5[3,2,2,1,1,1]|6,4)1$
with the one parameter subspace of $\cicy{-144}(\IP^4[4,3,2,2,1]|12)2$
\hkty~ holds at the one loop level.
E.g. the lowest invariants of elliptic curves
for the latter model are $n^e_{1,0}=-2$, $n^e_{1,1}=6$
and $n^e_{2,0}=762$, $n^e_{2,1}=-3060$, $n^e_{2,2}=18918$,
$n^e_{2,3}=225096$ with the general symmetry $n_{i,j}=n_{i,3i-j}$
and $n_{i,j}=0$ for $j>3i$; they reproduce the
first two entries for the $X_{6,4}$ model above.

In summary the topological one loop partition function
for the one parameter models is given by
\eqn\zzz{
F_1^{top}=\log\left[{a^{{\sum_i k_i\over 12}\int c_2 J}
\over w_0^{4-{\chi+4\over 12}}}
\left({1\over a}{\partial a\over \partial t}\right)^{1/2}
K_{ttt}^{1\over 6}\right].}

Next we treat a hypersurface
$\cicy{-240}(\IP^4[7,2,2,2,1]|14)2$ with two moduli. The singular locus is,
in addition to the lines $z_1=0$, $z_2=0$, given by
\eqn\singI{\eqalign{
\Delta_0&=
1 +
27 z_1 -
63 z_1 z_2+
56 z_1 z_2^2-
112 z_1 z_2^3-
(7-4 z_2)^4 z_1^2 z_2^3\cr
\Delta_1&= 4 z_2-7}.}
Equation \goodvar~and the analog of \linearrelations~ from
\hkty~ $l^{(1)}=(-7;0,1,1,1,-3,-7)$ and $l^{(2)}=(0;1,0,0,0,1,-2)$
define\foot{Note the  scale factor $7$ introduced in \hkty~in order to
simplify \singI.} the relation $z_1=a_2^7/a_1^7$ and $z_2= 7/a_2^2$
by which we transform the expressions of singular components \singI~to
the $a_i$ variables. Beside this components we have to care about
the sets $\delta_1=a_1=0$ and $\delta_2=a_2=0$. For the further
calculation it turns out to be advantegous to get rid of the
denominators in the transformed expressions \singI~by rescaling
$\Delta_1\rightarrow a^{14}\Delta_1$ and
$\Delta_2\rightarrow b^2 \Delta_2$.

The Euler number is $\chi=-240$ the
Hodge numbers are $h^{1,1}=2, h^{2,1}=122$. We calculate
$\int c_2 \wedge J_1 =44$ and $\int c_2\wedge J_2=126$,
where $J_1,J_2$ are the basis which generates
$H^2(X,\ZZ)$. It is  connected with the $(1,1)$ forms dual
to natural basis of the divisors in the polyheder
construction $J,D$ used in \hkty, by $J_1=J$ and
$J_2={1\over 2}(7J-D)$.
{}From \oneloopasym~we get two equations
\eqn\aaaa{
r_0=-{s_1\over 14}-{1\over 42},
\quad r_1=-{s_2\over 2}-{17\over 6}.}
The vanishing of the numbers of curves
$n^e_{0,1}=s_2$ and $n^e_{1,0}={27\over 28}\,(2- s_1)$
enforces $s_1=2$ and $s_2=0$. Other numbers of elliptic
curves $n^e_{i,j}$ with $i<3$ are then given in the following table,
where we list for convenience also the number of rational curves
$n^r_{i,j}$
$$
\vbox{\offinterlineskip\tabskip=0pt
\halign{\strut\vrule#
&\hfil~$#$
&\vrule#&~
\hfil ~$#$~
&\hfil ~$#$~
&\hfil $#$~
&\hfil $#$~
&\hfil $#$~
&\hfil $#$~
&\hfil $#$~
&\hfil $#$~
&\vrule#\cr
\noalign{\hrule}
&n_{i,j}&& j=0& j=1 &  j=2  &  j=3  &  j=4 & j=5 & j=6& j=7& \cr
\noalign{\hrule}
&n^r_{0,j} && 0& 28&  0&\forall j>1&&&&&\cr
&n^e_{0,j} && 0&\forall j&&&&&&&\cr
\noalign{\hrule}
&n^r_{1,j}&& 3&-56&378&14427&14427&378&-56&3&\cr
&n^e_{1,j} && 0&\forall j&&&&&&&\cr
\noalign{\hrule}
&n^r_{2,j}&&-6&140&-1512&9828&-69804&500724&29683962&68588248&\cr
&n^e_{2,j}&& 0&  0&    0&   0&     0&   378&6496&27564&\cr
\noalign{\hrule}}
\hrule}$$

Here we have a symmetry $n_{i,j}=n_{i,7 i-j}$ and $n_{i,j}=0$
for $j>7i$. The identification of the one parameter
subspace of this model via $n_i=\sum_j n_{i,j}$
with the hypersurface $X_8=\cicy{-256}(\IP^4[4,1,1,1,1]|8)1$ observed
in \hkty~ can also be checked at one-loop level.

Next we treat a hypersurface in a product of two
projective spaces
$$
\cicy-162(\IP^2\cr \IP^2|3\cr 3)2.
$$
The number $n^r_{d_1,d_2}$ of rational curves of bidegree $(d_1,d_2)$
was obtained in \hkty. The evaluation of the second
Chern class on $J_1,J_2$ can be read off from \top;
$\int c_2\wedge J_1=\int c_2 \wedge J_2=36$. Because
of the symmetry we have $s_1=s_2$. In this
case we have to consider only one component
\eqn\bbbb{
\Delta_0 = 1- (1-z_1)^3+(1-z_2)^3+ 3 z_1 z_2 ( z_1+ z_2+7)}
of the discriminant of the complete intersection,
where the connection with the parameters $a_1,a_2$ is
encoded in the $l^{(i)}$ given by \linearrelations,
$z_1=3^3/a_1^3$ and $z_2=3^3/a_2^3$.
Again after clearing the denominator, the comparison with the
large radius limit gives us the relation
$r_0=-{7\over 6}-{s_1\over 9}$, leaving
us with one unknown constant $s_1$, which is determined
by requiring $n^e_{0,1}={9\over 2}\,(9 + s_1)=0$. The following
table contains the numbers of rational and elliptic curves
up to bidegree $d_1+d_2\le 6$. Because of the exchange symmetry we
list only $n^i(d_1,d_2)$ with $d_1\le d_2$.
$$
\vbox{\offinterlineskip\tabskip=0pt
\halign{\strut\vrule#&~$#$~~&$#$~&\vrule$#$&\qquad\hfil$#$~&\vrule$#$\cr
\noalign{\hrule}
&(0,1)&n^r &&  189 &\cr
&&    n^e &&    0 &\cr
\noalign{\hrule}
&(0,2)&n^r &&  189 &\cr
&& n^e &&    0 &\cr
\noalign{\hrule}
&(0,3)&n^r &&  162 &\cr
&&    n^e &&    3 &\cr
\noalign{\hrule}
&(0,4)&n^r &&  189 &\cr
&& n^e &&    0 &\cr
\noalign{\hrule}
&(0,5)&n^r &&  189 &\cr
&&    n^e &&    0 &\cr
\noalign{\hrule}
&(0,6)&n^r &&  162 &\cr
&& n^e &&    0 &\cr
\noalign{\hrule}
}
\hrule}
\quad
\vbox{\offinterlineskip\tabskip=0pt
\halign{\strut\vrule#&~$#$~~&$#$~&\vrule$#$&\qquad\hfil$#$~&\vrule$#$\cr
\noalign{\hrule}
&(1,1)&n^r &&  8262 &\cr
&&    n^e &&      0 &\cr
\noalign{\hrule}
&(1,2)&n^r &&  142884 &\cr
&& n^e &&           0 &\cr
\noalign{\hrule}
&(1,3)&n^r &&  1492290 &\cr
&&    n^e &&    -378 &\cr
\noalign{\hrule}
&(1,4)&n^r &&  11375073 &\cr
&& n^e &&      -16524&\cr
\noalign{\hrule}
&(1,5)&n^r &&  69962130 &\cr
&&    n^e &&   -285768 &\cr
\noalign{\hrule}
}
\hrule}
\quad
\vbox{\offinterlineskip\tabskip=0pt
\halign{\strut\vrule#&~$#$~~&$#$~&\vrule$#$&\quad\hfil$#$~&\vrule$#$\cr
\noalign{\hrule}
&(2,2)&n^r && 1310892  &\cr
&&    n^e &&  8262  &\cr
\noalign{\hrule}
&(2,3)&n^r &&  516953097 &\cr
&& n^e &&      1519434 &\cr
\noalign{\hrule}
&(2,4)&n^r &&  12289326723 &\cr
&&    n^e &&   71809416 &\cr
\noalign{\hrule}
height-5pt&\omit&\omit&\omit&\omit&\omit\cr
\noalign{\hrule}
&(3,3)&n^r && 55962304650  &\cr
&&    n^e &&  818388234 &\cr
\noalign{\hrule}
}
\hrule}
$$
Here we observe $n^e_{1,3}=-2\, n^r_{0,1}$, $n^e_{2,2}=n^r_{1,1}$
and $n^e_{1,d}=-2\, n^r_{1,d_2-3}$ for $d_2\ge 4$.

For the three-generation complete intersection case
({\it 5 i}) the relevant components of the discriminant surface
were obtained by different methods in section 4 and 5:
$\Delta_0=(1-27{z_1})^3-27{z_2}$, $\Delta_1=(1-27 z_2)$.
Proceeding as before we get
$r_0=-{1\over 6}-{s_2\over 9}$, $ r_1=3 -s_1+s_2$
and the vanishing of $n^e_{1,0}={9\over 2} s_2$ requires
$s_2=0$ and then from
$n^e_{1,1}={3\over 2} (9 s_1-27)=0$ we get $s_1=3$. The corresponding
predictions for $n^e_{d_1,d_2}$ are
$$
\vbox{\offinterlineskip
\hrule
\halign{ &\vrule# & \strut\quad\hfil$#$\quad\cr
\noalign{\hrule}
height1pt&\omit&   &\omit&\cr
&(1,0)&& 0  &\cr
&(2,0)&& 0 &\cr
&(3,0)&& 3 &\cr
&(4,0)&& 0  &\cr
&(5,0)&& 0  &\cr
&(6,0)&& 0  &\cr
height1pt&\omit&   &\omit&\cr
\noalign{\hrule}}
\hrule}
\,\,
\vbox{\offinterlineskip
\halign{ &\vrule# & \strut\quad\hfil$#$\quad\cr
\noalign{\hrule}
\noalign{\hrule}
height1pt&\omit&   &\omit&\cr
&(0,1)&& 0  &\cr
&(0,2)&& -27  &\cr
&(0,3)&& 81 &\cr
&(0,4)&& -324  &\cr
&(0,5)&& 1728  &\cr
&(0,6)&& -8955  &\cr
height1pt&\omit&   &\omit&\cr
\noalign{\hrule}}
\hrule}
\quad
\vbox{\offinterlineskip
\halign{ &\vrule# & \strut\quad\hfil$#$\quad\cr
\noalign{\hrule}
\noalign{\hrule}
height1pt&\omit&   &\omit&\cr
&(1,1)&&  0 &\cr
&(2,2)&&  -16028  &\cr
&(3,3)&&  -124719 &\cr
height1pt&\omit&   &\omit&\cr
\noalign{\hrule}
\noalign{\hrule}
height1pt&\omit&   &\omit&\cr
&(2,1)&& 0  &\cr
&(4,2)&& -924372  &\cr
\noalign{\hrule}
height1pt&\omit&   &\omit&\cr
&(3,1)&&  -126  &\cr
height1pt&\omit&   &\omit&\cr
\noalign{\hrule}
\noalign{\hrule}
height1pt&\omit&   &\omit&\cr
&(4,1)&&  -1944  &\cr
height1pt&\omit&   &\omit&\cr
&(5,1)&&  -30618  &\cr
height1pt&\omit&   &\omit&\cr
\noalign{\hrule}}
\hrule}
\,\,
\vbox{\offinterlineskip
\halign{ &\vrule# & \strut\quad\hfil$#$\quad\cr
\noalign{\hrule}
\noalign{\hrule}
height1pt&\omit&   &\omit&\cr
&(1,2)&&   972 &\cr
&(2,4)&&   -426222 &\cr
height1pt&\omit&   &\omit&\cr
\noalign{\hrule}
\noalign{\hrule}
height1pt&\omit&   &\omit&\cr
&(3,2)&&  159678 &\cr
height1pt&\omit&   &\omit&\cr
\noalign{\hrule}
\noalign{\hrule}
height1pt&\omit&   &\omit&\cr
&(1,3)&&  -486 &\cr
height1pt&\omit&   &\omit&\cr
\noalign{\hrule}
\noalign{\hrule}
height1pt&\omit&   &\omit&\cr
&(2,3)&&  27945  &\cr
height1pt&\omit&   &\omit&\cr
\noalign{\hrule}
\noalign{\hrule}
height1pt&\omit&   &\omit&\cr
&(1,4)&&   22356 &\cr
height1pt&\omit&   &\omit&\cr
\noalign{\hrule}
\noalign{\hrule}
height1pt&\omit&   &\omit&\cr
&(1,5)&&  -72900 &\cr
height1pt&\omit&   &\omit&\cr
\noalign{\hrule}}
\hrule}
$$

For the Tian-Yau manifold example (5 {\it ii}) the discriminant
can be read off from the Yukawa couplings
$\Delta_0=1-27\,z_1 - 27\,z_2$,
$\Delta_1= (1-27\,z_1)$, $\Delta_2= (1-27\,z_2).$
Obviously $r_1=r_2$ and $s_1=s_2$ and \oneloopasym~ yields
$r_0={1\over 2}+ {s_1\over 3}-r_1$.
The following predictions are obtained by imposing
$n^e_{0,1}={9\over 2}(s_1-2)=0$ and $n^3_{1,1}={243\over 2}(3 r_1-4)=0$ :

%lec = {nec[0, 1] -> 0, nec[0, 2] -> -27, nec[0, 3] -> 81, nec[0, 4] ->  -324,
%   nec[0, 5] -> 1728, nec[0, 6] -> -8955, nec[1, 0] -> 0, nec[1, 1] ->  0,
%   nec[1, 2] -> 324, nec[1, 3] -> 1458, nec[1, 4] -> 4374,
%   nec[1, 5] -> 15066, nec[2, 0] -> -27, nec[2, 1] -> 324,
%   nec[2, 2] -> -2916, nec[2, 3] -> -13176, nec[2, 4] -> -31104,
%   nec[3, 0] -> 81, nec[3, 1] -> 1458, nec[3, 2] -> -13176,
%   nec[3, 3] -> 108180, nec[4, 0] -> -324, nec[4, 1] -> 4374,
%   nec[4, 2] -> -31104, nec[5, 0] -> 1728, nec[5, 1] -> 15066,
%   nec[6, 0] -> -8955}

$$
\vbox{\offinterlineskip
\hrule
\halign{ &\vrule# & \strut\quad\hfil$#$\quad\cr
\noalign{\hrule}
height1pt&\omit&   &\omit&\cr
&(0,1)&& 0 &\cr
&(0,2)&& -27 &\cr
&(0,3)&& 81 &\cr
&(0,4)&& -324 &\cr
&(0,5)&& 1728 &\cr
&(0,6)&& -8955 &\cr
height1pt&\omit&   &\omit&\cr
\noalign{\hrule}}
\hrule}
\quad
\vbox{\offinterlineskip
\halign{ &\vrule# & \strut\quad\hfil$#$\quad\cr
\noalign{\hrule}
\noalign{\hrule}
height1pt&\omit&   &\omit&\cr
&(1,1)&&  0 &\cr
&(2,2)&&  -2916 &\cr
&(3,3)&&   108180&\cr
height1pt&\omit&   &\omit&\cr
\noalign{\hrule}
\noalign{\hrule}
height1pt&\omit&   &\omit&\cr
&(1,2)&& 324  &\cr
&(2,4)&& -31104 &\cr
height1pt&\omit&   &\omit&\cr
\noalign{\hrule}}
\hrule}
\quad
\vbox{\offinterlineskip
\halign{ &\vrule# & \strut\quad\hfil$#$\quad\cr
\noalign{\hrule}
\noalign{\hrule}
height1pt&\omit&   &\omit&\cr
&(1,3)&&  1458 &\cr
height1pt&\omit&   &\omit&\cr
\noalign{\hrule}
\noalign{\hrule}
height1pt&\omit&   &\omit&\cr
&(1,4)&&  4374&\cr
height1pt&\omit&   &\omit&\cr
\noalign{\hrule}
\noalign{\hrule}
height1pt&\omit&   &\omit&\cr
&(1,5)&& 15066  &\cr
height1pt&\omit&   &\omit&\cr
\noalign{\hrule}
\noalign{\hrule}
height1pt&\omit&   &\omit&\cr
&(2,3)&& -13176  &\cr
\noalign{\hrule}}
\hrule}
$$
Note that invariants $n^e_{0,i}$ of example (5 {\sl i})
coincide with $n^e_{i,0}=n^e_{0,i}$ of the present one.

The three moduli example in section
(5 {\sl v}) has the general discriminant
\eqn\cccc{
\eqalign{\Delta_0=&
1-108 z_1+ 4374 z_1^2-78732 z_1^3+ 531441 z_1^4 - 8 z_2+432 z_1 z_2 \cr
&-5832 z_1^2 z_2+16 z_2^2 - 8 z_3+ 432 z_1 z_3-5832 z_1^2 z_3 -23 z_2
z_3+
16 z_3^2}
}
and a second component of the discriminant locus
\eqn\dddd{
\Delta_1=1-8 z_2 + 16 z_2^2- 8 z_3 -32 z_2 z_3 + 16 z_3^2.}
As a slight technical simplification
we replace $\left(a_1\ldots a_k\over w_0\right)$ by ${1\over w_0}$ in
\onelooptoplim~, set $\delta_i=z_i$ in \ansatz~
and work in the following throughoutly with the large
complex structure parameters $z_i$,
which is possible as the $z_i$ are the good coordinates on
${\cal M}$ (cf. section 3).

By \oneloopasym~we can fix in this case $s_1=-4$,
$s_2=s_3=-3$ and from  $n^e_{0,0,1}=-9(1+6 r_0)=0$
we have again $r_0=-1/6$. Enforcing also $n^e_{1,0,0}=-4(r_0+r_1)=0$
we get the following predictions for the non-zero
invariants of the elliptic curves up to bidegree 7
$$
\vbox{\offinterlineskip
\halign{ &\vrule# & \strut\quad\hfil$#$\quad\cr
\noalign{\hrule}
\noalign{\hrule}
height1pt&\omit&   &\omit&\cr
&(0,1,1)&&    -12    &\cr
&(0,2,2)&&     15  &\cr
height1pt&\omit&   &\omit&\cr
\noalign{\hrule}
\noalign{\hrule}
height1pt&\omit&   &\omit&\cr
&(1,1,1)&&    288 &\cr
&(2,2,2)&&   -13284 &\cr
height1pt&\omit&  &\omit&\cr
\noalign{\hrule}
\noalign{\hrule}
height1pt&\omit&   &\omit&\cr
&(1,1,2)&&    2888 &\cr
height1pt&\omit&  &\omit&\cr
\noalign{\hrule}
\noalign{\hrule}
height1pt&\omit&   &\omit&\cr
&(1,2,2)&&    2160 &\cr
height1pt&\omit&  &\omit&\cr
\noalign{\hrule}
\noalign{\hrule}
height1pt&\omit&   &\omit&\cr
&(1,2,3)&&    2160  &\cr
height1pt&\omit&   &\omit&\cr
\noalign{\hrule}
\noalign{\hrule}
height1pt&\omit&   &\omit&\cr
&(1,3,3)&&    11664 &\cr
height1pt&\omit&  &\omit&\cr
\noalign{\hrule}
\noalign{\hrule}
height1pt&\omit&   &\omit&\cr
&(2,1,1)&&    -3024 &\cr
height1pt&\omit&   &\omit&\cr
\noalign{\hrule}
\noalign{\hrule}
height1pt&\omit&   &\omit&\cr
&(2,1,2)&&    -4320  &\cr
height1pt&\omit&   &\omit&\cr
\noalign{\hrule}}
\hrule}
\,\,
\vbox{\offinterlineskip
\halign{ &\vrule# & \strut\quad\hfil$#$\quad\cr
\noalign{\hrule}
\noalign{\hrule}
height1pt&\omit&   &\omit&\cr
&(2,1,3)&&    -3024 &\cr
height1pt&\omit&   &\omit&\cr
\noalign{\hrule}
\noalign{\hrule}
height1pt&\omit&   &\omit&\cr
&(2,2,3)&&    -2052 &\cr
height1pt&\omit&   &\omit&\cr
\noalign{\hrule}
\noalign{\hrule}
height1pt&\omit&   &\omit&\cr
&(3,0,0)&&      4 &\cr
height1pt&\omit&   &\omit&\cr
\noalign{\hrule}
\noalign{\hrule}
height1pt&\omit&   &\omit&\cr
&(3,0,1)&&    -36 &\cr
height1pt&\omit&   &\omit&\cr
\noalign{\hrule}
\noalign{\hrule}
height1pt&\omit&   &\omit&\cr
&(3,0,2)&&    -36 &\cr
height1pt&\omit&   &\omit&\cr
\noalign{\hrule}
\noalign{\hrule}
height1pt&\omit&   &\omit&\cr
&(3,0,3)&&    85212 &\cr
height1pt&\omit&   &\omit&\cr
\noalign{\hrule}
\noalign{\hrule}
height1pt&\omit&   &\omit&\cr
&(3,1,1)&&    17580 &\cr
height1pt&\omit&   &\omit&\cr
\noalign{\hrule}
\noalign{\hrule}
height1pt&\omit&   &\omit&\cr
&(3,1,2)&&     48024 &\cr
height1pt&\omit&   &\omit&\cr
\noalign{\hrule}
\noalign{\hrule}
height1pt&\omit&   &\omit&\cr
&(3,1,3)&&     48024 &\cr
height1pt&\omit&   &\omit&\cr
\noalign{\hrule}
\noalign{\hrule}
height1pt&\omit&   &\omit&\cr
&(3,2,2)&&     991536 &\cr
height1pt&\omit&   &\omit&\cr
\noalign{\hrule}}
\hrule}
\,\,
\vbox{\offinterlineskip
\halign{ &\vrule# & \strut\quad\hfil$#$\quad\cr
\noalign{\hrule}
\noalign{\hrule}
height1pt&\omit&   &\omit&\cr
&(4,0,1)&&     -432 &\cr
height1pt&\omit&   &\omit&\cr
\noalign{\hrule}
\noalign{\hrule}
height1pt&\omit&   &\omit&\cr
&(4,0,2)&&      -4212 &\cr
height1pt&\omit&   &\omit&\cr
\noalign{\hrule}
\noalign{\hrule}
height1pt&\omit&   &\omit&\cr
&(4,0,3)&&     -432   &\cr
height1pt&\omit&   &\omit&\cr
\noalign{\hrule}
\noalign{\hrule}
height1pt&\omit&   &\omit&\cr
&(4,1,1)&&    -464296 &\cr
height1pt&\omit&   &\omit&\cr
\noalign{\hrule}
\noalign{\hrule}
height1pt&\omit&   &\omit&\cr
&(4,1,2)&&    -330300 &\cr
height1pt&\omit&   &\omit&\cr
\noalign{\hrule}
\noalign{\hrule}
height1pt&\omit&   &\omit&\cr
&(5,0,1)&&    -4212   &\cr
height1pt&\omit&   &\omit&\cr
\noalign{\hrule}\noalign{\hrule}
height1pt&\omit&   &\omit&\cr
&(5,0,2)&&    -191484 &\cr
height1pt&\omit&   &\omit&\cr
\noalign{\hrule}
\noalign{\hrule}
height1pt&\omit&   &\omit&\cr
&(5,1,1)&&    -46008 &\cr
height1pt&\omit&   &\omit&\cr
\noalign{\hrule}
\noalign{\hrule}
height1pt&\omit&   &\omit&\cr
&(6,0,1)&&    -35820   &\cr
height1pt&\omit&   &\omit&\cr
\noalign{\hrule}
\noalign{\hrule}}
\hrule}
$$

We finally evaluate the invariants of the elliptic curves
for the hypersurface and the complete intersections in one
singular projective space discussed in (5 $viii$). The topological
data can be found in the table of section (5 $viii$) and the
discriminant may be read of from \uuu :
$$
\vbox{\offinterlineskip\tabskip=0pt
\halign{\strut\vrule#
&\hfil ~~$#$~~
&\hfil $#$~~
&\hfil $#$~~
&\vrule#
&\hfil ~~$#$~~
&\hfil $#$~~
&\hfil $#$~~
&\hfil $#$~~
&\vrule#\cr
\noalign{\hrule}
&{\rm Model}
   &r_i                       &s_i       && n^e_{i,j} & j=0 & j=1 &j=2&\cr
\noalign{\hrule}
&A:&r_0=-{1\over 6}           &s_1=-{17\over 6}&&    &     &n^e_{i,j}=0,\,i<3 &
 &\cr
&  &r_1=-{5\over 6}           &s_2=-3    &&i=3 & -1280&2560   &2560   &\cr
&  &                          &          &&i=4 & -317864&1047280&15948240&\cr
\noalign{\hrule}
&B:&r_0=-{1\over 6}&s_1=-6     &&    & &n^e_{i,j}=0,\,i<2 &  &\cr
&  &r_1=-1                   &s_2=-3     &&i=2 &-16 &48   &-16 &\cr
&  &                         &           &&i=3 &-5364& 18972 &237897&\cr
\noalign{\hrule}
&C:&r_0=-{1\over 6}          &s_1=-{19\over 6}&&   & &n^e_{i,j}=0,\,i<4 &
&\cr
&  &r_1=-{7\over 6}          &s_2=-3     &&i=4 &-280  &1120   & 13072   &\cr
&  &                         &           &&i=5 &-20992&119808 & 429608  &\cr
\noalign{\hrule}}
\hrule}$$
The invariants for case $A$ are in accordance with \cdfkm.
Also for the complete intersections we obtain for the
invariants $n^e_{i,j}=0$ for $j>i$ as well as the symmetry
$n^e_{i,j}=n^e_{i,j-i}$. Comparison with
the models $\cicy{-176}(\IP^5|4,2)1$, $\cicy{-144}(\IP^6|3,2,2)1$
and $\cicy{-120}(\IP^7|2,2,2,2)1$ reveals the expected relation
$n^e_i=\sum_{j} n^e_{i,j}$, which is an affirmative consistency
check of the calculations at the one-loop level.

In all our examples we observed that the holomorphic anomaly
is of index $r_0=-1/6$ at the general component of the
discriminant. This also holds true for examples with
higher-dimensional moduli spaces and for models with a
somewhat different type of singularity such as
$\cicy{-540}(\IP^4[9,6,1,1,1]|18)2$ or
$\cicy{-480}(\IP^4[12,8,2,1,1]|24)3$. This is  related to the
fact that the manifolds approach always a nodal configuration
along this component of the discriminant. The exponent of the
holomorphic ambiguity seems to be universal for this type
of singularity.

%%%%%%%%%%%%%%%%%%%%%%%%%%%%%%%%%%%%%%%%%%%%%%%%%%%
\newsec{Discussion}
%%%%%%%%%%%%%%%%%%%%%%%%%%%%%%%%%%%%%%%%%%%%%%%%%%%

To extend the discussion of mirror symmetry to CICY manifolds
with higher-dimensional moduli spaces, we have described how to set
up the Picard-Fuchs equations and specified the point of maximal
unipotent monodromy. We have developed a convenient way to
construct all its solutions around this point by showing the
equivalence of the solutions with the elements of special
representatives of the ring ${\cal R}$. In fact, the top element
of $\cal R$ corresponds to the cubic monomials of
intersection numbers for the generating elements of $H^2(\hat X,\ZZ)$.
We have found very simple formulas for the instanton-corrected
intersection numbers from which the number of rational curves
can be obtained, always assuming that the mirror symmetry is correct.
In this paper we have focused on the region of the moduli space of
large K\"ahler and complex structure.
The extension to the whole moduli space requires
an analytical continuation of the periods. This is in principle
straightforward using an integral representation of
the solutions of the Picard-Fuchs equations.
Technically this is however rather
involved and was performed so far only for one modulus cases
\cdgp\ktone\font~ and for two types of two moduli models \cdfkm.

The form of the topological partition function is fixed by the
holomorphic anomaly equation up to a holomorphic function.
To specify the latter we had to analyse their asymptotic
behaviour at the singular locus of the Picard-Fuchs system.
In general we had to use the vanishing of the elliptic
curves of low degree to provide this information.

{}From a more technical point of view the method described here
requires, for the prediction of the instanton expansion in the
large K\"ahler structure limit, only the generators of the Mori
cone $l^{(i)}$ and the associated intersection numbers. Given these
data the expansion for the corrected Yukawa couplings and the
prepotential can be simply obtained via \icy~\prepot.
As it bypasses the evaluation of
the Yukawa couplings on the complex structure side
it is applicable to higher-dimensional moduli spaces,
where the evaluation of these complicated algebraic
expressions is extremely tedious. It should be clear by example
(5 {\it vii}) and (5 {\it viii}) that the data mentioned above
can be provided more generally for the moduli associated to algebraic
deformations in the general class of Calabi-Yau
manifolds representable as hypersurfaces or complete intersections
in toric varieties\foot{In fact, the methods can be easily
extended to evaluate three-point functions on higher-dimensional
Calabi-Yau spaces with higher dimensional moduli spaces
\ref\htku{S. Hosono, A. Klemm and S. Theisen: unpublished}.
Some results for one parameter families
were obtained in \ref\gpm{B. Greene, R. Plessner and
D. Morrison, {\it Mirror Manifolds in Higher Dimension},
CLNS-93/1253, IASSNS-HEP-94/2, YCTP-P31-92 (hep-th/9402119)}}.
It would be interesting to see if these data can also be
obtained for the twisted sectors of the
models in the above class and for those of the
3345 Landau-Ginzburg models with more then five fields
constructed in \ref\ksks{M. Kreuzer and H. Skarke, \npb388(1993)113;
A. Klemm and R. Schimmrigk, \npb411(1994)559}, which can not be
reduced to CY-threefolds as the examples in section 6 and others in \fkss,
but have only an interpretation as higher dimensional manifolds with $c_1>0$.

The calculation of the prepotential was discussed.
This problem can also be considered from the point
of view of topological field theory. In fact, in many
cases the prepotential can be obtained from the
axioms of topological field theory contained in the
Witten-Dijkgraaf-Verlinde-Verlinde equations
\ref\dubrovin{B. Dubrovin, Nucl. Phys. B379 (1992) 627}.
However for this approach the Calabi-Yau manifolds are a critical
case because here the operator algebra of topological
field theory is nilpotent, at least if we restrict ourselves
to the massless perturbations.
For the threefolds the information from the WDVV equations just
defines special geometry, but do not give further information on the
prepotential.

{}From the identification of the two hypersurfaces
$\cicy{-168}(\IP^4[2,2,2,1,1]|8)2\,\,$ and \hfill\break
$\cicy{-252}(\IP^4[6,2,2,1,1]|12)2$
with complete intersections, which was also confirmed by comparison with
the Gepner models, we are  taught to view the system $L_1,\ldots,L_k$,
which contains (at the point $z=0$) the information about
the ring ${\cal R}$, as the object of primary interest.

Going one step further back, one may see as the basic input the data of
a Riemann-Hilbert problem with sympletic integral representations
of the monodromy and singular points, where the solutions
are characterised by a graded ring whose homogeneous subspaces are
of a suitable type. In fact these data seem to encode all topological
data required for the classification of the homotopy type of families
of Calabi-Yau manifolds by the theorem of Wall and might lead to
a refined classification of $N=2$ vacua of string theory and Calabi-Yau
threefolds.

\vskip 5 mm
\noindent{\bf Acknowledgements :} We would like to thank
V. Batyrev, M. Berschadsky, B. Lian, J. Louis, S.S. Roan and A. Todorov
for discussions.

\vfill\eject

%%%%%%%%%%%%%%%%%%%%%%%%%%%%%%%%%%%%%%%%%%%%%%%%%%%%
%%%%%%%%%%%%%%%%%%%%%%%%%%%%%%%%%%%%%%%%%%%%%%%%%%%%
\appendix{A}{The pole structure in the coefficients
of the logarithmic solutions to the Picard-Fuchs equation}
\def\a{\alpha}
\def\b{\beta}
\def\c{\gamma}
\def\d{\delta}
\def\la{l^{(\a)}}
\def\lb{l^{(\b)}}
\def\lc{l^{(\c)}}
\def\ld{l^{(\d)}}
\def\p{\partial}
\def\{{\lbrace}
\def\}{\rbrace}
\def\({\lbrack}
\def\){\rbrack}
\overfullrule0pt

In this appendix we exhibit the pole structures of the
logarithmic solutions of the Picard-Fuchs equations.

The starting point are the generators of the Mori cone
$\vec\la$, which are all of the form\foot{We have changed the sign of the
components $\la_{0j}$ as compared
to \linearrelations~for notational convenience.}
$$
\eqalign{
\vec\la=(-\{\la_{0j}\},\{\la_i\})\,\quad
&j=1,\dots,{\rm number\,of\,polynomials}\cr
&\a=1,\dots,h^{1,1}}
$$
where
$$
\eqalign{
-\sum_j\la_{0j}&+\sum_i\la_i=0\cr
&\la_i\in{\bf Z}\cr
&\la_{0j}\in{\bf Z}_{\geq}\, .}
$$
The fundamental period is
$$
w_0(z)=\sum_n c(n) z^n,
$$
where the sum is over all non-negative integers $n_\a$ and the expansion
coefficients are
$$
c(n)={\prod_j(\sum_\a\la_{0j}n_\a)!\over\prod_i(\sum_\a\la_i n_\a)!}
$$
We then define
$$
w_0(z,\rho)=\sum_n c(n,\rho)z^{n+\rho}
$$
with
$$
c(n,\rho)={\prod_j\Gamma(\sum_\a\la_{0j}(n_\a+\rho_\a)+1)\over
\prod_i\Gamma(\sum_\a\la_i(n_\a+\rho_\a)+1)}
$$
The logarithmic solutions contain the coefficients
$$
\eqalign{({\rm I})
\quad  \p_\b c(n)&\equiv{\p\over\p_\b}c(n+\rho)\Big|_{\rho=0}\cr
({\rm II}) \quad \p_\b\p_\c c(n)\,
\qquad&{\rm and}\,\qquad ({\rm III})\quad\p_\b\p_\c\p_\d c(n)}
$$
The following definitions will become useful:
$$\eqalign
{A^{(\b)}(n)&=\sum_j\lb_{0j}\psi(\sum_\a\la_{0j}n_\a+1)
             -\sum_i\lb_i\psi(\sum_\a\la_i n_\a+1)\cr
B^{(\b\c)}(n)&=\sum_j\lb_{0j}\lc_{0j}\psi'(\sum_\a\la_{0j}n_\a+1)
             -\sum_i\lb_i\lc_i\psi'(\sum_\a\la_i n_\a+1)\cr
C^{(\b\c\d)}(n)&=\sum_j\lb_{0j}\lc_{0j}\ld_{0j}\psi''(\sum_\a\la_{0j}n_\a+1)
             -\sum_i\lb_i\lc_i\lc_i\psi''(\sum_\a\la_i n_\a+1),}
$$
where $\psi$ is the logarithmic derivative of the gamma function.
We then get the expressions
$$\eqalignno{
\p_\b c(n)&=c(n) A^{(\b)}(n)&(1)\cr
\p_\b\p_\c c(n)&=c(n)\Bigl\{A^{(\b)}(n)A^{(\c)}(n)+B^{(\b\c)}(n)\Bigr\}&(2)\cr
\p_\b\p_\c\p_\d c(n)&=c(n)\Bigl\{A^{(\b)}(n)A^{(\c)}(n)A^{(\d)}(n)
+A^{(\b)}(n)B^{(\c\d)}(n)+A^{(\c)}(n)B^{(\d\b)}(n)&\cr
&\qquad\qquad\qquad\qquad +A^{(\d)}(n)B^{(\b\c)}(n)+C^{(\b\c\d)}(n)\Bigr\}&(3)}
$$
whose pole structures we have to examine.
Before doing this we define
$$
A^{(\b)}_{k_1,\dots,k_p}(n)=\sum_j\lb_{0j}\psi
\left(\sum_\a\la_{0j}n_\a+1\right)
-\sum_{i\neq k_1,\dots,k_p}\lb_i\psi\left(\sum_\a\la_i n_\a+1\right)
$$
and l
$C^{(\b\c\d)}_{k_1,\dots,k_p}(n)$.
For
$\sum_\a\la_k n_\a<0$ we set
$$
\sum_\a\la_k n_\a\equiv -m_k
$$
and
$$\eqalign{
\tilde A^{(\b)}_{k_1,\dots,k_p}&= A^{(\b)}_{k_1,\dots,k_p}
-\sum_{i=1}^p\lb_{k_i}\psi(m_{k_i})\cr
\tilde B^{(\b\c)}_{k_1,\dots,k_p}&= B^{(\b\c)}_{k_1,\dots,k_p}
-\sum_{i=1}^p\lb_{k_i}\lc_{k_i}\bigl(\pi^2-\psi'(m_{k_i})\bigr).}
$$
The last ingredients we need are the pole structures of $\Gamma$ and $\psi$
for $m\in{\bf Z}_>$:
$$\eqalign{
\Gamma(1-m)&={1\over\Gamma(m)}{\pi\over\sin(\pi m)}\cr
\psi(1-m)&=\psi(m)+\pi\cot(\pi m)\cr
\psi'(1-m)&=-\psi'(m)+\pi^2(1+\cot^2(\pi m))\cr
\psi''(1-m)&=\psi''(m)+2\pi^3\cot(\pi m)+2\pi^3\cot^3(\pi m)}
$$
We now consider the cases (I), (II) and (III) in turn:

\item{(I)} If we have more than one $-m_k<0$, then $c(n)$ has a double
zero.  However $A^{(\b)}$ only has a simple pole, i.e. we have to consider
only the case where $-m_k<0$ for one $k$ only. Then
$$\eqalign{
c(n)&={\prod_j(\sum_\a\la_{0j}n_\a)!\over\prod_{i\neq k}(\sum_\a\la_i n_\a)!}
(m_k-1)!{1\over\pi}\sin(\pi m_k)\cr
A^{(\b)}(n)&=-\lb_k\psi(1-m_k)+{\rm finite}\cr
&=-\pi\lb_k\cot(\pi m_k)+{\rm finite}}
$$
and we get
$$
\p_\b c(n)=-(-)^{m_k}\lb_k{\prod_j(\sum_\a\la_{0j}n_\a)!(m_k-1)!
\over\prod_{i\neq k}(\sum_\a\la_i n_\a)!}\,\qquad m_k\in{\bf Z}_>
$$

\item{(II)} Since $A^2$ and $B$ have double poles, we have to distinguish
two possibilities:
\itemitem{(i)} $-m_k<0$ for one $k$ only
\itemitem{(ii)} $-m_k<0$ for $k_1$ and $k_2$

\item{} For the two cases we find:
\itemitem{(i)}
$$
\p_\b\p_\c c(n)=-(-)^{m_k}{\prod_j(\sum_\a\la_{0j}n_\a)!(m_k-1)!
\over\prod_{i\neq k}(\sum_\a\la_i n_\a)!}
(\tilde A_k^{(\b)}\lc_k+\tilde A_k^{(\c)}\lb_k)
$$
\itemitem{(ii)}
$$
\p_\b\p_\c c(n)=(-)^{m_{k_1}+m_{k_2}}{\prod_j(\sum_\a\la_{0j}n_\a)!
(m_{k_1}-1)!(m_{k_2}-1)!
\over\prod_{i\neq k_1,k_2}(\sum_\a\la_i n_\a)!}
(\lb_{k_1}\lc_{k_2}+\lc_{k_1}\lb_{k_2})
$$

\item{(III)} We now have to distinguish three cases:
\itemitem{(i)} $-m_k<0$ for one $k$ only
\itemitem{(ii)} $-m_k<0$ for $k_1$ and $k_2$
\itemitem{(iii)} $-m_k<0$ for $k_1,\,k_2$ and $k_3$

\item{} For these three cases we find:
\itemitem{(i)}
$$\eqalign{
\p_\b\p_\c\p_\d c(n)&=-(-)^{m_k}{\prod_j(\sum_\a\la_{0j}n_\a)!(m_k-1)!
\over\prod_{i\neq k}(\sum_\a\la_i n_\a)!}\cr
&\quad\times\Bigl\{2\pi^2\lb_k\lc_k\ld_k
+\bigl\(\tilde B_k^{(\b\c)}\ld_k+\tilde B_k^{(\c\d)}\lc_k+
\tilde B_k^{(\d\b)}\lc_k\bigr\)\cr
&\qquad\quad+\bigl\(\tilde A_k^{(\b)}\tilde A_k^{(\c)}\ld_k
+\tilde A_k^{(\c)}\tilde A_k^{(\d)}\lb_k
+\tilde A_k^{(\d)}\tilde A_k^{(\b)}\lc_k\bigr\)\Bigr\}}
$$
\itemitem{(ii)}
$$\eqalign{
\p_\b\p_\c\p_\d c(n)&=(-)^{m_{k_1}+m_{k_2}}
{\prod_j(\sum_\a\la_{0j}n_\a)!(m_{k_1}-1)!(m_{k_{2}}-1)!
\over\prod_{i\neq k_1,k_2}(\sum_\a\la_i n_\a)!}\cr
&\quad\times\Bigl\{
\tilde A^{(\b)}_{k_1,k_2}(\lc_{k_1}\ld_{k_2}+\ld_{k_1}\lc_{k_2})
+\tilde A^{(\c)}_{k_1,k_2}(\ld_{k_1}\lb_{k_2}+\lb_{k_1}\ld_{k_2})\cr
&\qquad\qquad\qquad\qquad\qquad
+\tilde A^{(\d)}_{k_1,k_2}(\lb_{k_1}\lc_{k_2}+\lc_{k_1}\lb_{k_2})
\Bigr\}}
$$
\itemitem{(iii)}
$$\eqalign{
\!\!\!\!\p_\b\p_\c\p_\d c(n)&=-(-)^{m_{k_1}+m_{k_2}+m_{k_3}}
{\prod_j(\sum_\a\la_{0j}n_\a)!(m_{k_1}-1)!(m_{k_{2}}-1)!(m_{k_3}-1)!
\over\prod_{i\neq k_1,k_2,k_3}(\sum_\a\la_i n_\a)!}\cr
&\qquad\quad\times\Bigl\{
\lb_{k_1}\lc_{k_2}\ld_{k_3}
+\lc_{k_1}\ld_{k_2}\lb_{k_3}
+\ld_{k_1}\lb_{k_2}\lc_{k_3}\cr
&\qquad\qquad\qquad+\lc_{k_1}\lb_{k_2}\ld_{k_3}
+\lb_{k_1}\ld_{k_2}\lc_{k_3}
+\ld_{k_1}\lc_{k_2}\lb_{k_3}
\Bigr\}.}
$$

\vfill\break

%%%%%%%%%%%%%%%%%%%%%%%%%%%%%%%%%%%%%%%%%%%%%%%%%%%%%
%%%%%%%%%%%%%%%%%%%%%%%%%%%%%%%%%%%%%%%%%%%%%%%%%%%%%
\appendix{B}{Predicted numbers of lines for complete
intersections in $\displaystyle{\IP^3\times\IP^3}$ and
$\displaystyle{\IP^3\times \IP^2}$}
%%%%%%%%%%%%%%%%%%%%%%%%%%%%%%%%%%%%%%%%%%%%%%%%%%%%%

For complete intersections in $\IP^3\times \IP^3$,
\eqn\TY{
\cicy{}({\IP}^3  \cr {\IP}^3 | s_1 \; s_2 \; s_3 \cr
                                    t_1 \; t_2 \; t_3 \cr ){}\;,  }
there are eight possible (non-trivial) configurations:
$(s_1,s_2,s_3|\,t_1,t_2,t_3)=
(3,0,1|0,3,1)$,
$(0,2,2|2,2,0)$,
$(2,1,1|2,1,1)$,
$(2,1,1|1,2,1)$,
$(2,1,1|0,3,1)$,
$(2,1,1|0,2,2)$,
$(3,1,0|0,2,2)$,
$(2,2,0|0,0,4)$.
For all of these, the predicted numbers of lines with bi-degree
$(n_1,n_2)$ with respect to the K\"ahler forms $J_1$ and $J_2$
from $\IP^3$ and $\IP^3$ are given generally by
\eqn\TYNO{
\eqalign{
N(1,0)
&
=
10 { t_1} { t_2} { t_3} { s_1}! { s_2}! { s_3}!
\cr&
-
{ s_1}!\,{ s_2}!\,{ s_3}!
\biggl\{
{1\over 2}\left(
{ s_2} { s_3} { t_1}^3
+ 3 {  s_1} {  s_3} {  t_1}^2 {  t_2}
+ 3 {  s_1} {  s_2} {  t_1}^2 {  t_3}
+ 3 {  s_1}^2 {  t_1} {  t_2} {  t_3}
\right)
   \left( \sum_{r = 1}^{{  s_1}}{1 \over r^{2}} \right)
\cr&
+ 4 \left(
{  s_3} {  t_1}^2 {  t_2}
+ {  s_2} {  t_1}^2 {  t_3}
+ 2 {  s_1} {  t_1} {  t_2} {  t_3}
\right)
\left( \sum_{r = 1}^{{  s_1}}{1\over r} \right)
\cr&
-
{1 \over 2} \left(
{  s_2} {  s_3} {  t_1}^3
+ 3 {  s_1} {  s_3} {  t_1}^2 {  t_2}
+ 3 {  s_1} {  s_2} {  t_1}^2 {  t_3}
+ 3 {  s_1}^2 {  t_1} {  t_2} {  t_3}
\right)
    \left( \sum_{r = 1}^{ s_1} {1\over r} \right)^2
%%%
\cr&
-
\left(
 2 {  s_2} {  s_3} {  t_1}^2 {  t_2}
+ 2 {  s_1} {  s_3} {  t_1} {  t_2}^2
+   {  s_2}^2 {  t_1}^2 {  t_3}
+ 4 {  s_1} {  s_2} {  t_1} {  t_2} {  t_3}
+   {  s_1}^2 {  t_2}^2 {  t_3}
\right)
   \left( \sum_{r = 1}^{{  s_1}}{1\over r} \right)
    \left( \sum_{r = 1}^{{  s_2}}{1\over r} \right)
%%%
\cr&
+({\rm cyclic}\;{\rm permutations:}\;\;
   ((s_1,t_1)\rightarrow (s_2,t_2) \rightarrow (s_3,t_3))\; ) \;\;
\biggr\} \;,
%\cr
%&\cr
%N(0,1)&=N(1,0)\bigm|_{(\; s_i \leftrightarrow t_i \;)}
\cr}  }
and by $N(0,1)=N(1,0)\bigm|_{(\; s_i \leftrightarrow t_i \;)}$.

The predictions for seven possible non-trivial examples in
$\IP^3\times \IP^2$,
$(s_1,s_2|\,t_1,t_2)=
(2,2|0,3)$,
$(3,1|0,3)$,
$(2,2|1,2)$,
$(4,0|1,2)$,
$(1,3|2,1)$,
$(3,1|2,1)$,
$(0,4|3,0)$, can be extracted from the above general formulas via
the manifold identity,
\eqn\mfdid{
\cicy{}({\IP}^3  \cr {\IP}^2 | s_1 \;  s_2 \cr
                                    t_1 \; t_2 \cr ){}
\cong
\cicy{}({\IP}^3  \cr {\IP}^3 | s_1 \; s_2 \; 0 \cr
                                    t_1 \; t_2 \; 1 \cr ){}\; .
}

Further reduction simply reproduces the result for the
bicubic model in $\IP^2\times\IP^2$ and $(4|2)$ in $\IP^3\times\IP^1$
which are treated in section 7.

\listrefs
\bye